\documentclass[a4paper,11pt]{article}
\pdfoutput=1 
\usepackage{jcappub}
\usepackage{amsmath}
\usepackage{amssymb}
\usepackage{rotating}
\usepackage{tabularx}
\usepackage{subcaption}
\usepackage{hyperref}

\def\({\left(}
\def\){\right)}
\def\[{\left[}
\def\]{\right]}
\def\be{\begin{equation}}
\def\ee{\end{equation}}
\def\beq{\begin{eqnarray}}
\def\eeq{\end{eqnarray}}

\title{\boldmath Morphology of 21cm brightness temperature during the Epoch of Reioinization using Contour Minkowski Tensor}
\author[a,b]{Akanksha Kapahtia,} 
\author[a]{Pravabati Chingangbam,}
\author[c]{Stephen Appleby}

\affiliation[a]{Indian Institute of Astrophysics, Koramangala II Block,
	Bangalore  560 034, India}
\affiliation[b]{Joint Astronomy Program, Department of Physics, Indian Institute of Science, C. V. Raman Ave., Bangalore  560 012, India} 
\affiliation[c]{Korea Institute for Advanced Study, 85 Hoegiro, Dongdaemun-gu,  
	Seoul 02455, Korea}

\emailAdd{akanksha.kapahtia@iiap.res.in}
\emailAdd{prava@iiap.res.in}

\abstract{

We use morphological descriptors, Betti numbers and Contour Minkowski Tensor (CMT) on 21cm brightness temperature excursion sets, to study the ionization and heating history of the intergalactic medium (IGM) during  and before the Epoch of Reionization (EoR). The ratio of eigenvalues of the CMT denoted by $\beta$, gives shape information while it's trace gives the contour length of holes and connected regions. We simulate the matter density, neutral hydrogen fraction, spin temperature and brightness temperature field using the publicly available code 21cmFAST in a redshift range of $z=20.22$ to $z=6$. We study the redshift evolution of three quantities - the Betti number counts $N_{con,hole}$, the characteristic size $r^{ch}_{con,hole}$ and shape anisotropy parameter $\beta^{ch}_{con,hole}$ of connected regions and holes for these fields and investigate the different physical origins of their evolution. We make a qualitative comparison of different models of heating and ionization during the EoR. We obtain different regimes of morphological evolution of brightness temperature, depending upon how the shapes and sizes of connected regions and holes change with redshift for different astrophysical settings affecting the ionization and heating history of the IGM during and before the EoR. We find that the morphology of the brightness temperature field traces the morphology of ionized regions below a certain redshift value depending upon the model, where $\Delta r^{ch}_{hole}<10 \%$ and $\Delta \beta^{ch}_{hole}<1 \%$  relative to the $x_{HI}$ field. This difference decreases with redshift. Therefore, the ionization history of the IGM can be reconstructed using the morphological description of $\delta T_b$ in real space.
}

\begin{document}
	
	\maketitle
	\flushbottom
\section{Introduction} 
The baryonic component of the universe post recombination was dominated by neutral hydrogen, which governed the formation of the first luminous sources. The emission from these first sources of radiation, ionized the neutral hydrogen in the intervening medium to mark an important epoch in the history of the universe called the Epoch of Reionization (EoR). The epoch is characterized by the appearance of ionized regions around luminous sources, which gradually grow in size and merge until the entire universe is ionized. Observations of high redshift quasars using Ly $\alpha$ absorption constrain the end of reionization to $z \simeq 6$  \cite{Fan:2006dp}. Recent results from the Planck mission \cite{Aghanim:2018eyx} give the average redshift of EoR to be $z_{re} \sim 7 ~$to $8$, obtained from the measurements of optical depth to the last scattering surface for a parameterized ionization history. One promising observational probe of the EoR is the brightness temperature of the redshifted 21cm signal from neutral hydrogen atoms. The 21cm wavelength redshifts to the range of frequencies accessible by radio telescopes and is measured in terms of the sky averaged global signal or power spectrum of the fluctuations of the brightness temperature. The global signal was recently claimed to have been detected by the EDGES experiment \cite{Bowman:2018} while the 21cm power spectrum is being studied by various intereferometers such as PAPER, MWA, LOFAR, SKA-low and GMRT ~\cite{Parsons,Tingay et al.(2013),van Haarlem,Maartens et al.(2015),Paciga:2011}. The observations of the 21cm signal is challenging due to the fact that the signal is weak compared to noise and foreground levels.

Analyses of cosmological fields in real space are alternative to methods that are employed in Fourier space such as the power spectrum. 
Real space methods can probe the morphology of cosmological fields by analyzing the geometry and topology of their level sets. Examples of such methods are Minkowski Functionals (MFs) ~\cite{Tomita:1986,Gott:1990,Mecke:1994,Schmalzing:1997,Schmalzing:1998} which have been widely used in cosmology (see e.g. ~\cite{Ade:2015ava,Chingangbam:2012wp,Vidhya:2014,Chingangbam:2017sap, Buchert:2017uup}). Closely related to the MFs are the Betti numbers which are number counts of holes or connected regions. They have also been used for cosmological analysis~\cite{Chingangbam:2012,Park:2013dga,Pranav:2018pnu}. In the context of the EoR MFs have been used to study the redshift evolution of the processes of reionization ~\cite{Lee:2007dt,Friedrich:2010nq,Ahn:2010hg,Wang:2015dna,Gleser:2006su,Yoshiura:2015}. 
Related methods are based on percolation theory~\cite{Bag:2018zon,Bag:2018fyr,Furlanetto:2016}. 
MFs are scalars quantities, and hence insensitive to direction information. Their tensorial generalization, called Minkowski Tensors~\cite{McMullen:1997,Alesker:1999,Hug:2008},  provide richer morphological information regarding the shape and relative orientation of structures~\cite{Schroder2D:2009,Schroder3D:2013}. They  have been recently applied to cosmological data~\cite{Vidhya:2016,Chingangbam:2017, Appleby:2017uvb,Appleby:2018tzk,Joby:2018}. They have previously been used for analysis of galaxy morphology~\cite{Beisbart:2002,Rahman:2003,Rahman:2004}.

One way to study the ionization history of the EoR is to study the morphology of ionized regions. The radiation from the luminous sources ionizes their surrounding region. At the beginning stages of reionization many such ionized bubbles appear and grow in size until they start merging. Therefore as reionization progresses, the shape and size of these ionized regions also changes. The details of the redshift  evolution of the morphology of the fields is, moreover, sensitive to the physics of reionization and hence their precise determination can potentially be used to discriminate between different models. This paper is the second of a series of papers that aims to compehensively investigate the application of MTs to understanding the physics of the EoR and constraining models using future observational data.  
In the first paper~\cite{Kapahtia:2017qrg} the authors used the rank-2 {\em Contour Minkowski Tensor} (CMT), which is the 
tensor generalization of the second scalar MF, the total contour length, in conjunction with Betti numbers, to track the redshift evolution of the morphological properties of the ionization field. It was shown that the characteristic length and time scales associated with the evolution can be decoded from the behaviour of the morphological variables derived from the CMT and the Betti numbers.

The goal of this paper is to trace the history of the IGM during EoR for different astrophysical scenarios and to demonstrate how they can be discriminated by using the method developed in~\cite{Kapahtia:2017qrg}.  We track the morphological properties encoded in the Betti numbers and CMT of the fluctuations in the density, ionization and spin temperature fields. We qualitatively analyse the redshift evolution of the morphology of these fields and show how their evolution is traced by the brightness temperature field as different signatures in the redshift evolution of it's CMT and Betti numbers. We identify three regimes in the redshift evolution of Betti numbers and CMT of the brightness temperature field which encode the underlying evolution of the IGM in terms of it's heating and ionization history. We also show how different astrophysical scenarios leading to different IGM histories result in a shift of these regimes. Our analysis uses simulations of the EoR, obtained using the publicly available code  {\texttt{21cmFAST}}~\cite{Mesinger:2010ne}. 

The paper is organized as follows. In Sec.~2 we describe the brightness temperature field, review {\texttt{21cmFAST}} and describe our choice of astrophysical models. In Sec.~3 we give a description of the mathematical formalism of the CMT and Betti numbers, describe their numerical calculation and the quantities that are derived from them. In Sec.~4 to 6 we analyse simulated density, ionization and spin temperature fields. In Sec.~7 we describe how the morphology of brightness temperature encodes the morphological evolution of all other fields during EoR and hence the heating and ionization history of the IGM. We end with discussion of our results in Sec.~8.

\section{21-cm Brightness Temperature Simulation}
The hyperfine levels of neutral hydrogen in ground state have an energy difference which corresponds to an excitation temperature of $T_*= 0.068$~K or a frequency of 1420 MHz. The spin temperature $T_s$ determines the emission or absorption of this radiation. It is the temperature that describes the relative population of the two levels. The relative population of the two levels is given by the Boltzmann distribution, $n_{1}/n_{0}=3~$exp~$(-T_{*}/T_{s})$ when they are in equilibrium. Then the spin temperature, $T_s$ is the physical temperature of that equilibrium system. However, when the system is not in equilibrium for example in the case of neutral hydrogen clouds in the IGM, the spin temperature is calculated from the steady state population obtained by a balance between the radiation it emits and is incident on the the cloud  \cite{Field(1958)}.

 The redshifted frequency lies in the radio range of frequencies so that in the Rayleigh Jean's regime the intensity of this spectral line can be quantified by the brightness temperature $T_b$. This is observed as an offset from the CMB temperature and is called the differential brightness temperature $\delta T_{b}$. For an observed frequency $\nu$, corresponding to a redshift z and at a given point in space $x$ \cite{Furlanetto:2006jb}:

\begin{equation}
\delta T_b(\nu,x) \approx 27 \ x_{HI}(x)\left(1+\delta_{nl}(x)\right)\bigg( 1-\frac{T_\gamma(z)}{T_s(x)}\bigg) \nonumber \ \frac{\Omega_bh^2}{0.023}
    \bigg({\frac{1+z}{10}\ \frac{0.15}{\Omega_Mh^{2}}}\bigg)^{1/2} (\rm{mK})\\ \\
    \label{eqn:Tb} \\
\end{equation}

where $\delta_{nl}\equiv \rho/\overline{\rho} -1$ is the evolved density contrast, $H(z)$ is the Hubble parameter and $T_{\gamma}$ is the CMB (Cosmic Microwave Background) temperature. The spatial fluctuations in $T_{\gamma}$ and the effect of peculiar velocities are ignored in the above expression because the former is very small compared to those due to other fields while the peculiar velocities are very small in magnitude compared to the expansion rate of the universe at the redshifts of interest.

The spin temperature $T_s$ is related to other physical temperatures through the following expression  \cite{Field(1958),Pritchard:2012}:

\begin{equation}
T_s^{-1}=\frac{T_{\gamma}^{-1} + (x_{c} +x_{\alpha})~T_{k}^{-1}}{1+{x_c}+x_{\alpha}} \\ \\
\label{eqn:Ts} \\ \\ \\
\end{equation} 
where $T_{\gamma}$ is the CMB temperature and $T_k$ is the kinetic temperature of the gas. The coupling contants $x_c$ and $x_{\alpha}$ describe the coupling of spin temperature $T_s$ to $T_k$ due to collisions and due to Lyman-$\alpha$ transition respectively. The latter dominates at later stages once the first collapsed objects begin to appear and keeps $T_s$ coupled to $T_k$ thereafter. The evolution of $T_k$ with $z$ is dominated by adiabatic cooling due to expansion of the universe at early redshifts and is taken over by X-ray heating of the IGM due to the first sources of light. This results in a dip in the evolution of $\overline T_s$ (Sec.~\ref{sec:Ts}), which is coupled to the evolution of $T_k$ as elucidated above.
\subsection{Review of 21cmFAST}
\label{sec:21cmfast}
 We use the publicly available, semi-numerical code \texttt{21cmFAST v1.3} \cite{Mesinger:2010ne} to generate mock 21cm fields. The code generates Gaussian random initial density field and then evolves it using first order perturbation theory (Zel'Dovich approximation). It generates the density $\delta (\vec{x})$, spin temperature $~T_{s}(\vec{x})$, gradient of the peculiar velocity along the line of sight $d\vec{v}({\vec{x}})/dr$ and ionization field  $~x_{HI}(\vec{x})$ (neutral hydrogen fraction) at every grid point $\vec{x}$ and finally calculates the differential brightness temperature $\delta T_b (\vec{x})$ at that point at a redshift $z$.

In order to identify ionized regions the code uses an excursion set approach similar to the Press-Schechter theory of halo mass function and uses the following criteria for ionization:~\cite{Furlanetto:2004nh} 
\begin{equation}
\zeta f_{coll} (x,z,R)\geq 1
\label{eqn:ion}
\end{equation}

Here, $f_{coll}$ is the collapse fraction and it depends upon the minimum mass,  $M_{vir}$ required for a halo to virialize while $\zeta$ is the ionizing efficiency describing the number of ionizing photons per unit baryon, that escape a halo. The minimum virial mass can be expressed in terms of the minimum virial temperature $T_{vir}$ $\propto (M_{vir})^{2/3}$. We have chosen $\zeta$ to be a single number and ignored any detailed astrophysical modelling of the parameter. One can reconstruct similar ionization histories for multiple combinations of $T_{vir}$ and $\zeta$ i.e. the two parameters are degenerate. 

 A central pixel in 21cmFAST is flagged as ionized if the condition (\ref{eqn:ion}) is fulfilled at some filter scale while reducing from a maximum value $R_{max}$ to the pixel size in logarithmic steps. 
 The above prescription holds if one ignores the effect of inhomogenous recombination. If the effect of inhomogenous recombination is important then the prescription for ionization used in 21cmFAST is:
\begin{equation}
\zeta f_{coll} (x,z,R)\geq 1+\overline{n}_{rec}(x,z,R)
\label{rec}
\end{equation}
where $\overline{n}_{rec}(x,z,R)$ is the total number of recombinations and is modelled according to the cell's ionization history and density \cite{Sobacchi:2014rua}. It is averaged over the smoothing scale R corresponding to the step at which the condition is being checked.

The spin temperature $T_s$ is affected by various physical processes throughout its evolution ( See \cite{Field(1958)}, \cite{Pritchard:2012} and references therein). In our case we will work in a redshift regime where the first collapsed luminous objects have started to form. As Ly-$\alpha$ coupling dominates (i.e. $x_{\alpha} \gg x_c$), $T_s$ will couple to $T_k$. The coupling constant, $x_\alpha \propto J_{\alpha}(\textbf{x},z)$, which is the background flux of Ly-$\alpha$ at a given redshift. This background flux depends upon the redshift evolution of emissivity of sources contributing to Ly-$\alpha$ . This further depends upon the spectral model (in terms of the number of photons produced per Hz per stellar
baryon) for the collapsed objects and the rate of change of collapsed fraction at a given $z$ (See Eq.~25 of ~\cite{Mesinger:2010ne}). As Ly-$\alpha$ couples $T_s$ to $T_k$ in most of the IGM and saturates, X-ray heating starts to dominate. This increases $T_k$ which turns over from adiabatic cooling to X-ray heating regime. The heating due to X-rays in 21cmFAST, is described in terms of X-ray heating rate per unit baryon $\epsilon_{X} (\textbf{x},z')$ (see eq. 18 of \cite{Mesinger:2010ne}). It is proportional to the efficiency of X-ray emission $\zeta_{X}$, which is the number of X-ray photons per unit baryon in collapsed objects and redshift evolution of $f_{coll}$ (which depends upon $T_{vir}$), for a given value of the luminosity spectral index $\alpha$.

We generate the fields at 29 redshift values between $z=6$ and $z=20.22$, separated by a logarithmic interval of 1.0404 in (1+z). We carry out our analysis after smoothing the fields with a Gaussian smoothing kernel with scale $R_s=4.5$ Mpc. The important time and length scales of the EoR typically do
not depend on the smoothing scale, as shown in ~\cite{Kapahtia:2017qrg}. For a proper comparison with observed data, $R_s$ should be chosen based on the specifications of the instrument.
\subsection{Description of models of reionization}
\label{sec:model}
 In order to study the morphology of the epoch of reionization we have generated $\delta_{nl}$, $x_{HI}$, $T_s$ and $\delta T_b$ fields on a $512^3$ grid of a $(200 ~ \rm Mpc)^3$ box. This gives a pixel resolution of $\sim 0.4 ~ \rm Mpc$ on a side. The initial conditions were generated on a $1024^3$ grid at a redshift of $z=300$. Different parameter sets describe different ionization and heating scenarios which affect the fluctuations and global evolution of $x_{HI}$ and $T_s$ fields. It is to be noted that the evolution of the $\delta_{nl}$ field is only affected by the initial conditions and the cosmology adopted in 21cmFAST. 
 
 We choose a \textit{fiducial model} described by a fixed set of parameter values for $\zeta$, $T_{vir}$ and $\zeta_X$. We do not include inhomogenous recombination for the brightness temperature calculation of our \textit{fiducial model}.
 
 In order to compare different models, we change one or more of the parameters while keeping the others fixed, such that they describe a different astrophysical setting affecting one or more of the fields which determine the brightness temperature. This has been done to conveniently compare with the \textit{fiducial model} and easily extend to any complicated history. Our choice of models is as follows:
 \begin{itemize}
 	\item \textit{Fiducial model}: $\zeta=17.5$ , $\zeta_X=2\times10^{56}$, $T_{vir}=3\times10^4$ K and $\alpha=1.2$
 	\item \textit{Recombination}: Model with effect of inhomogenous recombination taken into account with the same fiducial set of parameters.
 	\item \textit{Model with less massive sources}: $\zeta=10.9$, $\zeta_X=2\times10^{56}$, $T_{vir}=1\times10^4$ K and $\alpha=1.2$
 	\item \textit{Model with more massive sources}: $\zeta=23.3$,  $\zeta_X=2\times10^{56}$, $T_{vir}=5\times10^3$ K and $\alpha=1.2$
 	\item \textit{Model with increased X-ray efficiency}: $\zeta=17.5$, $\zeta_X=1\times10^{57}$, $T_{vir}=3\times10^4$ K and $\alpha=1.2$
 \end{itemize}
 
 The models have been chosen to yield an end of reionization roughly at $z_e\sim6$ and to an optical depth to CMB $\tau_{re} \sim 0.05$ \cite{Aghanim:2018eyx}. We choose population 2 stars as the stellar population responsible for early heating. The models adopted in this work represent simplified, parameterized ionization histories. In actuality the efficiency of heating and ionization would depend upon finer details and evolution of the astrophysical objects during the epoch of reionization. However the parameterized models considered here do give a general picture of IGM history in terms of globally defined parameters. 
 
 The \textit{fiducial model} corresponds to $\tau_{re} \sim 0.054$. Including inhomogenous recombination delays the redshift at which reionization ends. Recombinations slow down the growth of ionized regions by depleting the number of photons available for ionizing. This depletion of photons is accounted for by $\overline{n}_{rec}$ in the criterion in Eq. (\ref{rec}) . The $\zeta$ values corresponding to $T_{vir}=1\times10^4 ~ K$ and $T_{vir}=5 \times 10^4 ~K$ give optical depth values of $\tau_{re} \sim 0.058$ and $\tau_{re} \sim 0.052$ respectively. The value $\zeta_X=2\times10^{56}$ for the \textit{fiducial model} and $\zeta_X=10^{57}$ correspond to 0.3 and 1 X-ray photon per baryon respectively. 
 For our analysis we chose the $\Lambda$CDM parameters as per Planck 2018 \cite{Aghanim:2018eyx}.
 
The choice of $T_{vir}$ determines the collapse fraction and hence would affect the $x_{HI}$ and $T_s$ evolution. $\zeta$ affects only the evolution of $x_{HI}$ field while $\zeta_X$ affects $T_s$ evolution and has very small effect on $x_{HI}$ which decreases at lower z values relevant to the EoR as X-rays contribute more to heating than to ionization there \cite{FurlanettoStov:2010}. The effect of inhomogenous recombination on $x_{HI}$ becomes prominent during late stages of reionization \cite{Sobacchi:2014rua}. 
 
\section{Contour Minkowski Tensor and Betti numbers}
\subsection{Definition}
For any field the set of all field values greater than or equal to a certain threshold, $\nu$ is called an {\it{excursion set}}. The boundary curves of these excursion sets in two dimensions enclose either a connected region (regions formed by bounded curves enclosing a set of values greater than or equal to $\nu$) or a hole (regions formed by bounded curves enclosing a set of values less than $\nu$). The number of connected regions, $n_{con}$ and holes, $n_{hole}$ at the threshold $\nu$ are called {\it{Betti numbers}} ~\cite{Chingangbam:2012,Park:2013dga}. Minkowski functionals describe the morphology and topology of excursion set regions for a given random field as a function of field threshold $\nu$. The morphology and number of these excursion set regions changes as $\nu$ is varied. For a gaussian random field the analytical forms for scalar Minkowski Functionals are known as a function of $\nu$ ~\cite{Tomita:1986}.
Minkowski Tensors~(MTs) are tensor generalization of the Scalar Minkowski Functionals. We will focus on the translation invariant symmetric rank two tensor, which we refer to as the {\em contour} MT (CMT), defined for a single boundary curve $C$  as:
\begin{equation}
\mathcal{W}_{1}=\int_C \hat{T}\otimes \hat{T} ~{\rm d}s
\label{eqn:W1}
\end{equation}
where $\hat{T}$ is the unit tangent vector at every point on the curve, $\otimes$ denotes the symmetric tensor product given by
\begin{equation}
\left(\hat{T}\otimes \hat{T}\right)_{ij} = \frac12\left( \hat{T}_i\hat{T}_j + \hat{T}_j\hat{T}_i\right),
\end{equation}
and ${\rm d}s$ is the infinitesimal arc length. Our notation follows \cite{Chingangbam:2017}  \footnote{Note that in \cite{Chingangbam:2017} we had defined $\mathcal{W}$ with a factor of 2 on the right hand side of Eq. \ref{eqn:W1} which was erroneously equated with the contour length. This resulted in an over-estimation of the size of structures in that paper by a factor of 2. There was also an error in the calculation of scale defined for the eigen values by a multiplicative factor of 0.4 Mpc. Therefore the $r^{ch}$ in that paper will have an overall multiplicative factor of 1.25.} where $\mathcal{W}_{1}$ is referred to as ${W}_2^{1,1}$ in \cite{Schroder2D:2009,Chingangbam:2017,Appleby:2017uvb}. $\mathbf{Tr}\left(\mathcal{W}_{1}\right)$ is two times the second scalar MF i.e. the total contour length denoted by $W_1$. 

Any anisotropy in the boundary curve will manifest as an inequality between the eigenvalues of the matrix ${\cal W}_{1}$. We define the eigenvalues in ascending order, $\lambda_1<\lambda_2$ and define the {\em shape anisotropy parameter} as $\beta\equiv\lambda_1/\lambda_2$. Hence for a generic curve, $\beta$ will have values between 0 and 1. The CMT also gives an estimate of the size of the area enclosed by a curve. If $\lambda\equiv \lambda_1+ \lambda_2$ denotes the perimeter of the closed curve and is equated to the circumference of a circle i.e. $2\pi r$, we determine $r$ to be
\begin{equation}
r\equiv\lambda/2\pi.
\end{equation}
$r$ will in general result in an overestimation of the size of the area enclosed by the curve due to the isoperimetric inequality. The overestimation will be larger for non-convex curves.
For a field, at a given threshold the excursion set will have many such boundary curves and the average value of $\beta$ gives the average shape of curves at that threshold. 
\subsection{Methodology}

 In order to carry out our analyses on the 3-D box, we subdivide it into 32 slices of thickness 6.25 Mpc each. We carry out our calculations on each 2-D slice. 
 Any field $u$ under consideration is redefined as:
$u\rightarrow \tilde{u} \equiv (u-\mu)/\sigma$, where $\mu$ is the mean and $\sigma$ is the standard deviation of $u$. This redefinition does not alter its geometrical and topological properties, but allows for a uniform choice of threshold values for different fields.

Since our field is dicretized into pixels, we shall use ${\cal{W}}_1$ for a polygon \cite{Schroder2D:2009} given by:
\begin{eqnarray}
({\cal{W}}_1)_{ij}= \sum_e |\vec{e}|^{-1} e_i e_j 
\end{eqnarray}
where $|\vec{e}|$ is the length of a two dimensional vector describing a discretized segment of the boundary curve between two vertices of the polygon.
The expression shows that the eigenvalues of the matrix will have the dimension of length. 
We now define the various quantities of interest. The threshold values $\nu$, of the standard normal field $\tilde u$ refers to the number of standard deviations of $u$ the field value is away from its mean $\mu$. 
 The suffix `con' ,`hole' and 'tot' refer to boundaries of connected regions, holes and total structures (i.e. both connected regions and holes), respectively. At each $\nu$ we denote the number of distinct curves enclosing connected regions, holes and total number of structures by $n_{\rm con}(\nu)$, $n_{\rm hole}(\nu)$ and $n_{\rm tot}(\nu)$ respectively. Then, at each redshift $z$ we define,
\begin{equation}
N_{\rm x}(z) \equiv \int_{\nu_{\rm low}}^{\nu_{\rm high}} {\rm d}\nu \,n_{\rm x}(\nu,z),
\label{eqn:Nx}
\end{equation}
where the suffix `x' denotes either `con', `hole' or 'tot'. We use sufficiently large sampling of the threshold range ($\tilde{u}=\nu$) from $\nu_{low}$ to $\nu_{high}$ so as to get convergent results. The number of thresholds used should enable one to sample very small peaks and shallow valleys  which do not vary much about the mean\footnote{ Such regions are encountered for the $x_{HI}$ and $\delta T_b$ field at $z$ values where reionization has just started} .

For well behaved smooth\footnote{Infinitely differentiable} random fields, $n_{\nu}$ goes to zero as  $\nu \to \pm\infty$. Therefore, the integral on the r.h.s of the above equation converges and $N_{\rm x}$ is finite when the cutoff thresholds are taken to $\pm \infty$. $N_{\rm x}(z)$ represents the ensemble of all curves within the chosen threshold range in the simulation box at a fixed redshift. We sample $\nu$ at a finite number of values and the integral is carried out using trapezoidal integration.

We reserve the symbols $\lambda_i$, $r$ and $\beta$ to denote the eigenvalues, characteristic radius and the ratio of the eigenvalues for a single curve. Let
\begin{eqnarray}
{\overline\lambda}_{i,\rm x}(\nu) &\equiv& \frac{\sum_{j=1}^{n_{\rm x}(\nu)} \lambda_{i,\rm x}(j)}{n_{\rm x}(\nu)}, \\
{\overline{r}}_{\rm x}(\nu) &\equiv& \frac{\sum_{j=1}^{n_{\rm x}(\nu)} r_{\rm x}(j)}{n_{\rm x}(\nu)}, \label{eqn:rx_nu}\\
\quad {\overline\beta}_{\rm x}(\nu) &\equiv&  \frac{\sum_{j=1}^{n_{\rm x}(\nu)} \beta_{\rm x}(j)}{n_{\rm x}(\nu)},
\label{eqn:betax_nu}
\end{eqnarray}
denote their averages over all curves at a given $\nu$ .In what follows all error bars correspond to the error in mean over the slices as $\sim \sqrt{\sigma^2 /32}$ where $\sigma^2$ is the variance of the statistics over the 32 slices. We use this measure of error rather than the standard deviation $\sigma$ because we are not comparing with actual observational data, in which case other systematics would also contribute to the uncertainity. Therefore our error bars quote the uncertainty in our reproduction of the mean value. Since our slices are extracted from the same cube, they are correlated and hence the size of error bars is marginally low. 
We define, 
\begin{eqnarray}
\lambda^{\rm ch}_{i,\rm x}(z) &\equiv& \frac{\int_{\nu_{\rm low}}^{\nu_{\rm high}} {\rm d}\nu \,n_{\rm x}(\nu,z) {\bar{\lambda}}_{i,\rm x}(\nu)}{N_{\rm x}(z)}, 
\label{eqn:lambdah} \\
r^{\rm ch}_{\rm x}(z) &\equiv& \frac{\int_{\nu_{\rm low}}^{\nu_{\rm high}} {\rm d}\nu \,n_{\rm x}(\nu,z) {\bar{r}}_{\rm x}(\nu)}{N_{\rm x}(z)}, \label{eqn:rch}\\
\beta^{\rm ch}_{\rm x}(z) &\equiv& \frac{\int_{\nu_{\rm low}}^{\nu_{\rm high}} {\rm d}\nu \,n_{\rm x}(\nu,z) {\bar{\beta}}_{\rm x}(\nu)}{N_{\rm x}(z)}.
\label{eqn:betach}
\end{eqnarray}
These integrals are convergent for the same reason as for $N_{\rm x}(z)$ and $\rm~'ch'$ respresents \textit{characteristic}. 
 Any difference in the morphology of two different models at a fixed redshift will manifest as a change in the area under the curves of the $\nu$ variation. Note that two different models of EoR  having different variation with threshold may have the same area under the curve of $\nu$ variation at a given redshift $z$. In that case the models should be compared by the threshold variation of the morphological quantities at a given redshift. However for the purpose of our analyses in this paper, we find that the redshift variation encapsulates any physical difference in the chosen models.
\subsection{Overview of morphology of Gaussian random fields}

Before we proceed to interpret the morphology of excursion sets for any given field we first provide a general description of how the excursion set changes when the threshold is varied. In principle one can identify three different regimes as the threshold is varied from the highest value and progressively lowered. Initially, there exists isolated small connected regions around the highest peaks of the field and their number would gradually increase as more peaks enter the excursion set when lowering the threshold. In the second regime as we further lower the threshold, some of these small connected regions merge thereby decreasing their number. Finally in the third regime, as the threshold is decreased further these connected regions all merge to form a single connected region with holes puncturing it. These holes eventually shrink in size and disappear as we go lower in threshold and finally a single connected region remains which spans the entire region over which the field is defined.

For Gaussian random fields the analytic expression for the total contour length of all boundary contours (connected regions and holes) is known and has a simple closed form expression given by $W_1=A e^{-\nu^2/2}$ \cite{Schmalzing:1997}, where the amplitude $A$ depends on the ratio of the variance of the gradient of the field to the variance of the field per unit area. However, closed form expressions for $n_{con,hole}$, $\bar{r}_{con,hole}$ and $\bar{\beta}_{con,hole}$, are not known. $n_{con,hole}$ has been calculated numerically in \cite{Chingangbam:2012}, while  $\bar{\beta}_{con,hole}$ has been studied extensively using numerical computation in \cite{Appleby:2017uvb}. $\bar{r}_{con,hole}$ has not been studied before. We can infer their behaviour at very high and positive and very low and negative thresholds. Since $n_{tot} \sim n_{con}$ for large positive thresholds, $\nu\gg 0$ we expect that in this regime $\bar{r}_{con} \propto e^{-\nu ^2/2}/n_{con}(\nu)$.  Similarly, $n_{tot} \sim n_{hole}$ for large negative thresholds, $\nu\ll 0$. We expect that  in that regime $\bar{r}_{hole} \propto e^{-\nu^2/2}/n_{hole}(\nu)$. At $\nu \gg 1$, $n_{con} \sim W_2$, where $W_2$ is the genus of the excursion set which is $\propto \nu e^{\nu^{-2}}$ \cite{Schmalzing:1997}. Therefore $n_{con} (\nu)\propto \nu e^{-\nu^2}$. Similarly $n_{hole} (\nu) \propto \nu e^{-\nu^2}$ at $\nu \ll1$. Therefore $\bar{r}_{con,hole} \propto \nu^{-1}$ at these thresholds. The units of scale will enter through the amplitude, which will be the ratio of the variance of the field to the variance of the gradient of the field.

As an initial test we quantify the morphology  of Gaussian random fields, as encapsulated in $n_{con,hole}$, $\bar r_{con,hole}$ and $\bar \beta_{con,hole}$. We simulate 100 realizations with input flat power spectrum on a $512 \times 512$ square pixel grid. Then we smooth the fields over 12 pixels and compute the morphological quantities using these simulations. Fig.~1 show the plots of $n_{con,hole}$ (top), $\bar r_{con,hole}$ (middle) and $\bar \beta_{con,hole}$ (bottom). All plots are averaged over 100 realizations. $n_{con}$ peaks at $\nu=1$, while $n_{hole}$ peaks at $\nu=-1$. 

We can see from the plots for  $\bar r_{con,hole}$ at threshold values where $n_{con,hole}$ are large, that the average size of the structures are small, and vice versa.
Further, $\bar r_{con,hole}$ have an artificial sharp drop at $|\nu|> 1$. This is because we use periodic boundary condition on our simulation box which generates large unphysical structures that do not have a boundary. In order to avoid this we have excluded regions having area $> 0.9$ times the area of the simulation box. 
The plots for  $\bar \beta_{con,hole}$ show that the average shape of the structures do not vary much across the threshold range (i.e., remains within a small range of around 0.65), except the few large connected regions at very high negative threshold values, and the few large holes at high positive threshold values which exhibit higher $\beta$ values.

The results obtained in this section will be useful as a benchmark for analyzing the behaviour of the morpology of the fields of the EoR at different redshifts in the subsequent sections.

\begin{figure*}
	\begin{center}
		
		\includegraphics[width=7.5 cm,height=8.8cm]{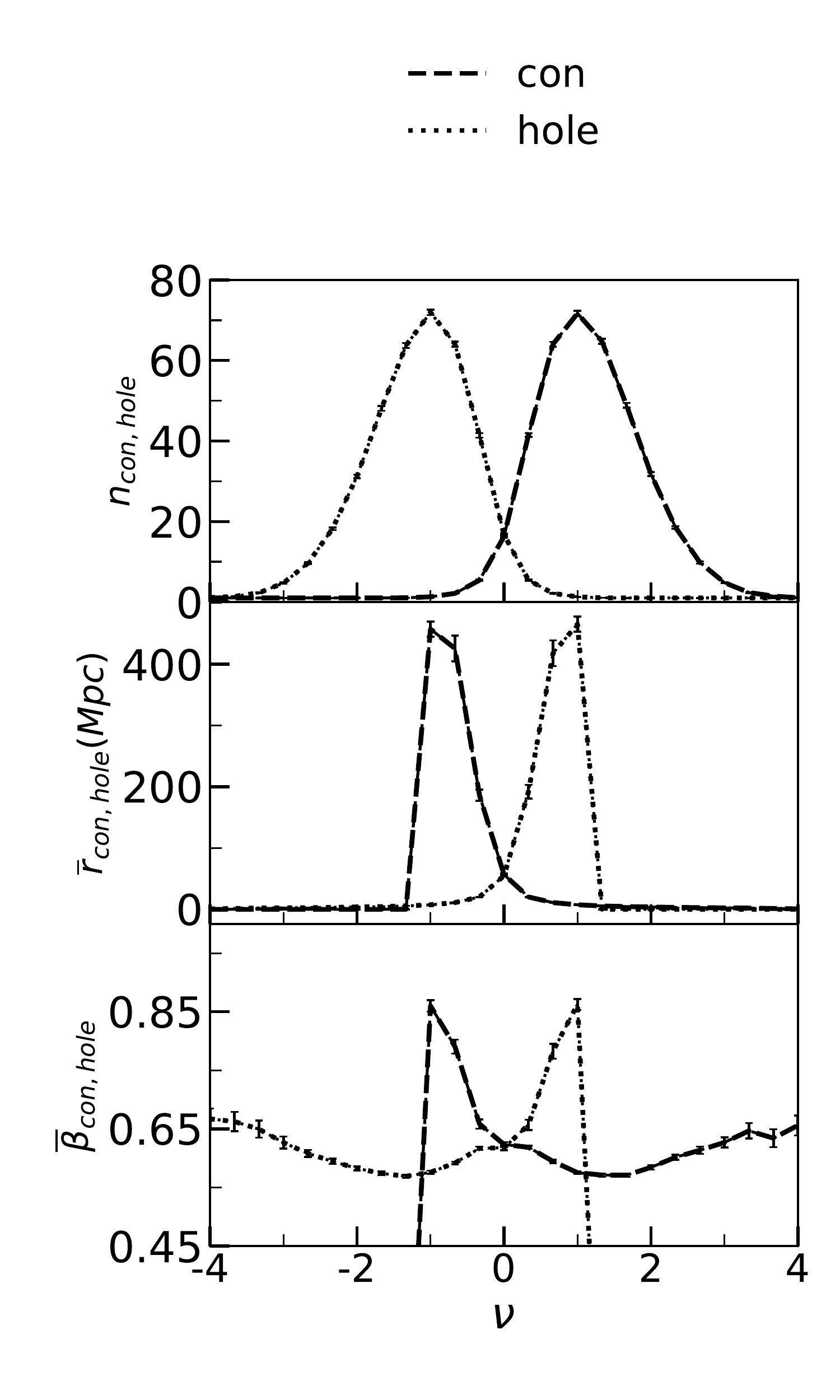}
	
	\end{center}
	
	\caption{Variation of $n_{con,hole}$, $\overline{r}_{con,hole}$ and $\overline{\beta}_{con,hole}$ with field threshold $\nu$ for a Gaussian random field constructed from 100 realizations of the density field on a 512 $\times$ 512 grid drawn from a flat power spectrum. The error bars denote the error in mean over 100 realizations of the field.}
	\label{fig:scale_inv}
	
\end{figure*}

\section{Morphology of Density field: $\delta_{nl}$}

As discussed in Sec.~\ref{sec:21cmfast}, \texttt{21cmFAST} simulates $\delta_{nl}$ using the Zel'dovich approximation. In this section we follow the redshift evolution of the field as manifested in its morphological properties. Holes at negative threshold values correspond to voids while connected regions at positive thresholds correspond to peaks.

\begin{figure*}
	\begin{center}
		\includegraphics[width=5.cm, height=4.5cm]{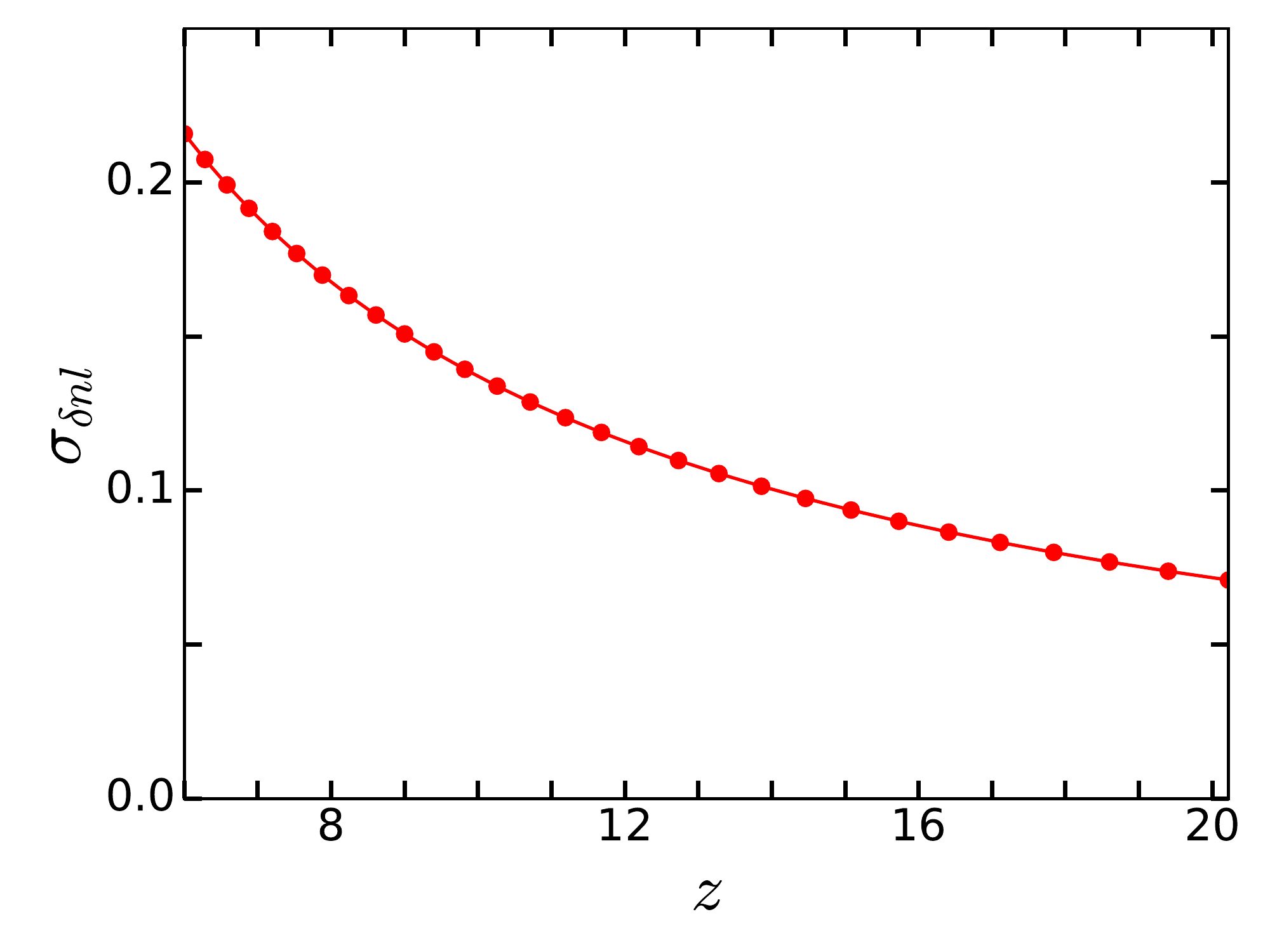} 
	\end{center}

	\caption{Redshift evolution of the standard deviation, $\sigma_{nl}$, of the density field.}
	\label{fig:delta_sd}
\end{figure*}

 The increase in the amplitude of fluctuations of $\delta_{nl}$ is captured by the redshift evolution of the variance of the field, denoted by $\sigma^2_{\delta_{nl}}$. Fig.  \ref{fig:delta_sd} shows $\sigma_{\delta_{nl}}$ versus redshift.
As the density perturbations grow, the high density peaks increase in height at the cost of low density regions which  become more under dense. In the linear regime, this growth is described as $\delta(z)= \delta_o/(1+z)$, where $\delta_o$ is the initial density contrast.  This leads to the increase in $\sigma_{\delta_{nl}}$ that we observe in the plot.
 The relatively large smoothing scale of 4.5 Mpc adopted for our analysis ensures $\delta_{nl}$ remains approximately linear for the redshift range that is under consideration.

\begin{figure*}
	\begin{center}
		
		\begin{subfigure}{0.48\textwidth}
			\resizebox{2.8in}{3.6in}{{\includegraphics{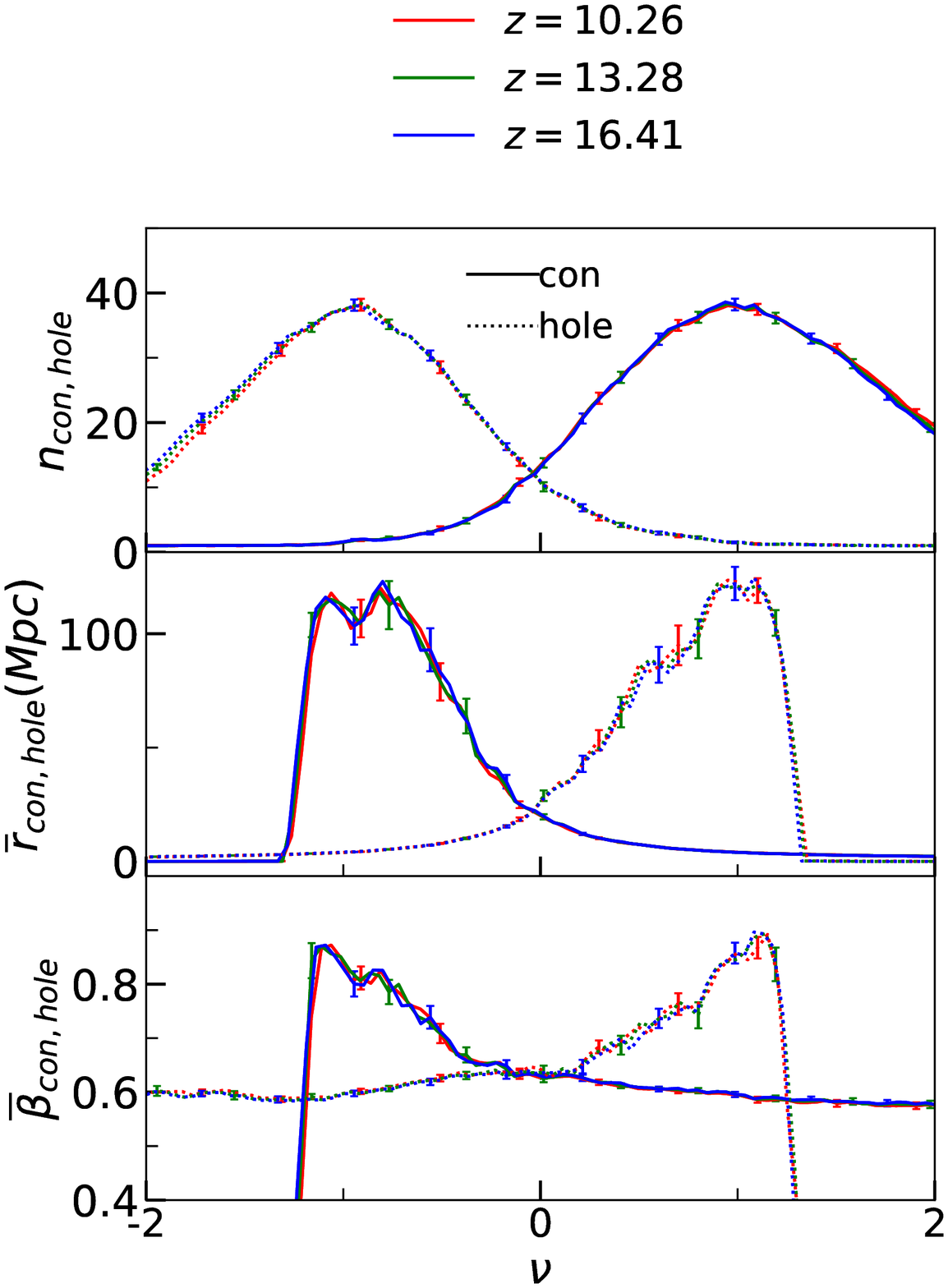}}}
			\caption{}
			\label{fig:delta_nu}
		\end{subfigure}
		\begin{subfigure}{0.48\textwidth}
			\resizebox{2.8in}{3.6in}{{\includegraphics{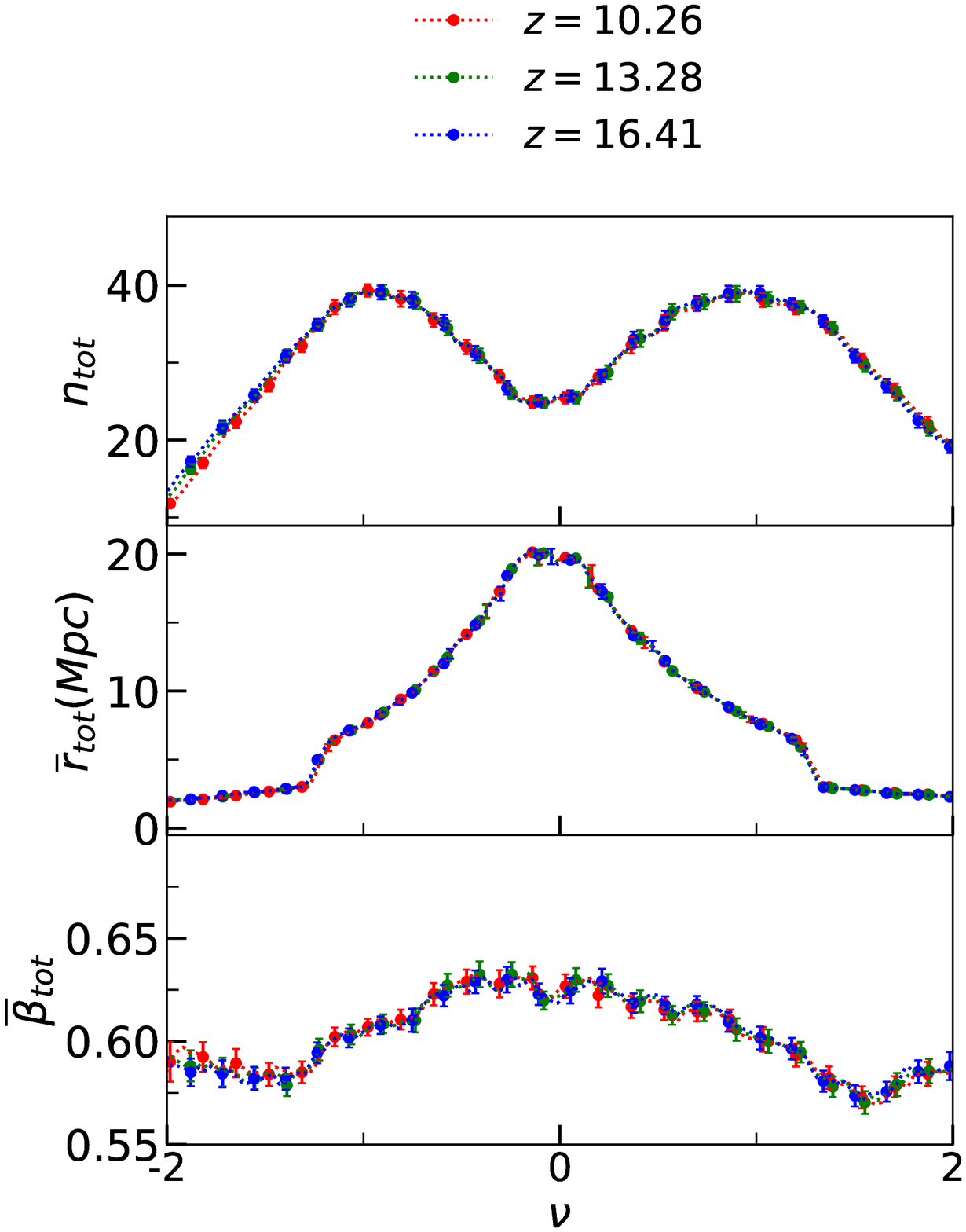}}}
			\caption{}
			\label{fig:delta_tot}
		\end{subfigure}
		\caption{(a) $n_{con,hole}$ (top), $\bar{r}_{con,hole}$ (middle) and $\bar\beta_{con,hole}$  (bottom) versus $\nu$ at redshifts 10.26 (red), 13.26 (green), and 16.41 (blue).  (b) $n_{tot}$ (top), $\bar{r}_{tot}$ (middle) and $\bar\beta_{tot}$ (bottom) with $\nu$, for the three redshift values as above. The error bars denote the error in mean over 32 slices.}
		
		\label{fig:delta_main}
	\end{center}
\end{figure*}	
		
\begin{figure*}
	\begin{center}
			\resizebox{3.in}{4.in}{{\includegraphics{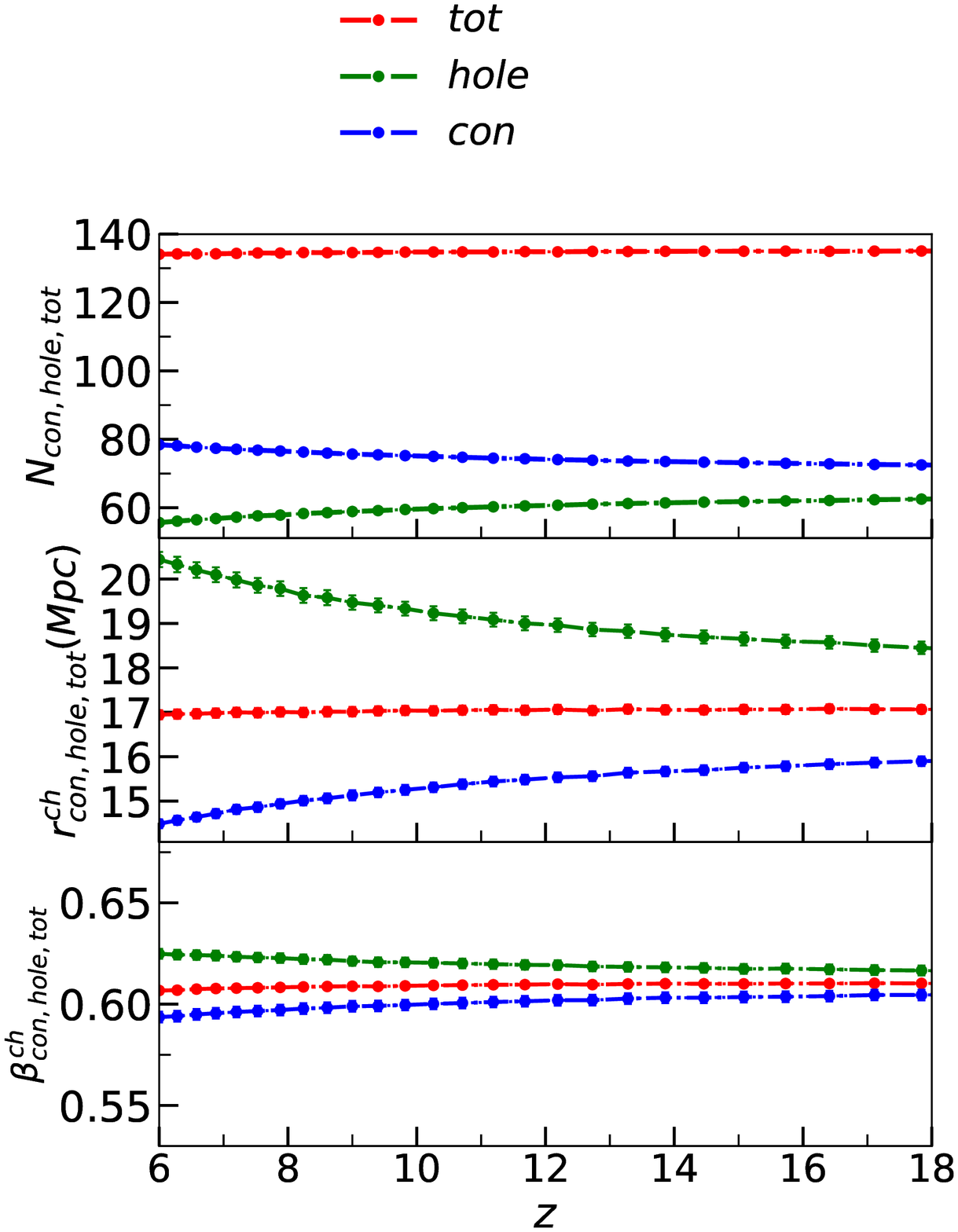}}}

			\caption{Redshift evolution for sum over all thresholds for connected regions (\textit{blue}), holes (\textit{green}) and for all structures (both connected regions (\textit{red})) described by $N_{con,hole,tot}$ (\textit{top}), $r^{ch}_{con,hole,tot}$ (\textit{middle}) and $\beta^{ch}_{con,hole,tot}$ (\textit{bottom}). The error bars denote the error in mean of the integrals in eq. \ref{eqn:lambdah}, \ref{eqn:rch} and \ref{eqn:betach} over the 32 slices.}
			\label{fig:delta_z}
		\end{center}
		\end{figure*}

In Fig.~\ref{fig:delta_main} we show the variation of the morphology of $\delta_{nl}$ with the field threshold $\nu$. The quantities are plotted as the mean over the 32 slices for a given $\nu$ value. 
Due to low statistics at very high and very low $\nu$ values we plot the variation up over the range $-2 < \nu < 2$. As  done in Sec.~3.3 we remove structures which have an area $>0.9$ times the area of the slice. In Fig.~\ref{fig:delta_nu} we  plot $n_{con},\ n_{hole}$ (top), $r^{ch}_{con},\ r^{ch}_{hole}$ (middle) and $\beta^{ch}_{con},\ \beta^{ch}_{hole}$ (bottom) as functions of threshold, at three redshift values $z=16.41, 13.28$ and $10.26$. Notice the shift in positions of error bars. This is due to the choice of our $\nu$ range between the maximum and minimum value of the field which changes with redshift. On visual comparison of $n_{con}$ and $n_{hole}$ with Fig.~\ref{fig:scale_inv} we find that all three redshifts roughly have the shape expected from a Gaussian random field. This is a consequence of the approximately linear evolution of density perturbations at the smoothing scale that we have chosen. Further, we find that the variation of both  $n_{con}$ and $n_{hole}$ with redshift is small. For a decreasing redshift, we can discern a small increase in $n_{con}$ for high positive thresholds $\nu \gtrsim 1$, while for $n_{hole}$ we find a small decrease towards high negative thresholds, $\nu\lesssim -1$. This implies that in the high density regions that correspond to large positive $\nu$, more sub-structure is forming as the redshift decreases. This is a consequences of peaks growing in height and hence a corresponding increase in $n_{con}$ at these high $\nu$ values relative to those at higher $z$ values. Peaks grow at the cost of low density voids making the density field positively skewed with decreasing redshift. At a given $\nu$ for a particular $z$ value we observe that $n_{con} \ne n_{hole}$ .The differences become more pronounced with decreasing redshift. It is visually discernible at $\nu \sim |2|$. This assmmetry is indicative of non-gaussianity introduced by gravity. For a Gaussian field the values are expected to be symmetric about $\nu=0$ (See Fig. \ref{fig:scale_inv}).

The middle panel of Fig.~\ref{fig:delta_nu} shows the variation of the sizes $\overline{r}_{con}$ and $\overline{r}_{hole}$ with threshold $\nu$. We find that the size (perimeter) of connected regions around $\nu\sim -1$ is statistically larger than that of holes around $\nu \sim +1$. This is an interesting feature in tracking the non-Gaussianity of perturbations induced by gravitational collapse, since for a Gaussian field the two statistics should be symmetric about $0$ (Fig. \ref{fig:scale_inv}).

The bottom panel of Fig.~\ref{fig:delta_nu} shows the variation of $\bar\beta_{con}$ and $\bar\beta_{hole}$ versus $\nu$. Again these plots are close to the expected shape for Gaussian fields (see Fig.~8  of~\cite{Appleby:2017uvb}). We can see very mild variation of the shape with redshift at intermediate $\nu$ values but differences at high and low $\nu$ values. The asymmetry between $\beta_{con}$ and $\beta_{hole}$ is not as pronounced as for $n_{con,hole}$ and $r_{con, hole}$.


In Fig.~\ref{fig:delta_tot} we plot the variation of $n_{tot}$, $\bar{r}_{tot}$ and $\bar\beta_{tot}$. These plots combine the information contained in \ref{fig:delta_nu} in such a way that most of the contribution for positive threshold values comes from connected regions, while for negative threshold values the contribution comes from holes. This is seen in the top panel of the figure for $n_{tot}$. At lower $\nu$ values $n_{hole}>n_{con}$ while $\bar{\beta}_{hole}<\bar{\beta}_{con}$. At these $\nu$ values the total morphology is a result of the morphology of the single large connected region punctured by numerous holes. Note that we have excluded the single large connected region at low thresholds and so the morphology is purely due to holes at $\nu \lesssim -1.5$. Opposite trend is expected for high $\nu$ values. The effect of the single large connected region and hole is very pronounced in the statistic $r_{tot}$ as it is a dimensional quantity unlike $\bar{\beta}_{tot}$. 

 We see a tilt in $\bar{r}_{tot}$ and $\bar{\beta}_{tot}$ towards higher $\nu$ values. Since $\bar{\beta}_{con}$ and $\bar{\beta}_{hole}$ are almost symmetric about $\nu =0$, the tilt in $\bar{\beta}_{tot}$ can be attributed to the asymmetry between $n_{con}$ and $n_{hole}$ at these thresholds.  The tilt is more pronounced at lower $z$ values as the difference between $n_{con}$ and $n_{hole}$ is more for lower redshifts.

In Fig.~\ref{fig:delta_z}  we plot the redshift evolution of $N_x$, $r^{ch}_{x}$ and $\beta^{ch}_x$, plotted as a mean of the integrals defined in Eqs.~\ref{eqn:Nx},~\ref{eqn:rch} and~\ref{eqn:betach} respectively over the 32 slices under consideration. The error bars are calculated as an error in mean over these 32 slices. Note that we follow the same methodology for calculation of the redshift evolution for all other fields in the subsequent sections. These plots contain the physical information encoded in Fig.~\ref{fig:delta_main} condensed into a single number at each redshift value. The limits of the $\nu$ integration, $\nu_{\rm high}$ and $\nu_{\rm low}$, are set to be the maximum and minimum values of the field.

The top panel indicates that the numbers of connected regions and holes integrated over all threshold values decreases as a function of redshift. The middle panels shows that the size of high density regions, integrated over all thresholds, shrink in size as the redshift decreases. In contrast the size of holes (voids) grow with decreasing redshift. This is due to the attractive nature of gravitational collapse. In the bottom panel we find that $\beta^{ch}_x$ does not show much variation with redshift $z$ and that the connected regions are more anisotropic than holes.

\section{Morphology of neutral hydrogen field for different models of EoR: $x_{HI}$}
\label{Sec:xh}
\begin{figure}[h]
	\begin{center}
		\resizebox{1.8in}{2.6in}{{\includegraphics{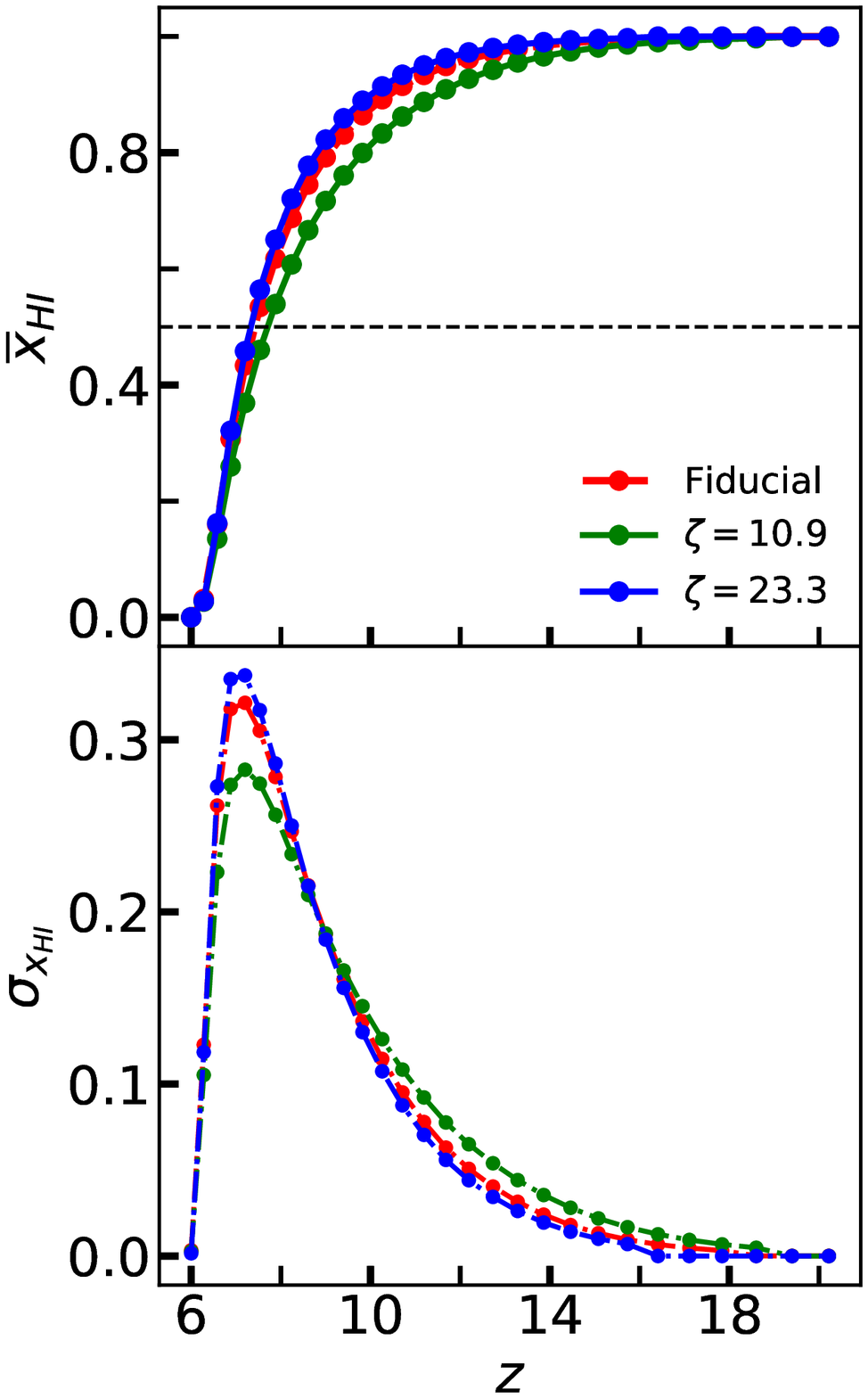}}} 
		\resizebox{1.8in}{2.6in}{{\includegraphics{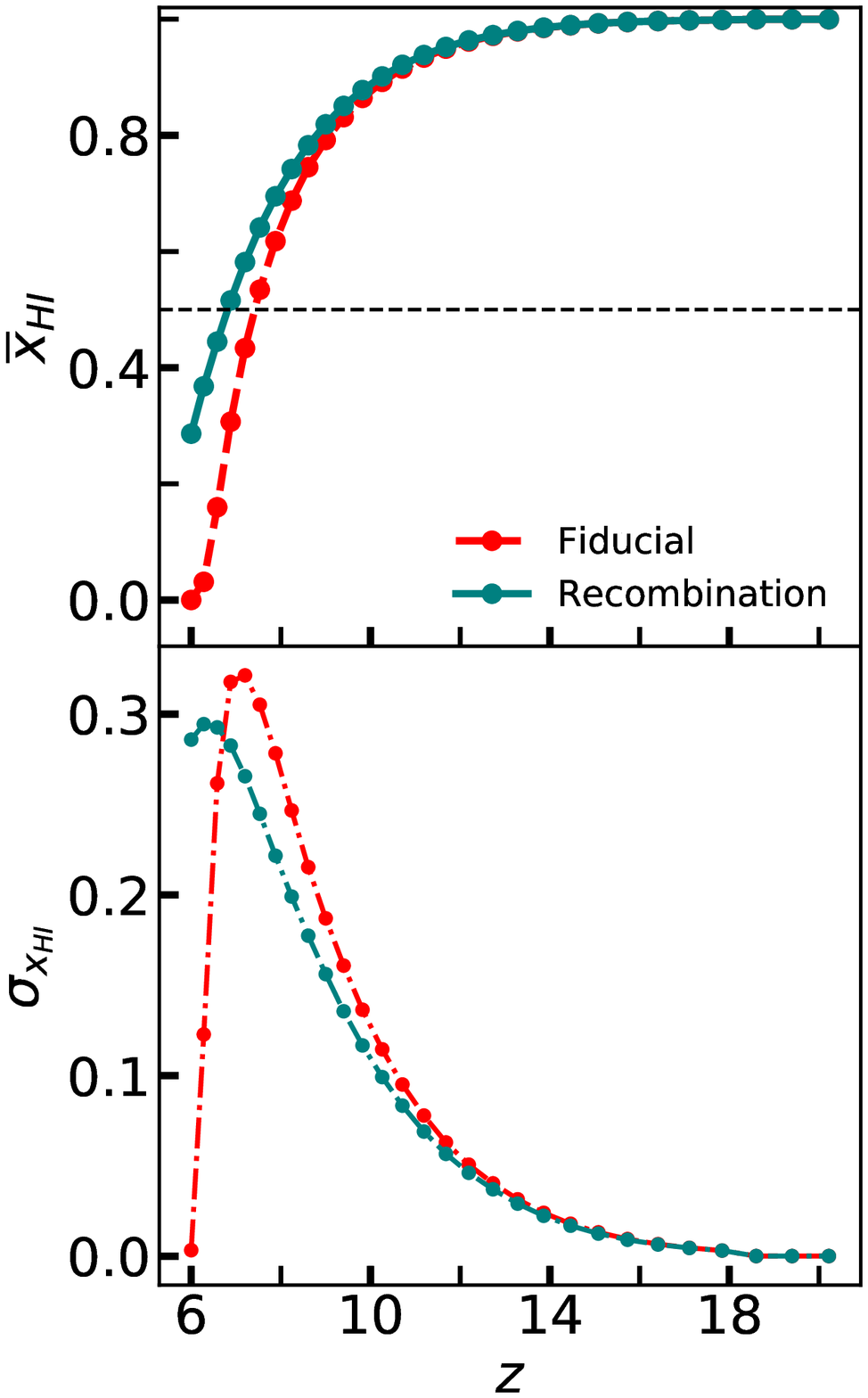}}} 
		\resizebox{1.8in}{2.6in}{{\includegraphics{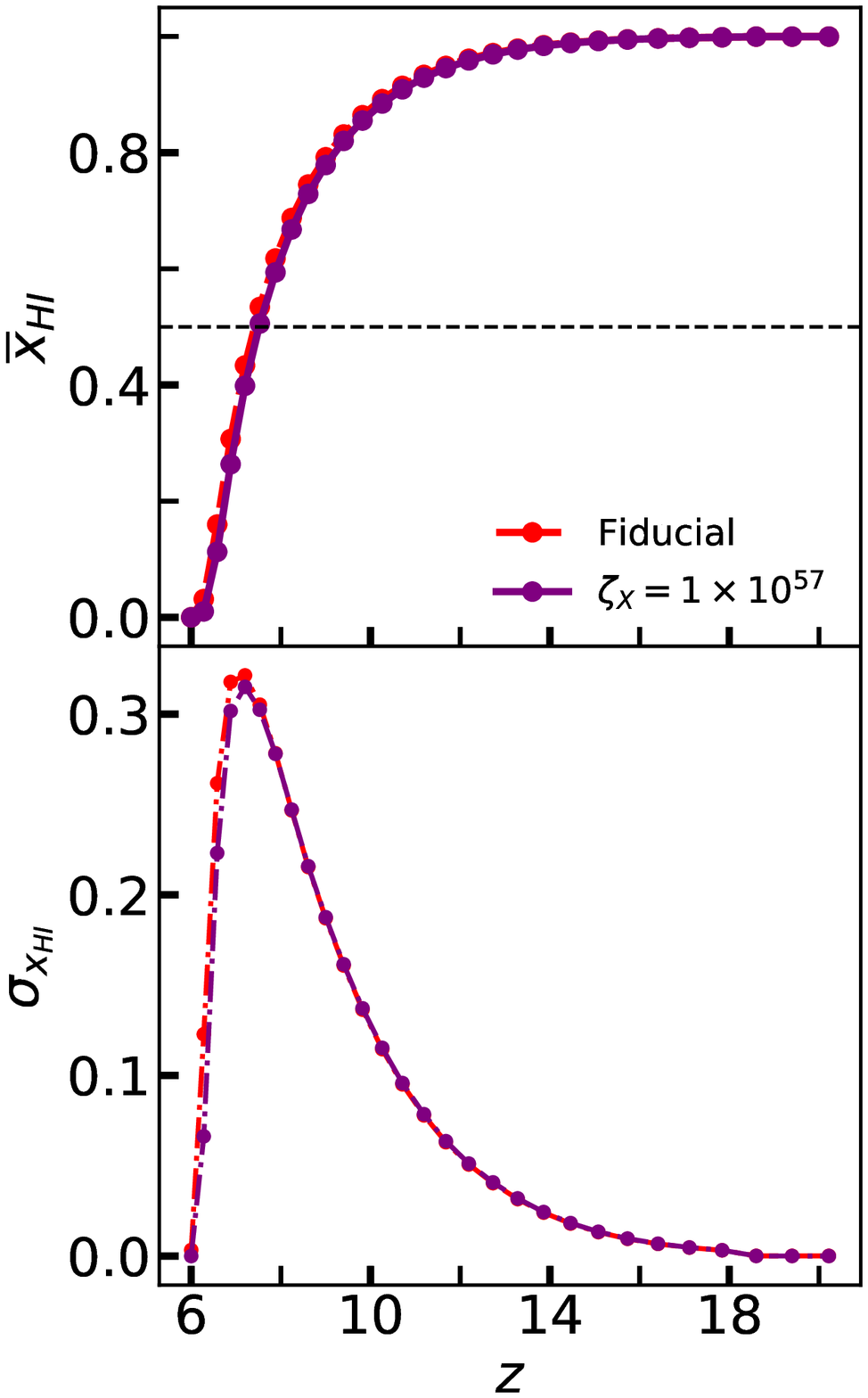}}}
	\end{center}
		\caption{Evolution of mean and standard deviation for neutral hydrogen field, $x_{HI}$ for different models, relative to the \textit{fiducial model} in red.}
		\label{fig:xH_avg}
\end{figure}
\begin{figure*}
	\begin{center}
		
		\resizebox{3.0in}{3.5in}{{\includegraphics{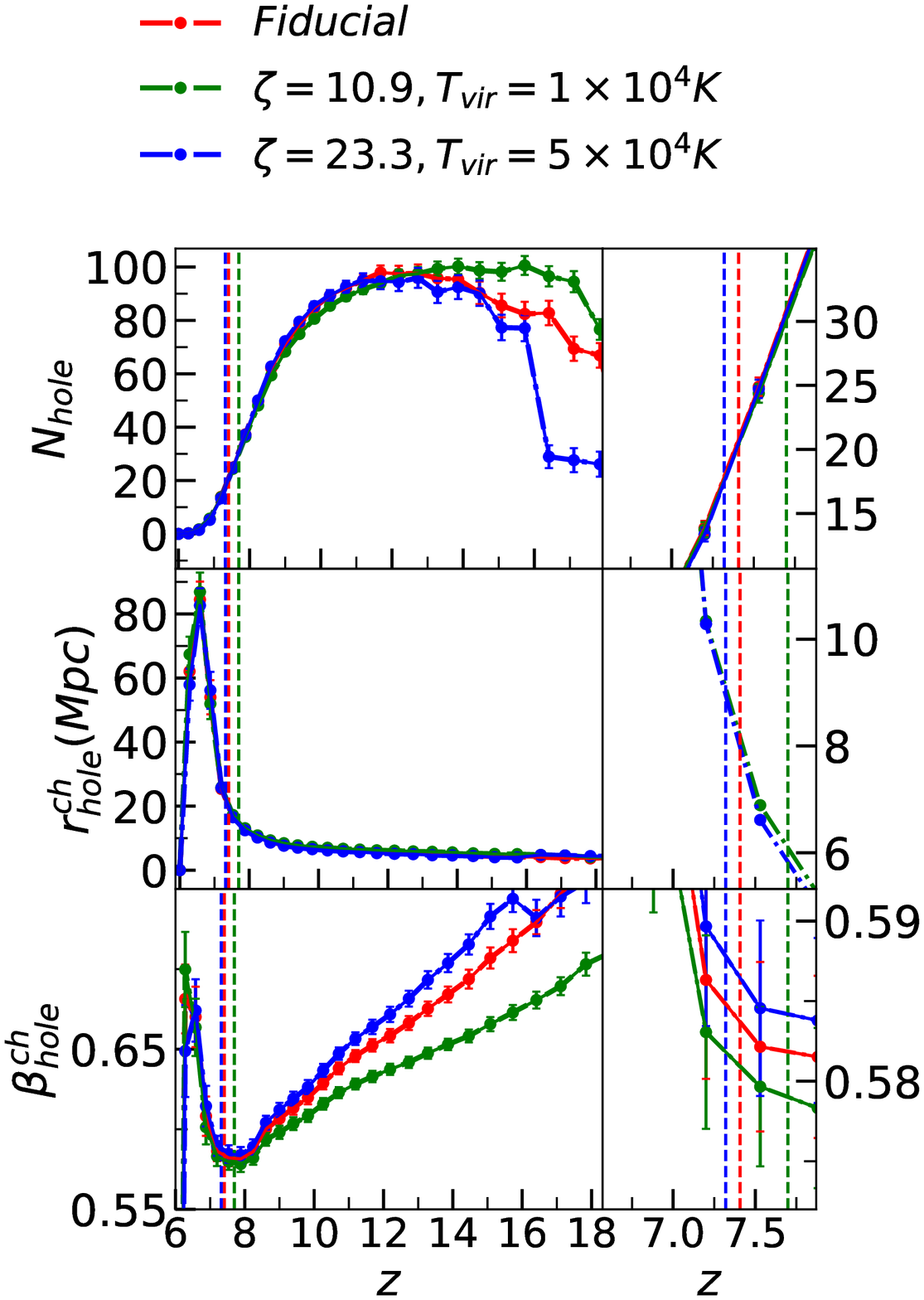}}}
		\resizebox{3.0in}{3.5in}{{\includegraphics{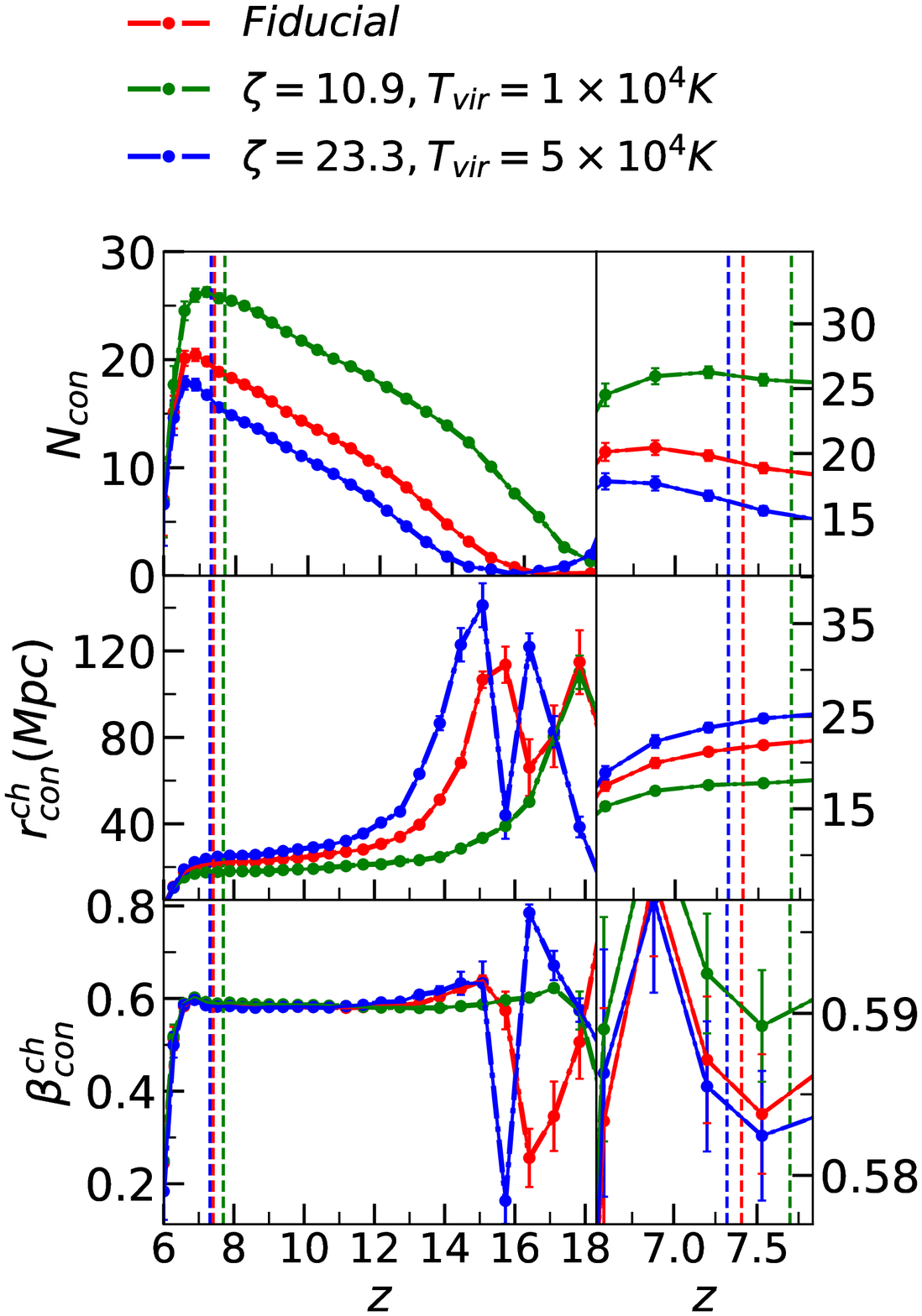}}}
		
	\end{center}
	\caption{The morphology of neutral hydrogen fraction for different values of $T_{vir}$ and $\zeta$ for holes (\textit{left panel}) and connected regions (\textit{right panel}), relative to the \textit{fiducial model}. The vertical lines show $z_{0.5}$ where $\bar{x}_{HI} =0.5$ and $N_{con}=N_{hole}$ for $\nu_{cut}=0$. The smaller panels on the right show a zoomed in version of the same plots to capture the variations around $z_{0.5}$.}
	\label{fig:xH_morph1}
\end{figure*}

\begin{figure*}
	\begin{center}
		\resizebox{3.0in}{3.5in}{{\includegraphics{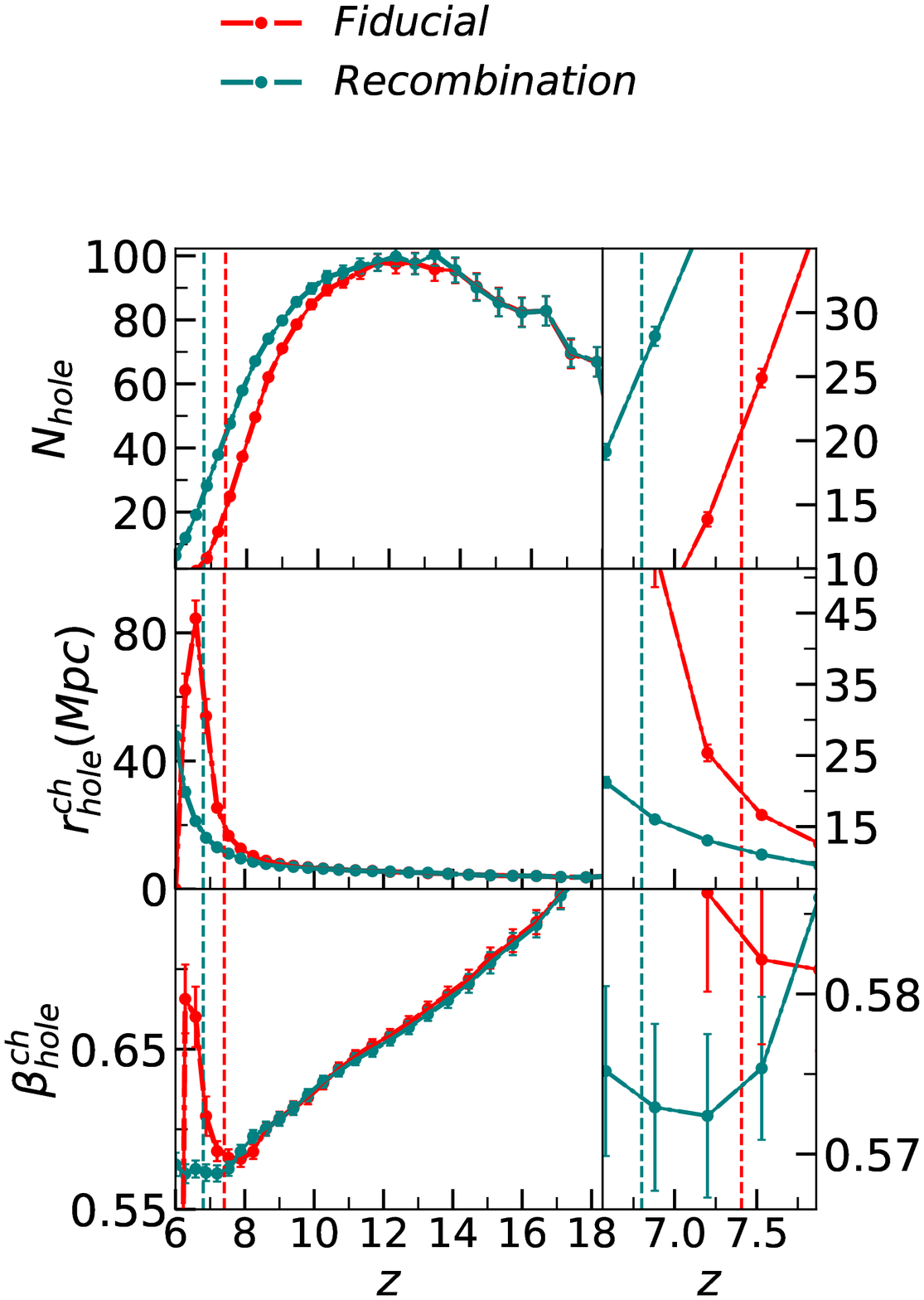}}}
		\resizebox{3.0in}{3.5in}{{\includegraphics{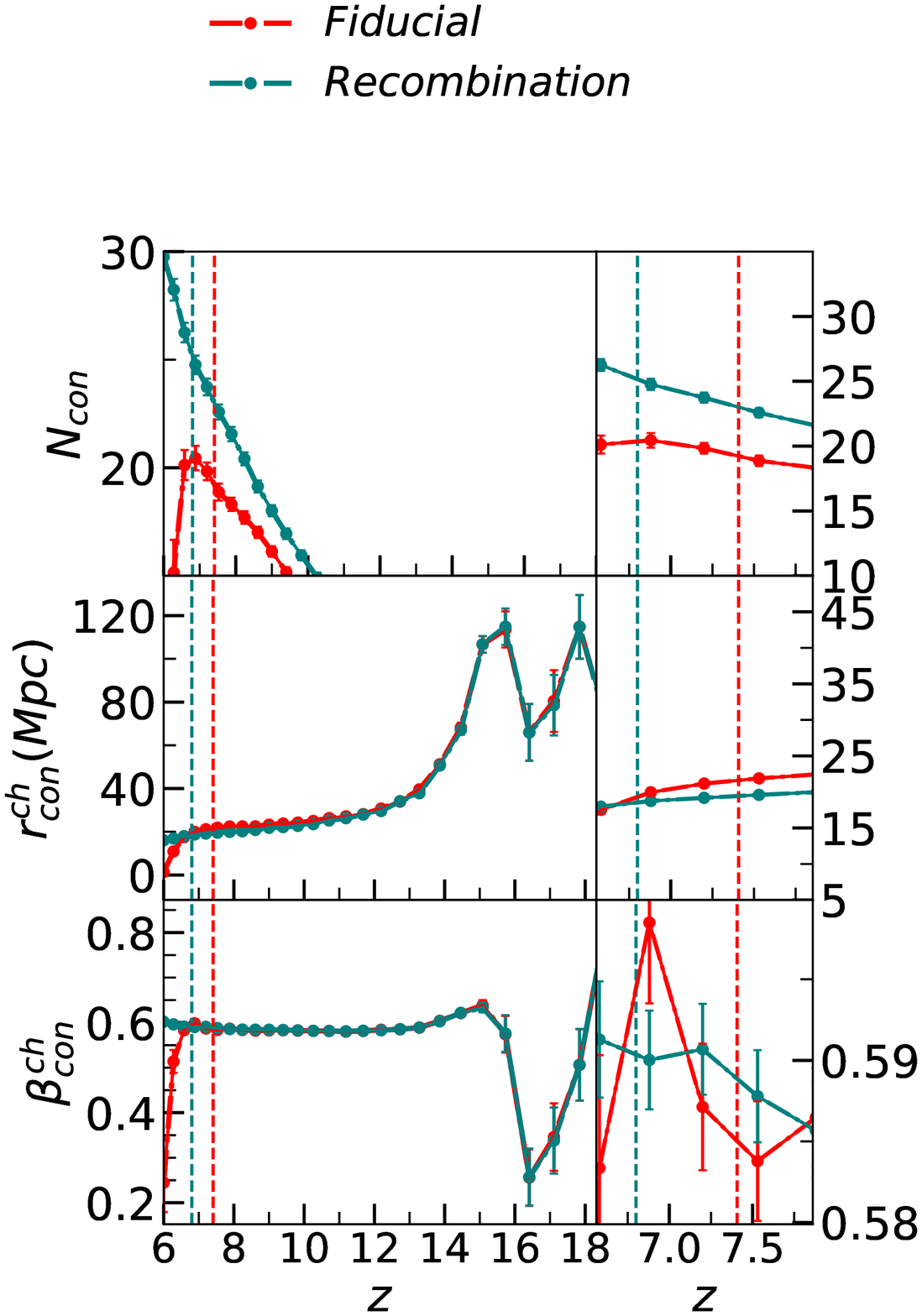}}}

	\end{center}
	\caption{The morphology of neutral hydrogen fraction with inhomogenous recombination relative to the \textit{fiducial model} without recombination. The vertical lines show $z_{0.5}$. The smaller panels on the right show a zoomed in version of the same plots to capture the variations around $z_{0.5}$ which is midway through ionization history and $N_{con}=N_{hole}$}
	\label{fig:xH_morph2}
\end{figure*}

The morphology of the neutral hydrogen field was studied in \cite{Kapahtia:2017qrg} for different smoothing scales and different values of $\nu_{cut}$. $\nu_{cut}$ refers to the value of threshold above (below) which a connected region (hole) is interpreted as a neutral (ionized) region. In this work we choose to work with $\nu_{cut}=0$, which is the mean value of the $x_{HI}$ field. The choice of $\nu_{cut}$ allows for the inclusion of extremely small peaks or shallow valleys at lower and higher $z$ values where the variance ($\sigma_{x_{HI}}$) of the $x_{HI}$ field is very small (Fig.~\ref{fig:xH_avg}). For this choice of $\nu_{cut}$, $N_{con} \sim N_{hole}$ at $x_{HI}=0.5$ \cite{Kapahtia:2017qrg}. 
 For the $x_{HI}$ field, a connected region corresponds to a neutral region and a hole corresponds to an ionized region. Ionized bubbles grow in size and merge. The rate of formation and growth of ionized bubbles, their sizes and the rate at which they merge depend upon the astrophysical properties of the collapsed objects and mean free path of ionizing photons. Statistically, mergers of ionized bubbles lead to an increase in anisotropy of the bubbles as expected and demonstrated in \cite{Kapahtia:2017qrg}. One would expect that locally apart from mergers, the anisotropy in the growth of a bubble could also depend upon the clumpiness of the density field around it. Therefore for two objects with the same astrophysical properties the bubble around one could be more anisotropic than the other because of more clumpiness in the distribution of neutral hydrogen around it. However for a given matter power spectrum the \textit{average} anisotropy of structures (as measured by $\overline{\beta}$) in our excursion set of ionized field can be attributed to mergers alone. The number of mergers depend upon the astrophysical properties of sources. 
 
We first review the expected morphology of the $x_{HI}$ field for the \textit{fiducial model} (see Fig.~3 of \cite{Kapahtia:2017qrg}). At high redshifts, as the ionized bubbles begin to appear around sources they are isolated and small with few mergers. In this regime holes dominate the morphology completely, across the entire range of threshold values. As reionization progresses ionized bubbles grow and new ionized regions begin to appear. Eventually further ionization leads to  increasing mergers of ionized regions. As mergers become more dominant the roughly smooth neutral region begins to break while ionized regions merge to form bigger and more anisotropic bubbles, leading to numerous, small connected regions and a decreasing number of holes which are larger and grow in size. Structures are highly non-convex, i.e. having regions of negative curvature,  in this regime. As mergers complete, nearly all of the ionized regions have merged into a single ionized area with a small number of neutral regions embedded. Therefore at this stage, the morphology of the ionized field is completely dominated by connected regions across the range of $\nu$ values. 

The evolution of the  morphology of the $x_{HI}$ field is dominated by three competing physical processes -- the rate of formation of collapsed objects capable of ionizing, the rate of growth of bubbles and the rate of mergers of ionized bubbles. For any general ionization history, initially the rate of formation of ionizing sources is greater than the rate of mergers of bubbles. The clustering of collapsed objects leads to mergers starting very early during EoR as was also observed in \cite{Kapahtia:2017qrg} where the value of $\beta$ was observed to decrease at high redshift. In this regime the value of $N_{hole}$ will increase as redshift decreases while there is a single connected region $N_{con} \sim 1$. At some $z=z_{frag}$ the rate of mergers begins to dominate over the rate of appearance of new sources. $N_{hole}$ begins to decrease in this regime, while $N_{con}$ starts to increase. There will be some value of z at which $N_{con}=N_{hole}$. If we choose $\nu_{cut} =0$ this equality occurs at $z=z_{0.5}$ which is defined as the redshift at which $\overline x_{HI}=0.5$. At a redshift $z_{e}$ mergers approach completion, $N_{con}$ begins to decrease.

\begin{table}[ht]
	
	\centering 
	
	\begin{tabular}{c c c c c c }
	
		\hline
		\hline                        
		Model &$z_{frag}$ & $z_{0.5}$ & $z_e$ & $z_{re}$ & $\tau_{re}$\\[0.5ex]
		\hline
		\\
		Fiducial &$\sim11.69$ & $\sim 7.407$ &  $\sim 6.58$ & $\sim 6.28$ & $\sim 0.054$ \\
		$T_{vir}=1\times 10^4 K$ &$\sim 13.857$ & $\sim 7.698$ &$\sim 6.58$ & $\sim6.00$ & $ \sim 0.058$\\
		$T_{vir}=5\times 10^4 K$ &$\sim 11.194$& $\sim 7.32$  &$\sim 6.58$ & $\sim 6.00$ & $\sim 0.052$ \\
		$\zeta_X=1 \times 10^{57}$ &$\sim 12.73$ & $\sim 7.5$ &$\sim 6.58$ & $\sim 6.28$ &$\sim 0.034$\\
		Recombination &$\sim 12.2$& $ \sim 6.8$&--&$<6.00$ & \\[1ex]
		\\
		\hline
		
	\end{tabular}
	\caption{Model histories chosen for our analysis. The table shows the redshift $z_{0.5}$ at which $\bar x_{HI}=0.5$, $z_{frag}$ at which fragmentation starts, $z_e$ where mergers complete and $z_{re}$ where reionization ends. The last column is the optical depth to the last scattering surface.}
	\label{table:evol}
	
\end{table}
The above regimes are reflected in the Betti numbers of the $x_{HI}$ field and the various transition redshifts are described as below: 

\begin{itemize}
	\item $z_{frag}$: Redshift at which $N_{hole}$ turns over to decrease from an initial stage of growth. It marks the value of $z$ where the bubble merger rate begins to dominate over rate of appearance of new collapsed sources.
	\item $z_{0.5}$: Redshift at which $\overline{x}_{HI}=0.5$.
	\item $z_{e}$:  Redshift at which $N_{con}$ turns over to decrease from an initial duration of increase marking the point where mergers approach end.
	\item $z_{re}$ : Redshift at which EoR ends, i.e. $\overline {x}_{HI} \sim 0$
\end{itemize}

The values of $z_{frag}, z_{0.5}, z_e$ and $z_{re}$ defined above will depend on the different physical processes of reionization, and hence on the model of EoR. Therefore, their values can be important characteristic features that can discriminate different models. 
Note that in \cite{Kapahtia:2017qrg} it was found that at $z_{0.5}$, $N_{con}=N_{hole}$. We will show this to be true for all models we have considered in our study. In table \ref{table:evol} we summarize the values for these important $z$ values for our choice of models. The values of redshifts obtained for these transition are not exact because the simulations generate fields at discrete $z$ values (logarithmic interval of 1.0404 in ($1+z$) for our case).

In Fig. \ref{fig:xH_avg}, we show the evolution of $\overline{x}_{HI}$ and the rms fluctuation of $x_{HI}$ denoted by $\sigma_{x_{HI}}$ for all models under consideration. 
We now interpret the morphology of the $x_{HI}$ field for these models as shown in Fig. (\ref{fig:xH_morph1} and \ref{fig:xH_morph2}). In order to obtain an ionization history of the IGM it would suffice to obtain the various transition redshifts of the evolution of morphology as described in Table $\ref{table:evol}$ and observe how they shift relative to the \textit{fiducial model}. However for a detailed astrophysical modelling, one would have to compare at redshift values corresponding to the same epoch in the ionization history as described by Table $\ref{table:evol}$. Therefore in addition to comparing the general shift in the values of $z_{frag}$, $z_{0.5}$ and $z_e$, we also compare the morphological descriptions specifically at these transition redshifts.

\subsection{Models with different $T_{vir}$ values}
\label{sec:Tvir}

As noted in Sec. \ref{sec:model}, different combinations of $T_{vir}$ and $\zeta$ can give similar ionization histories. However the fluctuations in $x_{HI}$ field are expected to differ. This is because a lower $T_{vir}$ value corresponds to less efficient sources as compared to higher $T_{vir}$ values. This is reflected in the respective $\zeta$ values required for reionization to end at the same $z_{re}$. 
The sources with lower $T_{vir}$ values would lead to a higher collapse fraction at a given redshift as compared to higher $T_{vir}$ and hence would be more numerous. Therefore reionization will start earlier for a lower $T_{vir}$ value. Such sources would lead to bubbles which are more numerous and smaller in size at a given redshift as compared to sources with higher $T_{vir}$ values. Fig. \ref{fig:xH_morph1} shows the redshift evolution of the morphology of the neutral hydrogen fraction field for different combinations of $T_{vir}$ and $\zeta$, as described by $N_{con,hole}$, $r^{ch}_{con,hole}$ and $\beta^{ch}_{con,hole}$.

The left panel of Fig. \ref{fig:xH_morph1} reflects our qualitative reasoning.  More numerous bubbles are reflected in the higher value of $N_{hole}$ for the lowest $T_{vir}$ value of $1 \times 10^4$ K until $z\sim z_{frag}$. We note that $z_{frag}$ is highest for the model with the lowest $T_{vir}$ value, i.e. mergers begin to dominate earlier. This leads to a shift of $z_{0.5}$ and $z_e$ to higher $z$ values for lower $T_{vir}$ values. This occurs because even though the sources are less efficient, they are more numerous. This leads to a correspondingly higher number of bubbles and hence merging begins to dominate at a $z$ value earlier than cases where $T_{vir}$ is greater. The differences in $N_{hole}$ for different models is less pronounced once mergers dominate the morphology as seen in the zoomed panel at $z=z_{0.5}$ for $N_{hole}$. However they differ in morphology. The plot of $r^{ch}_{hole}$ and the zoomed panel, show that the size of bubbles at $z_{0.5}$ is smallest for the lowest value of $T_{vir}$. Bubbles for lower $T_{vir}$ values are more anisotropic at $z_{0.5}$ as is seen for their $\beta^{ch}_{hole}$ values. The large bubbles formed as a result of mergers for the case of smaller $T_{vir}$ values is a consequence of more numerous and faster rates of merging as compared to larger $T_{vir}$ values. More mergers statistically increases anisotropy by $z=z_{0.5}$ as seen in the relatively smaller values of $\beta^{ch}_{hole}$ for smaller $T_{vir}$ values.

The right panel of Fig. \ref{fig:xH_morph1} shows the variation of $N_{con}, r^{ch}_{con}$ and $\beta^{ch}_{con}$. We notice that the connected regions are more numerous for smaller $T_{vir}$ values across the redshift range of interest. This is because they are less efficient sources and even if there are more mergers the number of efficient photons available to ionize the regions with same density is less than for the \textit{fiducial model} for which the sources are more efficient. The large neutral region that fragments will fragment into smaller sized neutral regions in the case of more mergers. Therefore an opposite trend is seen for $r^{ch}_{con}$. If a single large connected region is fragmented, then the model in which there are more fragments, the size of the fragments will be smaller. However we observe that the connected regions for lower $T_{vir}$ are less anisotropic as seen in the zoomed in panel at $z=z_{0.5}$. This is opposite to the trend for holes. It is not straightforward to anticipate this trend but it indicates that more mergers are leading to fragmentation of connected neutral regions into less anisotropic peices.

\subsection{Model with inhomogenous recombination}

The morphology of ionized fields when inhomogenous recombination is included in the excursion set formalism has been studied in \cite{Sobacchi:2014rua, Choudhury:2008aw} using the power spectrum of the 21cm brightness temperature, $\delta T_b$. Here we carry out a complementary study in real space. The prescription for incorporating inhomeogenous recombination is described in eq.(\ref{rec}). The rate of recombination in a region with number density of electrons $n_e$ is $\propto \langle n_{e}^2 \rangle$. The effect of recombination manifests some time after reionization begins ($z\lesssim12$ from visual inspection of Fig. \ref{fig:xH_avg}) and becomes more pronounced with decreasing redshift. At early stages the number of photons is insufficient to ionize hydrogen in high density regions. Therefore only the lower density regions are ionized, where ionization dominates over recombinations. Therefore at this stage, recombination is unimportant in both high and low density regions. At later times as the collapsed fraction increases, the photons are able to permeate higher density neutral regions and ionize.  But in those regions the rate at which recombination occurs is faster than the rate at which the photons are ionizing. The increased number of recombinations lead to decreased efficiency of ionization when compared with the \textit{fiducial model} due to a paucity of ionizing photons in high density regions. Therefore at these late redshifts some higher density regions which would have otherwise been ionized in case of the \textit{fiducial model} remain neutral. 

The important salient point is that when comparing with the \textit{fiducial model}, the rate of appearance of newer ionized regions is the same but the rate of growth and merger of ionized regions is different in the two cases. Inhomogenous recombinations slow down the entire process of growth and mergers. The inhomogenity in the density distribution introduces an additional anisotropy in the excursion set morphology beyond the anisotropy due to mergers alone. The redshift at which the EoR ends for the model with recombination is $z_{e} < 6$. But here we shall only analyse recombination until $z=6$ so that we can compare with the \textit{fiducial model}. In Fig. \ref{fig:xH_morph2} we show the effect of inhomogenous recombination relative to the \textit{fiducial model} and from Table \ref{table:evol} we see that the different transition redshifts in the evolution of $x_{HI}$ morphology are shifted to lower $z$ values relative to the \textit{fiducial model}.

The left panel of Fig.~\ref{fig:xH_morph2} shows the redshift evolution of $N_{hole}$, $\beta^{ch}_{hole}$ and $r^{ch}_{hole}$. We observe that the number of holes for the model with recombination is nearly the same as that of \textit{fiducial model} at very early redshifts until $z_{frag}$. At $z\sim12$ they start diverging i.e. number of holes for the model with inhomogenous recombination is more than that for the \textit{fiducial model}. This confirms that recombination has suppressed the number of mergers as compared to the \textit{fiducial model} . 

The variation of $r^{ch}_{hole}$ with redshift shows that there is no substantial difference in bubble sizes as a result of recombination until $z \sim z_{0.5}$. At $z=z_{0.5}$, the model with recombination has bubble sizes smaller than the \textit{fiducial model} with a difference in size $\sim 3~\rm Mpc$. This is again a result of a smaller number of mergers relative to the \textit{fiducial model}.

The variation of $\beta^{ch}_{hole}$ with redshift shows that the value is nearly equal to the \textit{fiducial model} until $z \gtrsim z_{0.5}$ where $\beta^{ch}_{hole}$ is lower for the model with recombination by $1\%$. Moreover the turnover is more gradual in the case of the model with recombination due to the slowing down of the entire process of reionization as discussed above. At $z_{0.5}$ the value of $\beta^{ch}_{hole}$ is less compared to that for the \textit{fiducial model}. The higher anisotropy seen for the model with recombination is because of the inhomogenity in the density field.  

The right panel of Fig. \ref{fig:xH_morph2} shows the variation of the morphology for connected regions. The number of connected regions $N_{con}$ is more for the model with recombination at $z<z_{frag}$ . This is because the neutral regions in high density regions which could get ionized in the case of  \textit{fiducial model} remain neutral when inhomogenous recombination is included. Moreover the mergers in the case of inhomogenous recombination lead to merged ionized regions which are smaller due to suppresion at high density regions. This leads to fragmentation of the neutral region into correspondingly higher number of fragments. A higher number of smaller fragments generates a smaller value of $r^{ch}_{con}$ compared to the \textit{fiducial model}.
 We do not observe much difference between the two models for $\beta^{ch}_{con}$ at $z=z_{0.5}$. At this $z$ value the connected regions of the \textit{fiducial model} are high density neutral regions which cannot be ionized due to insufficient photons to ionize them. For the model with recombination the connected regions are either the ones where ionization never occured like in case of \textit{fiducial model} or where ionization occured but recombination took over. The former regions are the same regions as in the case of the \textit{fiducial model} while the latter regions would be holes in the \textit{fiducial model} at the same $z$ values because the efficiency of ionizing sources is the same in both the cases. Since $N_{con}$ is different in the two cases at $z=z_{0.5}$, the $\beta$ values show that the the connected regions in the case of the model with recombination are dominated by regions which did not ionize and are the same regions as the high density neutral regions at $z=z_{0.5}$ for the \textit{fiducial model}.

We emphasise that the effect of including inhomogenous recombinations to our \textit{fiducial model} leads to a shift in the redshifts of transitions, towards lower z values.

In Table \ref{table:bubble} we summarize the characteristic bubble sizes for the different models at $z=z_{0.5}$. The characteristic bubble size at $z_{0.5}$ for our \textit{fiducial model} is $\sim 20.5$ Mpc. For a linear increase in $T_{vir}$, the bubble sizes show a somewhat linear increase. The size of bubbles is reduced to $ \sim 17.5$ Mpc once the effect of recombinations is accounted for. 
\begin{table}[ht]
	
	\centering 

	\begin{tabular}{c c}
		
		\hline
		\hline                        
		Model & ~$r^{ch}_{z_{0.5}} ~ (\rm Mpc$) \\[0.5ex]
		\hline
		\\
		Fiducial & $\sim 20.5\pm 0.78$ \\
		$T_{vir}=1\times 10^4 K$  & $\sim 15 \pm 0.424$\\
		$T_{vir}=5\times 10^4 K$ & $\sim 22.5\pm 0.96$\\
		$\zeta_X=1 \times 10^{57}$ & $\sim 20 \pm 0.689 $\\
		Recombination &  $\sim 17.5 \pm 0.548$ \\[1ex]
		\\
		\hline
		
	\end{tabular}
	\caption{The characteristic size of ionized regions at $z=z_{0.5}$ for the different EoR models under consideration. The error bars shown are the error on mean over the 32 slices.}
	\label{table:bubble}
\end{table}

\section{Morphology of Spin Temperature field: $T_s$}
\label{sec:Ts}

\begin{figure}[h]
	\begin{center}
		\resizebox{1.4in}{1.5in}{{\includegraphics{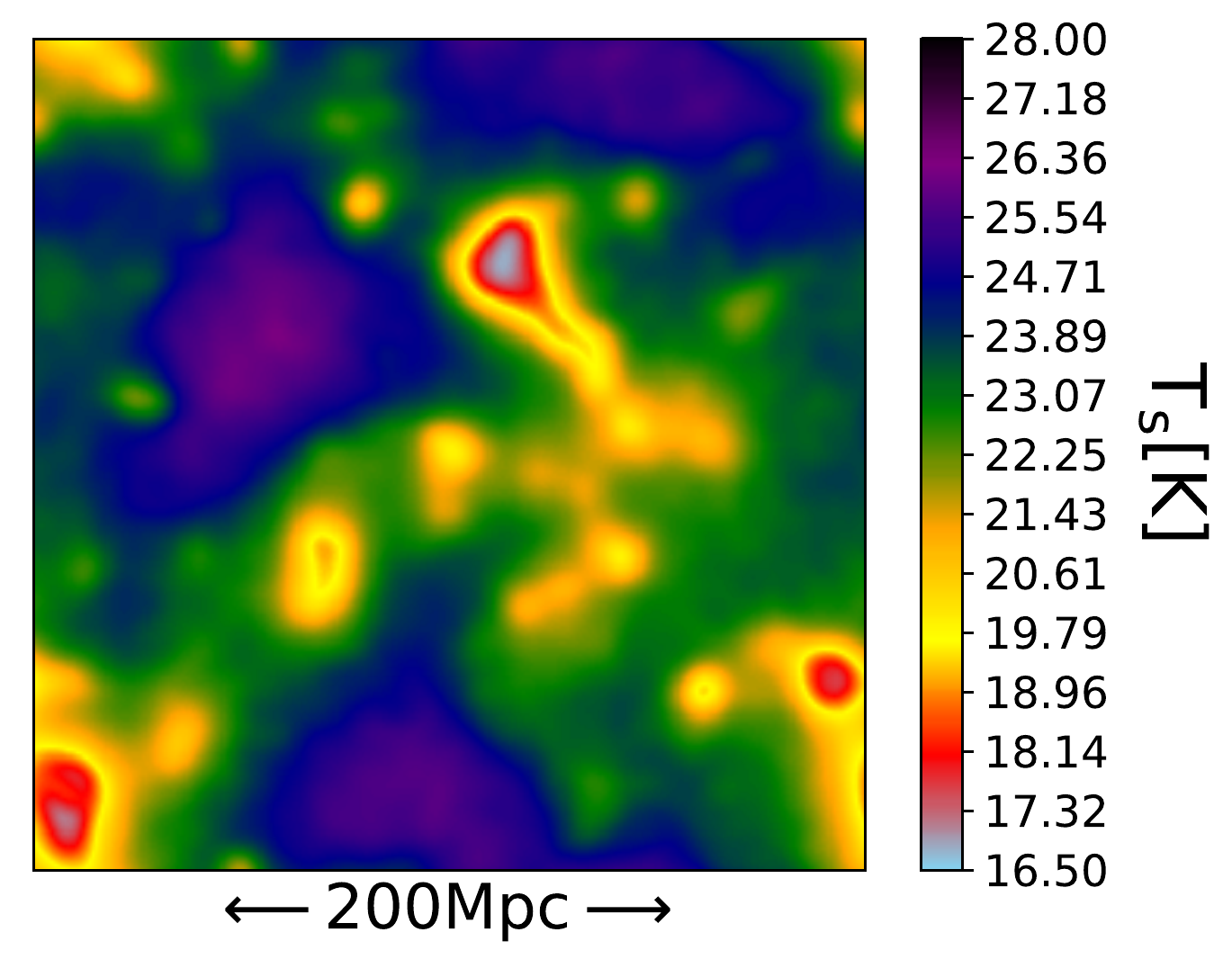}}}
		\resizebox{1.4in}{1.5in}{{\includegraphics{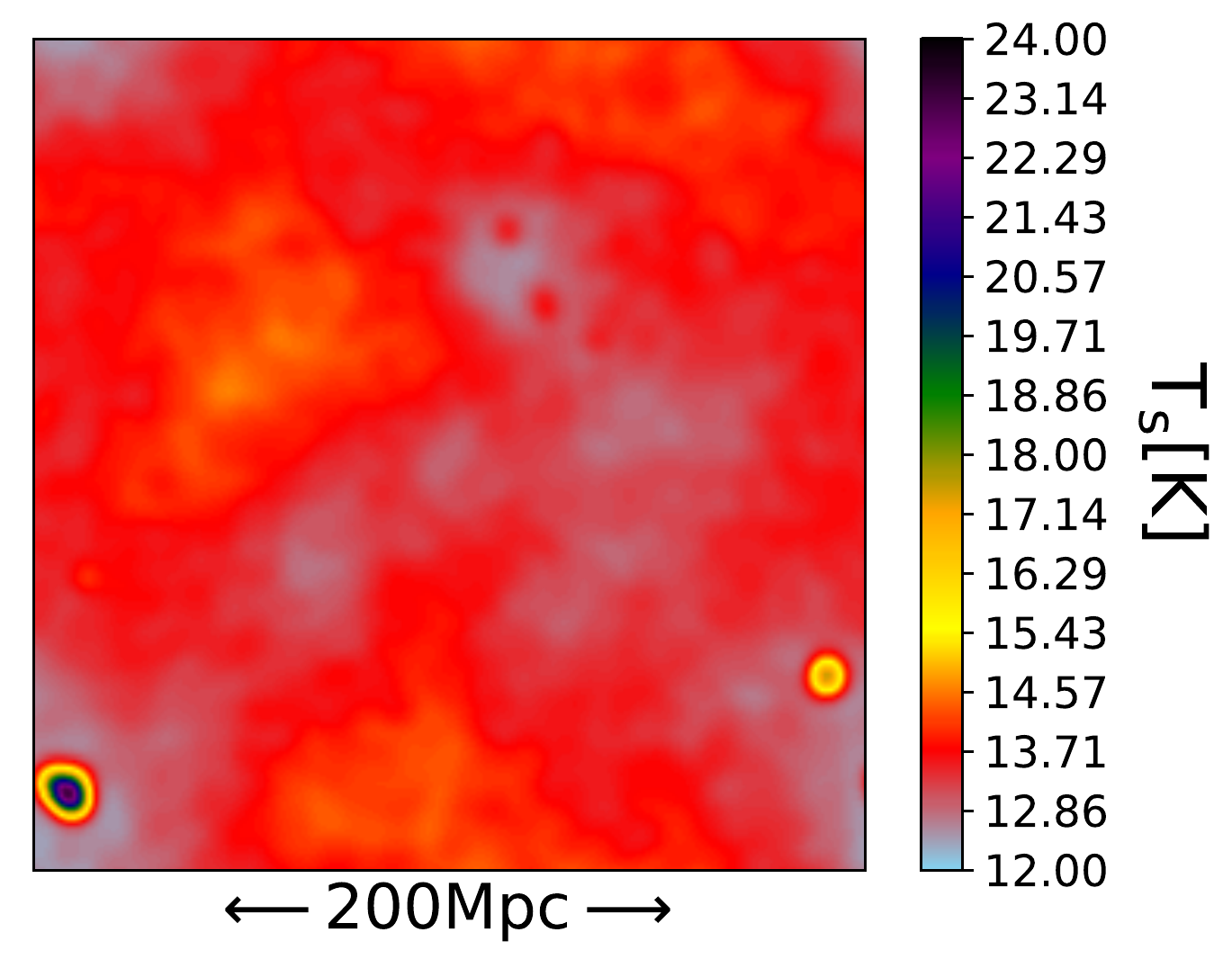}}} 
		\resizebox{1.4in}{1.5in}{{\includegraphics{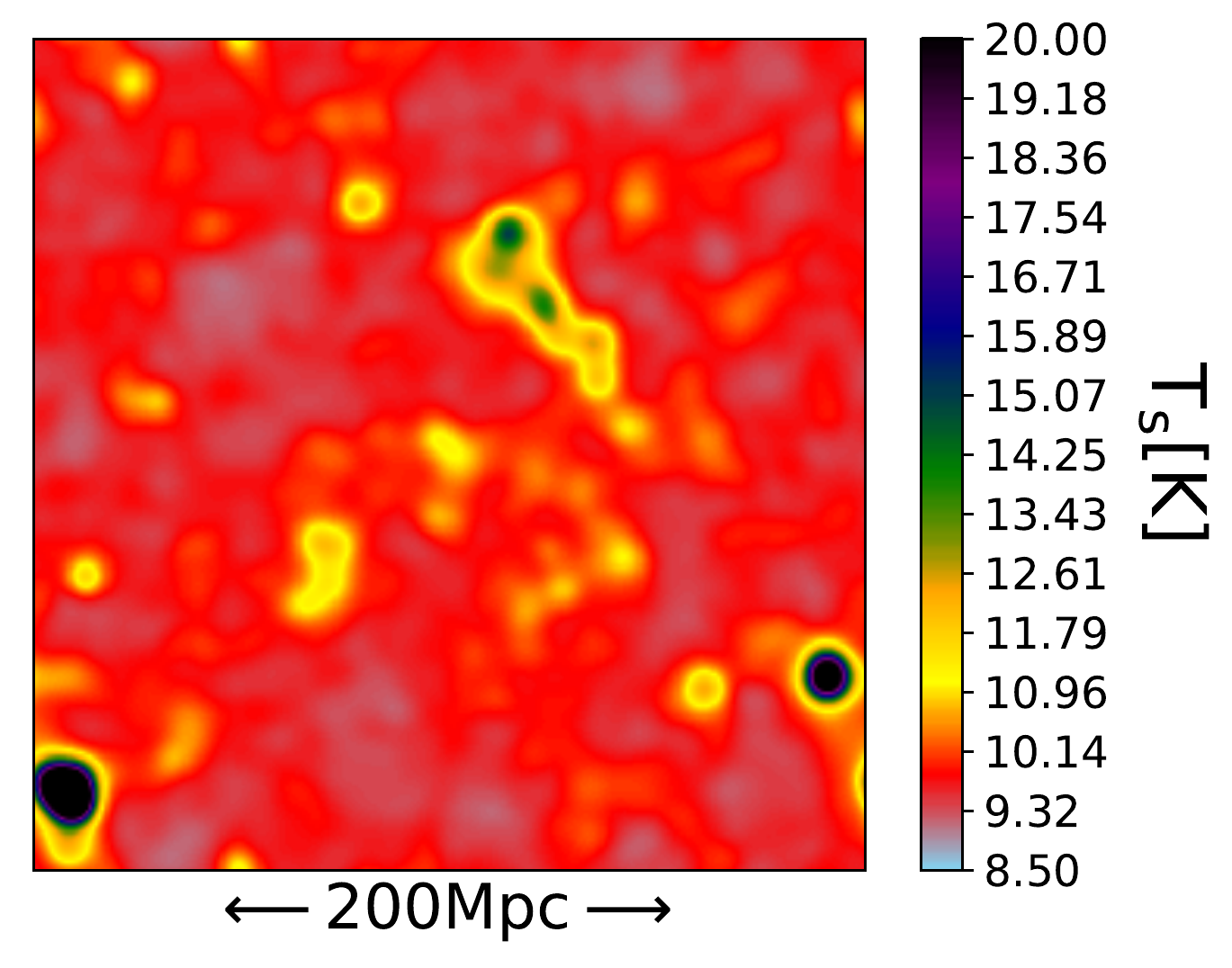}}}
		\resizebox{1.4in}{1.5in}{{\includegraphics{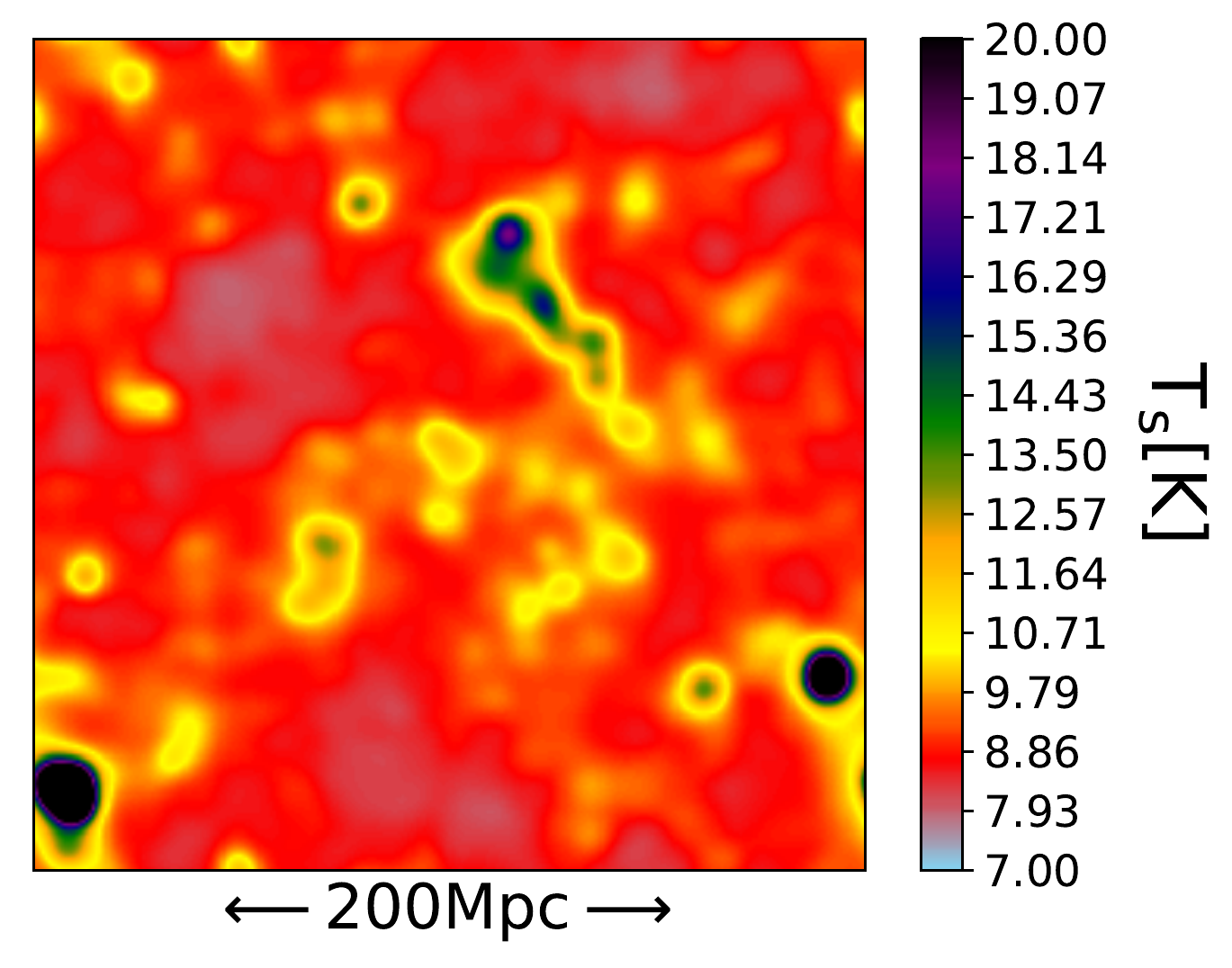}}}\\
		\resizebox{1.4in}{1.5in}{{\includegraphics{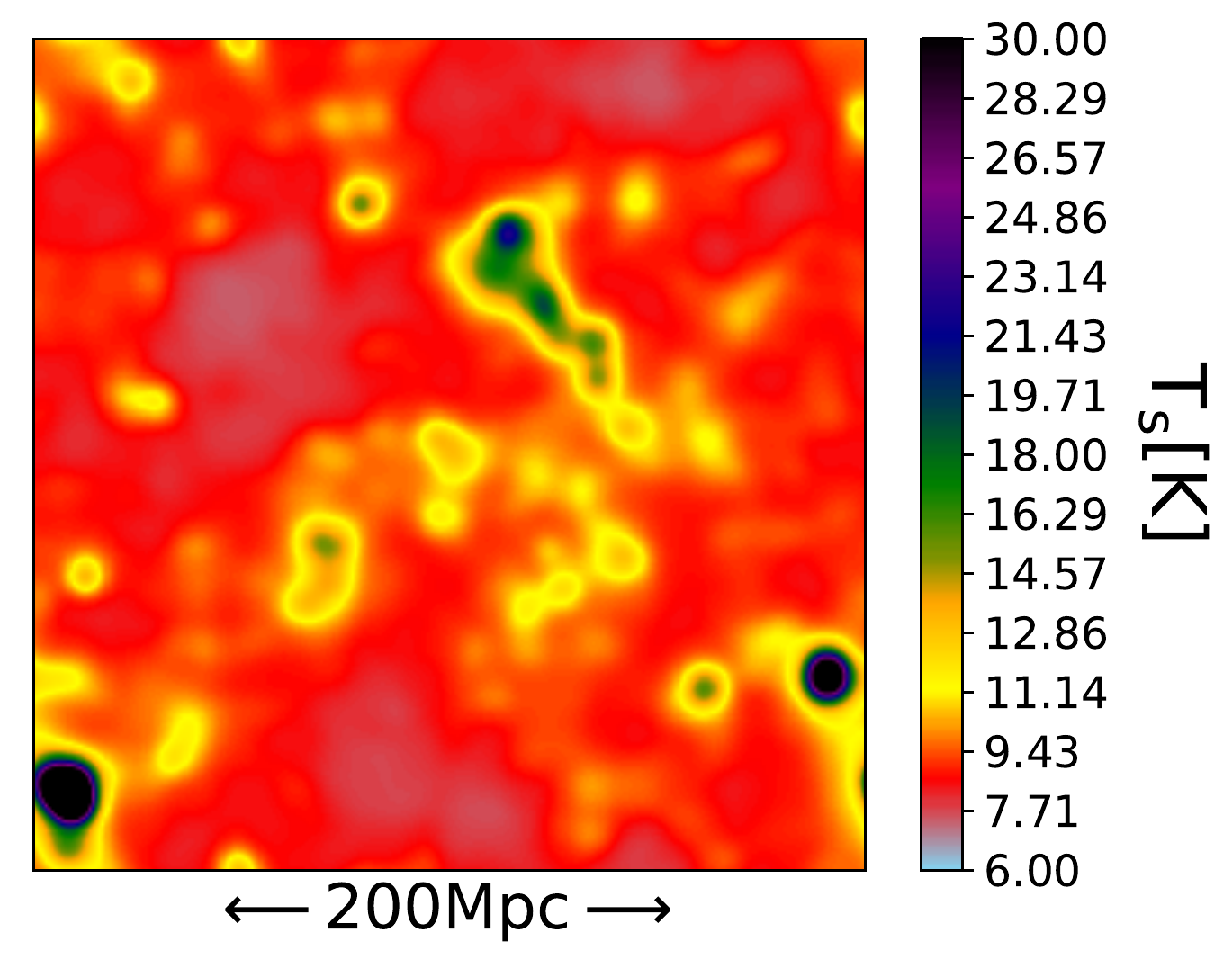}}}
		\resizebox{1.4in}{1.5in}{{\includegraphics{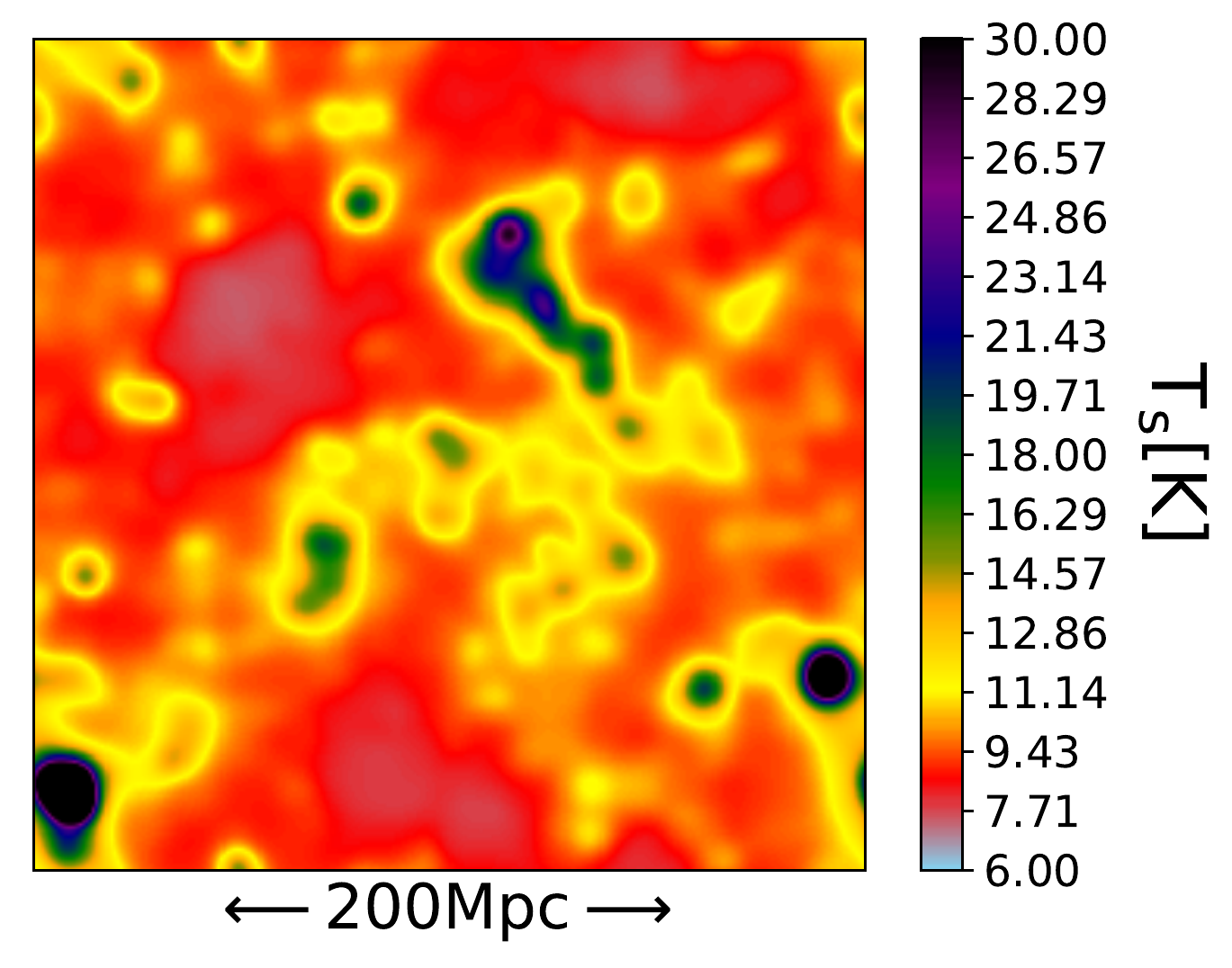}}}
		\resizebox{1.4in}{1.5in}{{\includegraphics{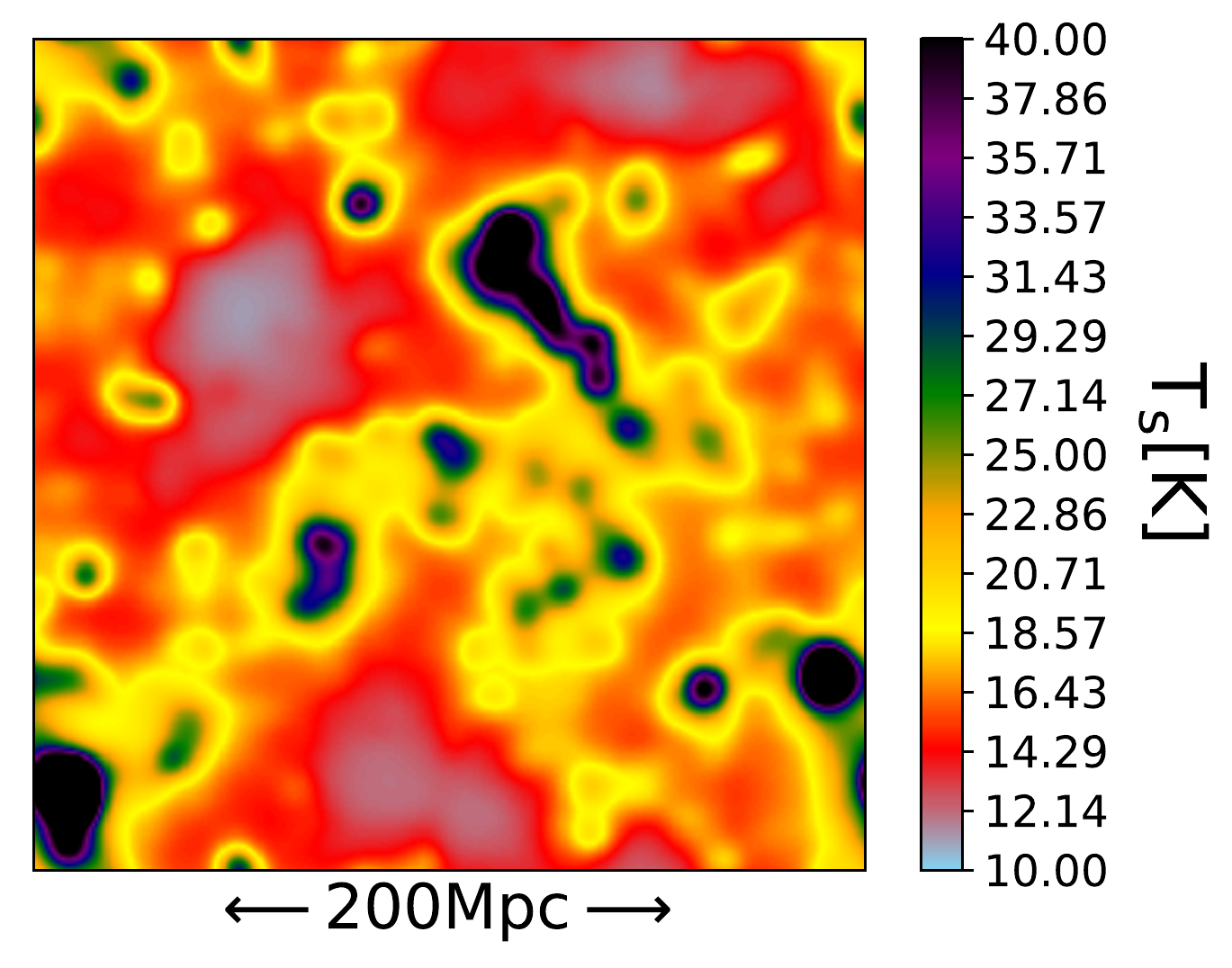}}}
		\resizebox{1.4in}{1.5in}{{\includegraphics{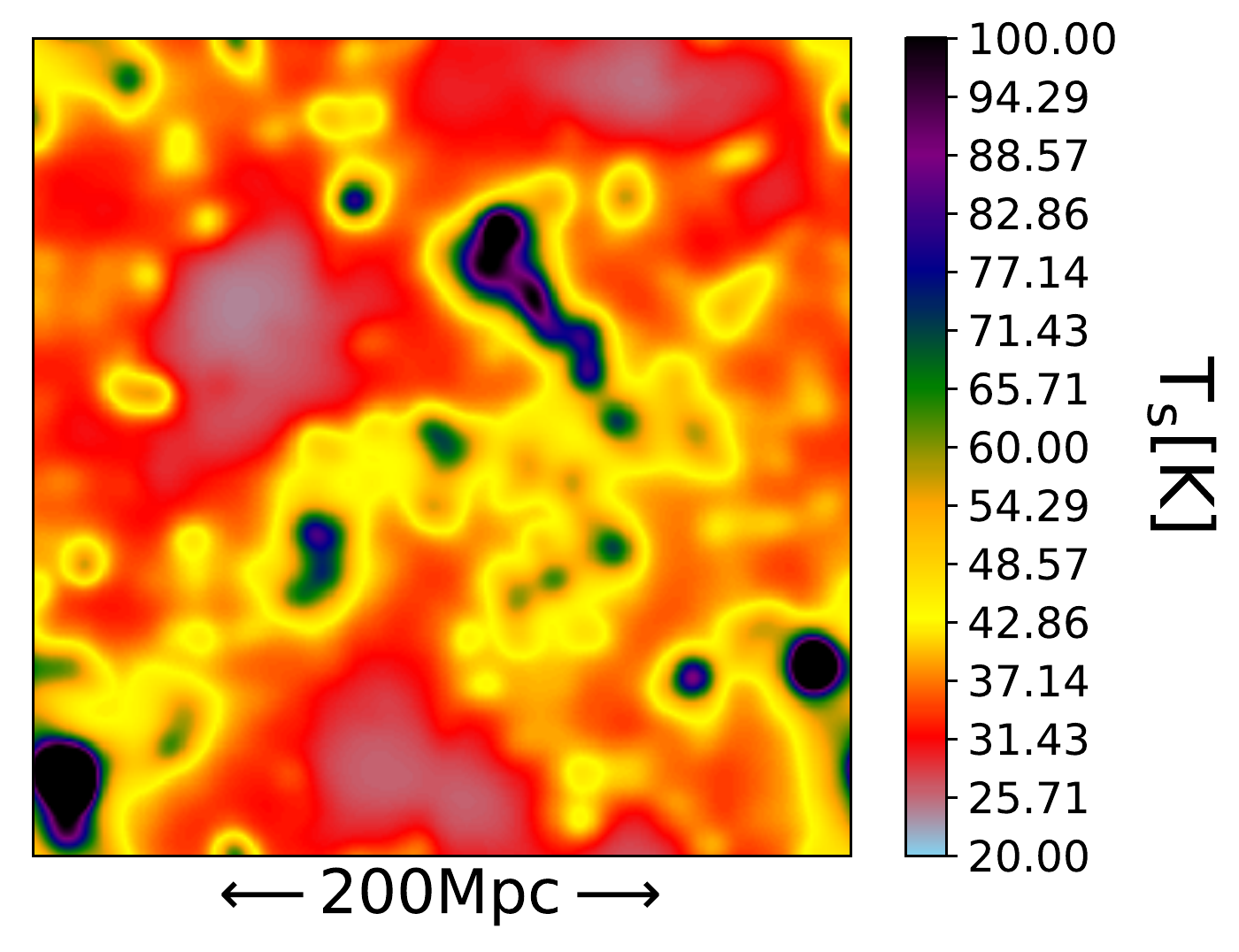}}}   
	\end{center}
	
	\caption{Evolution of spin temperature with redshift at different redshifts for our \textit{Fiducial Model}. \textit{Top row: (Left to Right)} z=20.22, 18.60, 17.11 and 16.41. \textit{Bottom row: (Left to Right)} z= 15.73, 15.08, 13.28 and 11.68. Note that since the range of field values vary with redshift, the color coding in the colorbar changes accordingly.} 
	\label{fig:Ts_maps_fid}
\end{figure}

\begin{figure}[h]
	\begin{center}

	\resizebox{2.in}{2.6in}{{\includegraphics{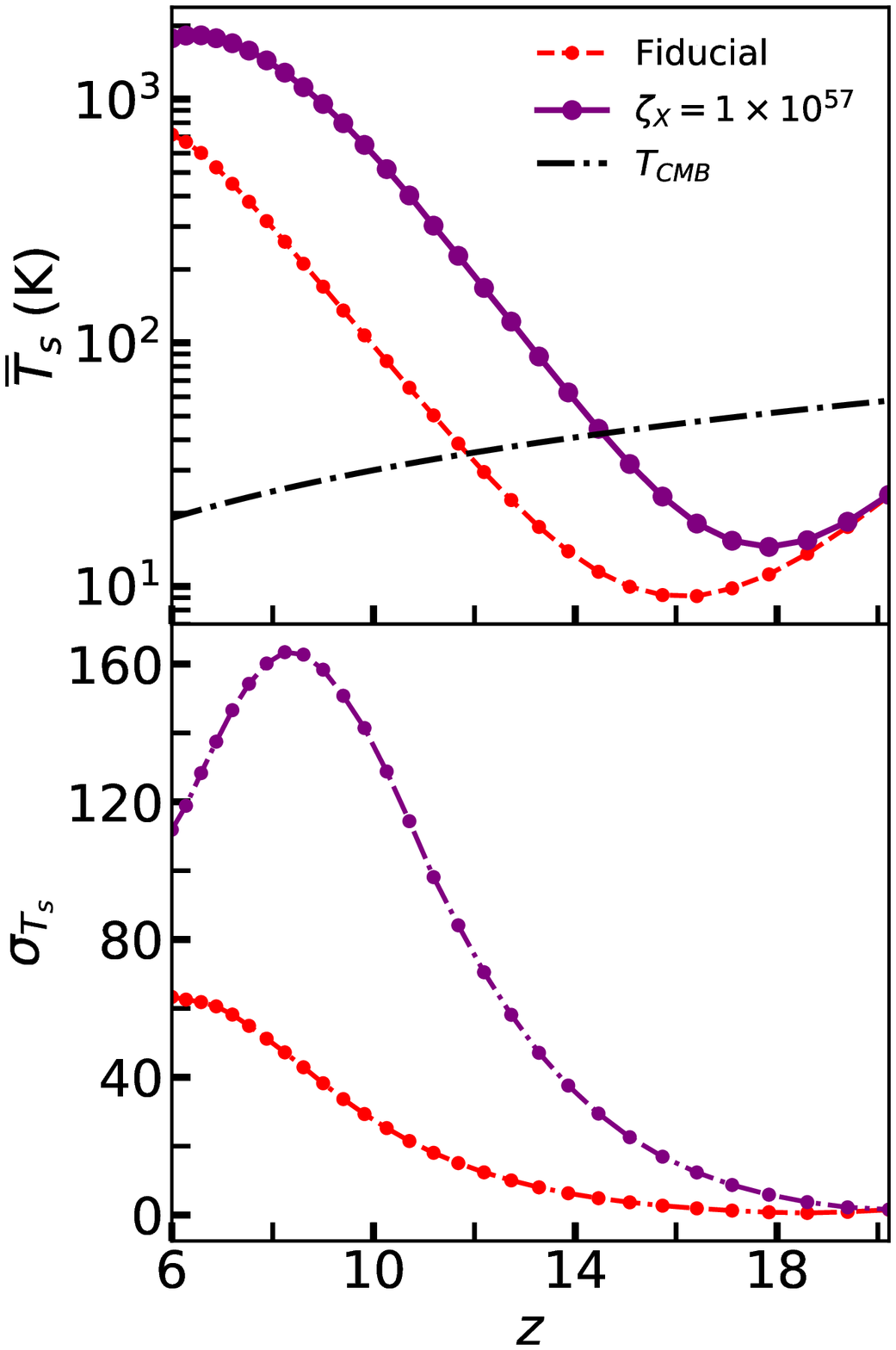}}}
	\resizebox{2.in}{2.6in}{{\includegraphics{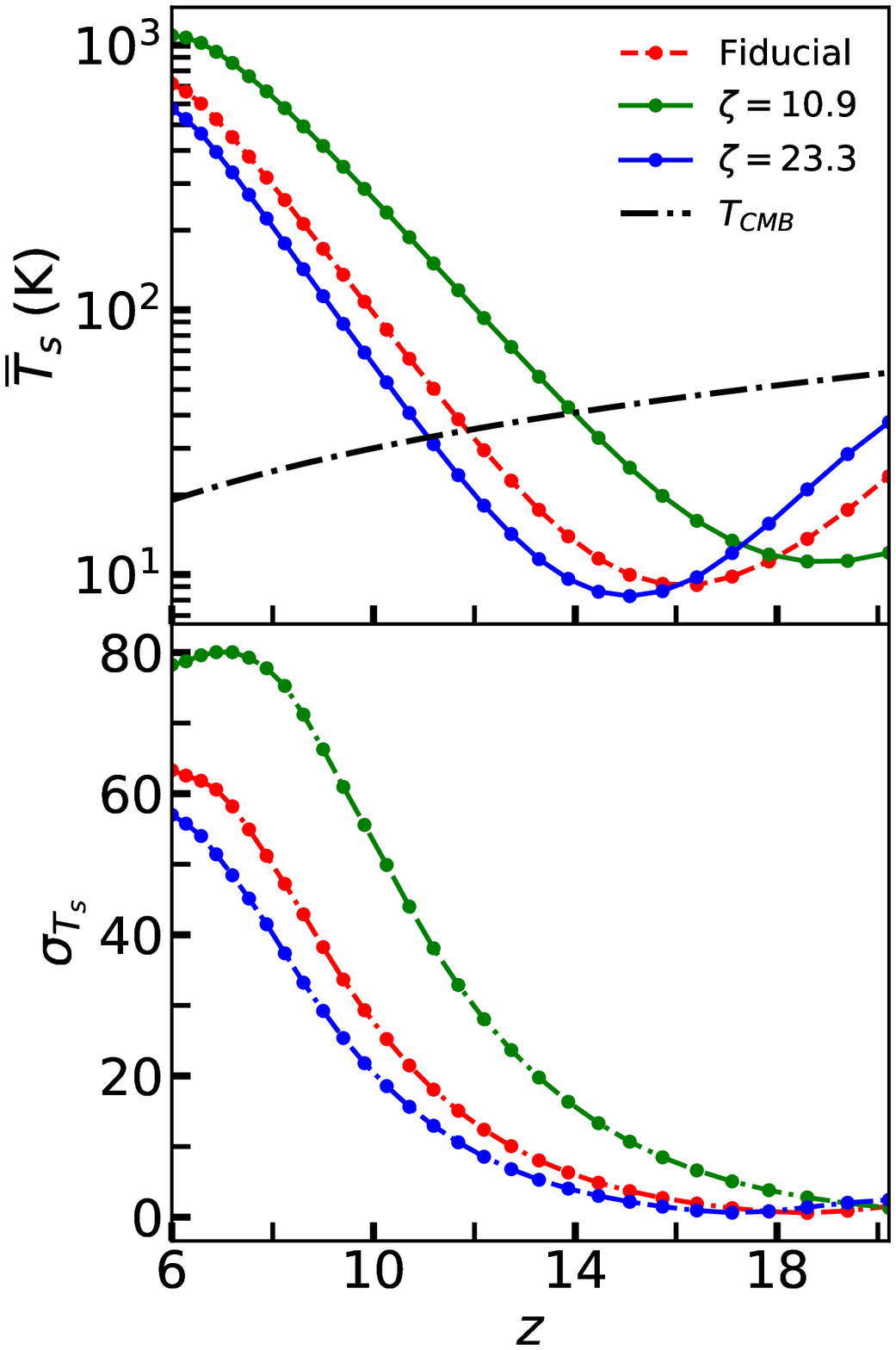}}}
\end{center}
\caption{Evolution of the mean and standard deviation for spin temperature with redshift, relative to the \textit{fiducial model} for an enhanced X-ray heating efficiency in the (\textit{left}) and different values of $T_{vir}$ (\textit{right}). The black dashed line marks the evolution of the temperature of the CMB.} 
\label{fig:Ts_avg_sd}
\end{figure}

In this section we analyse the morphology of the $T_s$ field. As described in Eq.~\ref{eqn:Ts}, the evolution of $T_s$ is a result of the evolution of $T_{\gamma}$, $T_k$, $x_c$ and $x_{\alpha}$. For the redshift range under study the collisional coupling constant $x_c$ satisfies $x_c \ll x_{\alpha}$ where $x_{\alpha}$ is the Ly-$\alpha$ coupling constant. This is because as the universe expands, the probability of collisions between $e^- e^-$, $e^- H$ and $H-H$ decreases. The Ly-$\alpha$ coupling constant $x_{\alpha}$  depends upon the emissivity of sources capable of producing Ly-$\alpha$ transitions. Ly-$\alpha$ excitations occur due to emission from the first collapsed objects. Since Ly-$\alpha$ is a lower energy transition, excitation is possible by low emissivity sources, unlike X-rays which requires more efficient sources. Therefore, Ly-$\alpha$ coupling will precede X-ray heating of the IGM (Sec.~3 of \cite{Mesinger:2010ne}). Ly-$\alpha$ does not contribute much to the heating of the IGM but couples $T_s$ to $T_k$ \cite{FurlanettoStov:2010}. Therefore, prior to X-ray heating while $T_k$ is still following adiabatic cooling due to the expansion of the universe, regions with higher matter density will have higher value of $x_{\alpha}$. Due to this reason the $x_{\alpha}T_k$ term dominates in the expression for $T_s$ in eq.~\ref{eqn:Ts}. In the redshift range under study $T_k < T_{\gamma}$. Therefore $T_k \le T_s \le T_{\gamma}$. Higher the value of $x_{\alpha}$ lower is the value of $T_s$ and it approaches $T_k$. Otherwise it approaches $T_{\gamma}$. If ${x_{\alpha} \gg 1}$ and $T_{\gamma}^{-1} \ll x_{\alpha} T_k^{-1} $ then Ly-$\alpha$ coupling saturates which means $T_s \sim T_k$. In regions where coupling due to Ly-$\alpha$ is still inefficient, $T_s$ will be higher than $T_k$ but less than $T_{\gamma}$, as can be seen from eq. \ref{eqn:Ts}. Therefore prior to X-ray heating, any fluctuation in the matter density field will lead to fluctuation in $f_{coll}$ and hence $x_{\alpha}$ which further leads to fluctuation in $T_s$. Note that in this regime, $T_k$ is not fluctuating but follows adiabatic cooling due to the expansion of the universe. Eventually Ly-$\alpha$ coupling saturates in most of the IGM, where $T_k$ is coupled to $T_s$. Meanwhile, X-ray heating starts in very high density regions and $T_k$ begins to rise. In these regions the fluctuations in $T_s$ are a result of the fluctuations in $T_k$. Since X-rays have large mean free path, the effect of X-ray heating is not localized around high density regions and soon permeates the entire IGM, until all of the IGM is under the influence of X-ray heating. Then the fluctuations in $T_s$ are completely determined by the fluctuations in heating due to X-rays.
 
 In Fig.~\ref{fig:Ts_maps_fid}, we exhibit the maps of $T_s$ for our \textit{fiducial model}. Note that the colour bars have different range for every map as the temperature ranges change with redshift. We describe the evolution from $z=20.22$ to $z=11.68$ and largely focus on interpreting the morphology at these redshift values. At lower redshifts fluctuations in $T_s$ are not reliable from 21cmFAST as the code does not take into account the effect of fluctuations in the ionization fields on the evolution of heating fluctuations. Therefore, we shall not interepret the $T_s$ field below redshift values where $x_{HI} \lesssim 0.8$.  Starting from top left map at z=20.22 we find that the high density regions are cooler regions, surrounded by lower density warmer regions where Ly-$\alpha$ coupling is inefficient. Heating due to X-rays has not begun at this redshift. In the next panel at $z=18.60$, some X-ray heated regions appear in places which correspond to the coolest regions in the maps at $z=20.22$ (c.f. violet regions in the map). These are regions of highest density where emmissivity of sources is sufficient for X-ray heating to start. The rest of the regions are still dominated by fluctuations in Ly-$\alpha$ coupling alone. Also notice that there is less scatter in the value of $T_s$ at these redshifts and all of the IGM has temperature less than that at the previous redshift of $z=20.22$, except for places where X-ray heating has started. The sky blue regions are the coolest, yet high density regions. In the next panel at $z=17.11$ we see that in some of the coolest regions at $z=18.60$, some new X-ray sources appear while the rest of the IGM decreases further in temperature (note the lower limit of the colour bar). On the other hand in regions where X-ray sources appeared at earlier redshifts, those heated regions have increased in size as the effect of X-rays starts permeating outwards. This same trend of newer X-ray sources appearing, the increase in size of older X-ray heated regions and the rest of the IGM decreasing in temperature, is seen until the map for $z=15.08$. Thereafter we see that even the lowest temperature regions are increasing in temperature. Therefore fluctuations in $T_s$ maps at $z\sim 13.28$ and $z\sim11.68$ are dominated by fluctuations in X-ray heating. Now the cooler regions are regions which are far away from the X-ray sources where only a few X-ray photons have reached. The effect of X-rays is not localized around its sources due to its higher mean free path. At the lower redshifts we see that heated regions grow in size and merge with nearby X-ray heated regions.

 The evolution of $\overline{T}_s$ with $z$ is shown in Fig.~\ref{fig:Ts_avg_sd}. The evolution until $z \sim 16$ is similar to that for adiabatic cooling. This is the regime where most of the IGM is coupled to $T_k$ and any fluctuations in $T_s$ will be dominated by fluctuations in $x_{\alpha}$ as $T_k$ is not fluctuating but uniformly decreasing as a result of adiabatic cooling. As X-ray heating starts to dominate over adiabatic cooling due to the expansion of the universe, the evolution of $T_s$ shows a turnover from an initial period of decrease (at $z \lesssim 16.4$ in the case of our \textit{fiducial model}). This appears as an absorption peak in the $\overline {\delta T_b}$ evolution ~\cite{Mesinger:2010ne}. Now $T_k$ is also a fluctuating component as heating starts around high density regions. It is the fluctuations in $T_k$ that dominate the fluctuations in $T_s$ at the redshifts where X-ray heating dominates.
 
\begin{figure}
	
	\begin{center}
		
		\resizebox{3.0in}{3.5in}{{\includegraphics{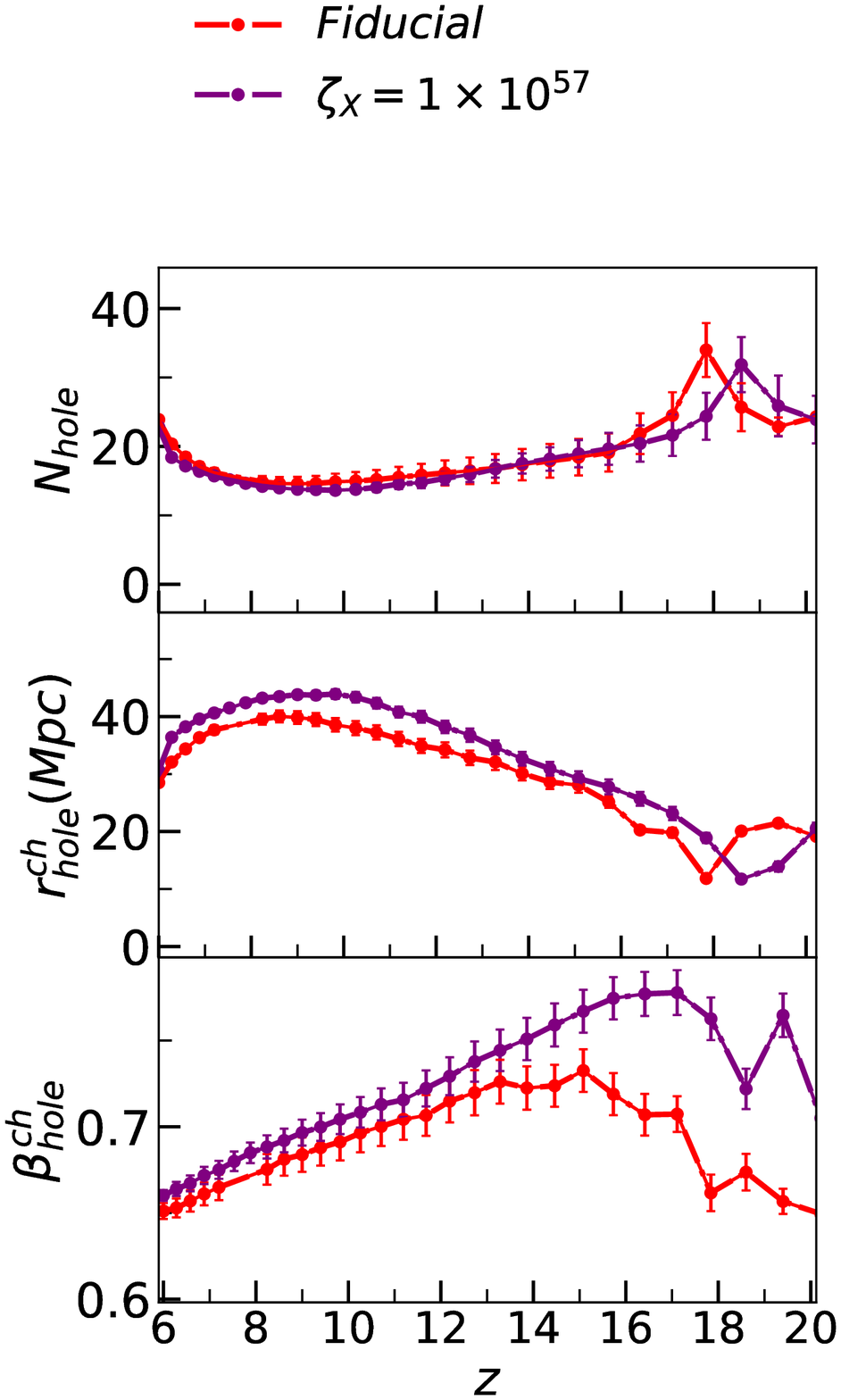}}}
		\resizebox{3.0in}{3.5in}{{\includegraphics{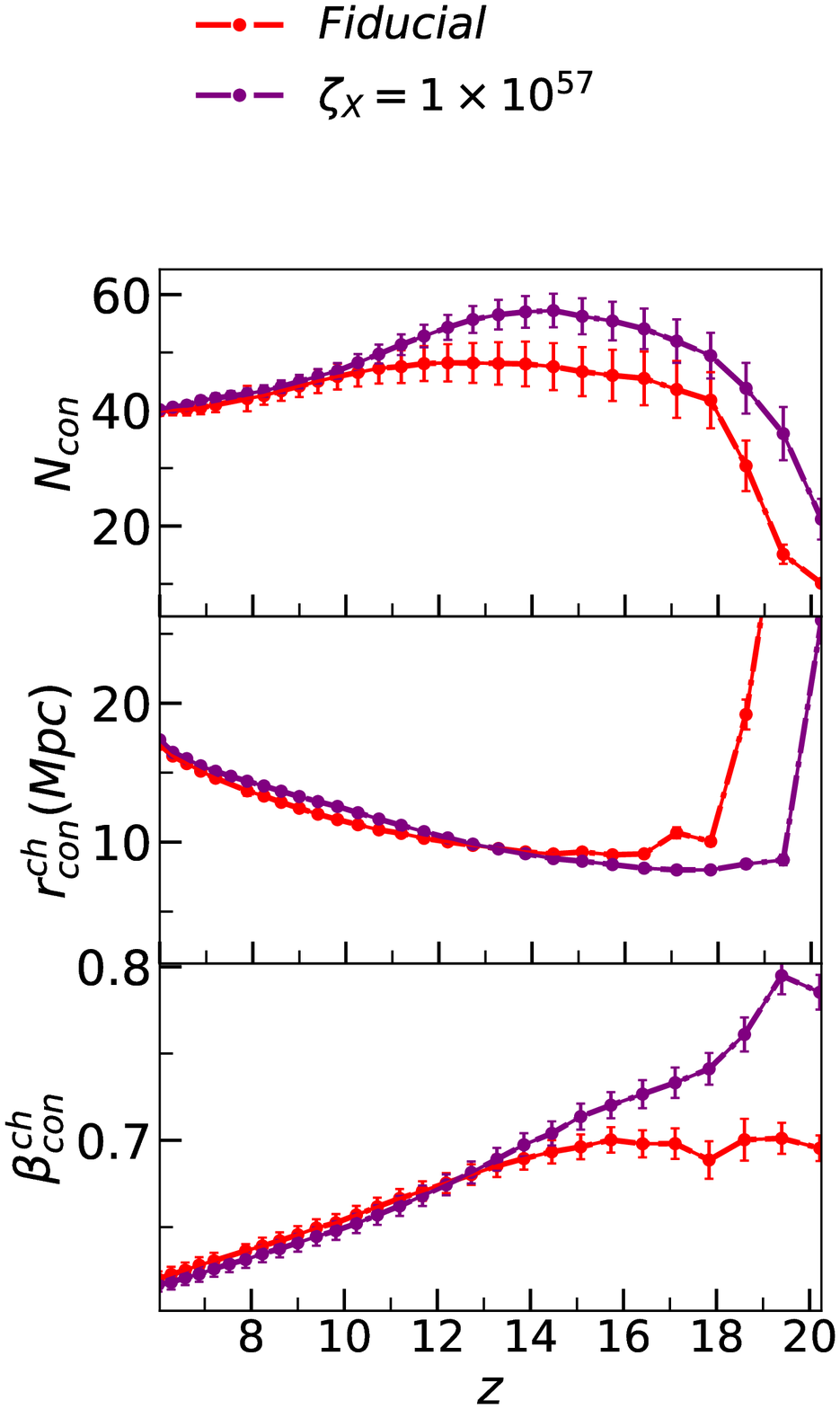}}}

	\end{center}
	\caption{The morphology of spin temperature $T_s$ for the \textit{fiducial model} relative to the model with an increased X-ray heating. }
	\label{fig:Ts_z_morph1}
\end{figure}

\begin{figure}
	
	\begin{center}
		
		\resizebox{3.0in}{3.5in}{{\includegraphics{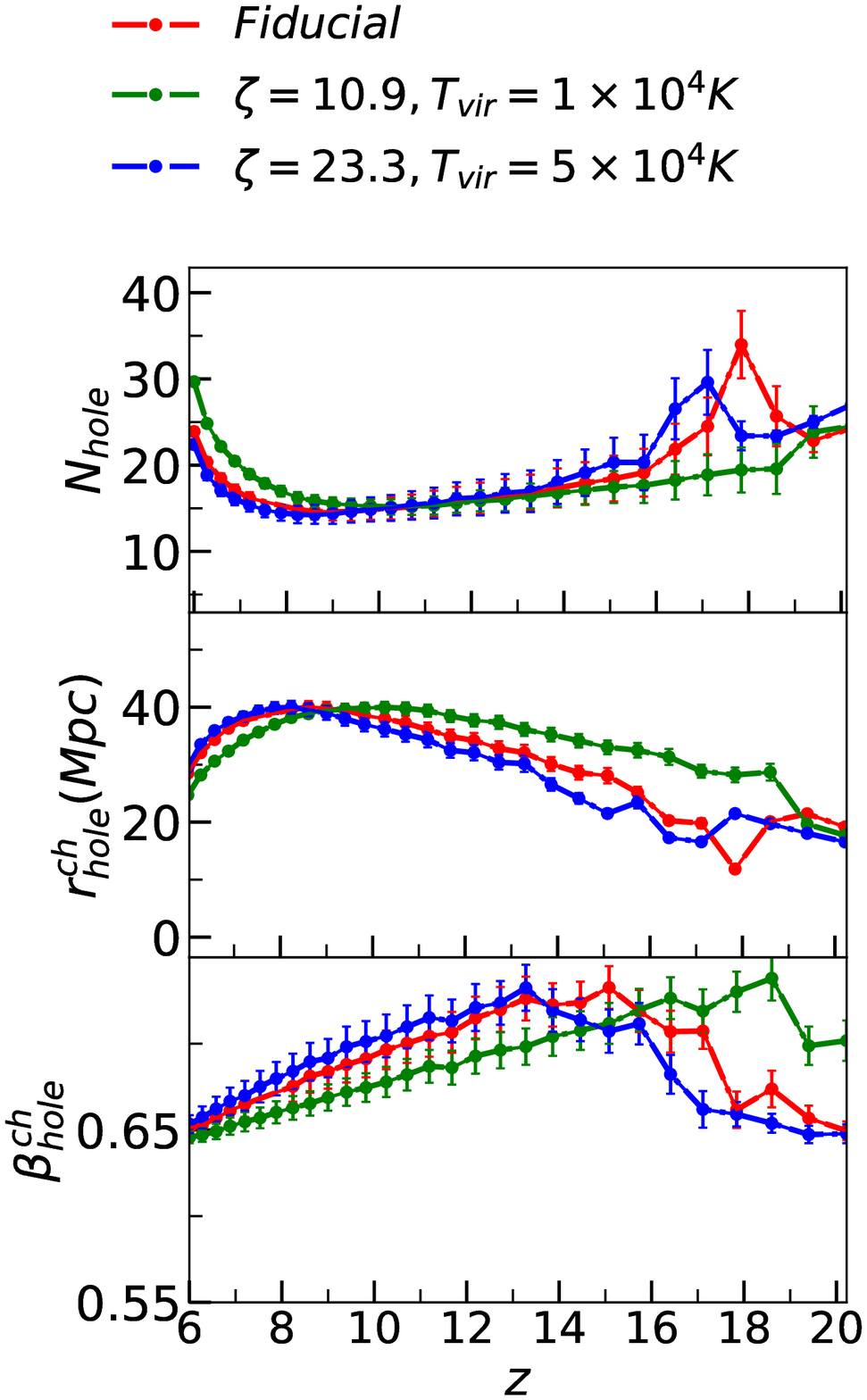}}}
		\resizebox{3.0in}{3.5in}{{\includegraphics{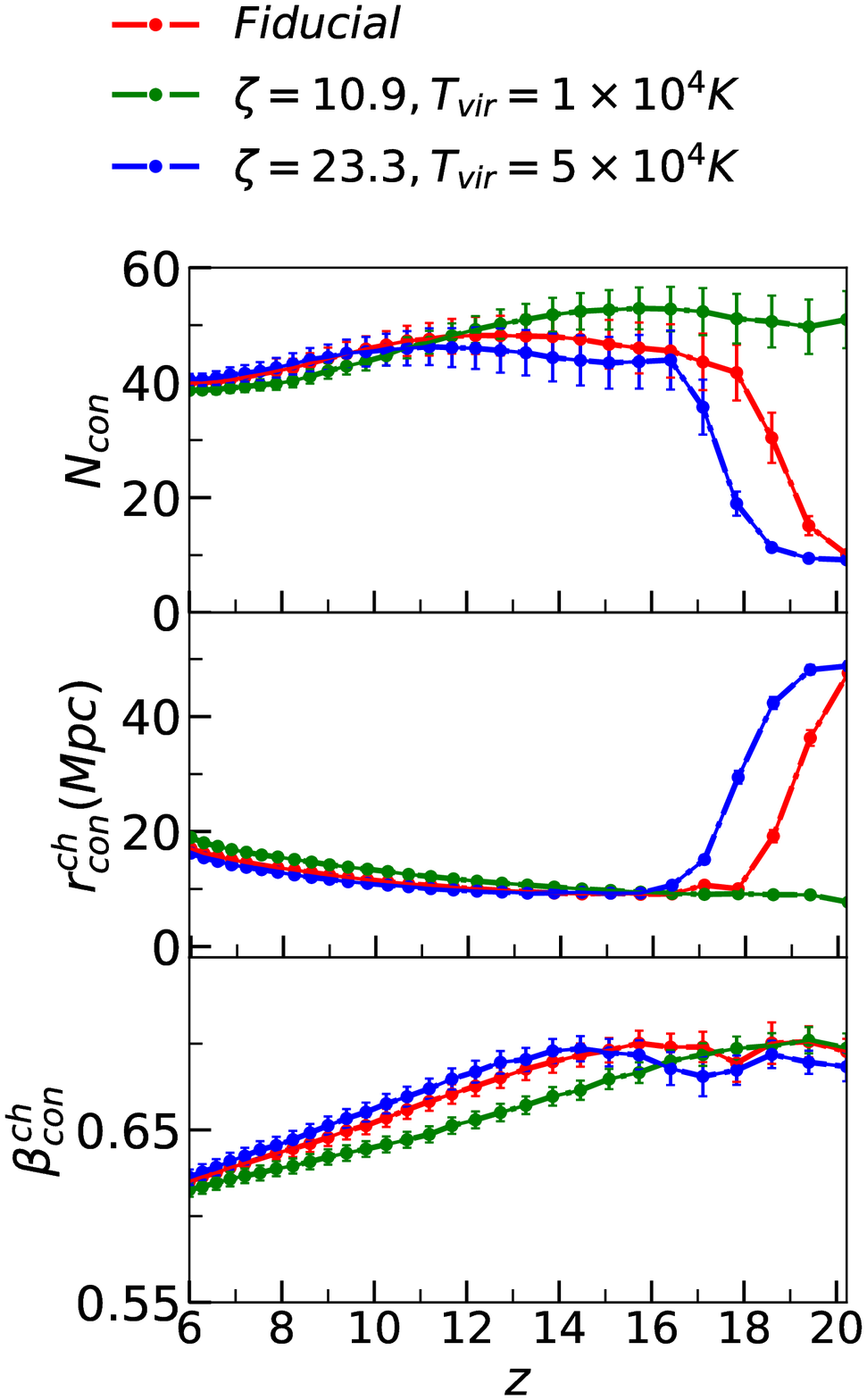}}}

	\end{center}
	\caption{The morphology of spin temperature $T_s$ for the \textit{fiducial model} relative to the models with different $T_{vir}$ values.}
	\label{fig:Ts_z_morph2}
\end{figure}

 In Fig.~\ref{fig:Ts_z_morph1}, we show the morphology of $T_s$ for models with a different X-ray heating efficiency, while Fig.~\ref{fig:Ts_z_morph2} shows models with different values of $T_{vir}$ in comparison to the \textit{fiducial model}. The connected regions correspond to hotter regions while holes correspond to low temperature valleys. We shall first focus on interpreting the $T_s$ morphology for our \textit{fiducial model} (plotted in red in Fig.~\ref{fig:Ts_z_morph1} and Fig.~\ref{fig:Ts_z_morph2}).
  We see from the plot in the top panel that initially until $z \gtrsim 18$ , $N_{hole}$ does not vary significantly while $N_{con}$ is increasing with decreasing redshift. We find $N_{hole}>N_{con}$. This shows that initially when Ly-$\alpha$ coupling dominates the field, the morphology is dominated by holes.  These holes are the cooler regions where Ly-$\alpha$coupling is more efficient, surrounded by higher temperature regions where the coupling is inefficient. These surrounding relatively higher temperature regions (which would be a single large connected region punctured by holes) and one or two scattered X-ray heated regions correspond to connected regions (which would be isolated small connected regions inside holes which are lying inside the single big connected region described above). As described in the maps above, the coolest regions at an early redshift become sites where X-ray sources appear at later redshifts. Therefore regions which correspond to holes switch over to connected regions later on. This leads to a decrease in the number of holes and an increase in the number of connected regions with redshift as more X-ray sources begin to appear. 
  
  The evolution of $r^{ch}_{con}$ and $r^{ch}_{hole}$ in the middle panel shows an initial drop  in $r^{ch}_{con}$ until $z\sim 18$. At these high redshifts (c.f. map for $z=20.22$ in Fig.~\ref{fig:Ts_maps_fid}), connected regions correspond to larger hotter regions adiabatically cooling in the low density voids. Later very small X-ray heated regions start appearing around sources. As more X-ray sources appear, the average of the sizes starts to be dominated by the connected regions corresponding to X-ray heated regions (isolated small connected regions inside holes). Therefore we observe a drop in $r^{ch}_{con}$. After $z \sim 17$ the morphological properties of connected regions are morphologies of X-ray heated regions. These X-ray heated regions grow and merge with nearby X-ray heated regions. Therefore there is an increase in $r^{ch}_{con}$ with redshift.
  
   As X-ray heating dominates, connected regions correspond to higher temperature regions concentrated around high density regions and holes are coooler regions far away from the X-ray sources. As heating proceeds, these X-ray heated regions grow in size. Therefore we get a mild increase in $r^{ch}_{con}$ . The evolution of $r^{ch}_{hole}$ also shows an increase with redshift. This is because initially the holes are those concentrated around Ly-$\alpha$ efficient sources. Inside these holes X-ray heating starts taking place and the inner regions of the holes now host connected regions. Since, $r^{ch}_{hole}$ is an average over the thresholds for holes, it would have contribution from the outer bigger contours of the holes and as the inner ones are now occupied by connected regions, they also have an inner boundary (the holes would be like a ring around connected regions due to X-ray heated regions on the inner boundary and inefficiently coupled relatively hotter regions on the outer boundary(c.f. the skyblue regions around X-ray heated sources in the panel for $z=18.60$) in Fig.~\ref{fig:Ts_maps_fid}). Therefore the overall size of the holes increases as X-ray heated regions concentrate in the inner regions of holes and expand. At later redshifts the holes are the coolest regions which are far away from X-ray sources and are influenced by few X-ray photons reaching them (c.f. the map for $z=13.28$ and 11.68 in Fig.~\ref{fig:Ts_maps_fid}). Therefore the evolution is not as rapidly changing at these later redshifts.

   The bottom panels  describe the evolution of $\beta^{ch}_{hole}$ and $\beta^{ch}_{con}$. The variation of $\beta^{ch}_{con}$ shows constant evolution for early redshifts and a steady decrease thereafter. The initial constant evolution is because initially the connected regions do not evolve much as these are in low density voids where Ly-$\alpha$ coupling is inefficient and the effect is that of uniform adiabatic cooling. The shape of these regions is not affected until the effect of X-ray heating reaches them. They may also correspond to scattered but few X-ray heated regions where X-ray heating has just started. These regions would be localized peaks around X-ray sources and would be isolated. Therefore other than a change in size there is no change in the shape of these regions. As X-ray heating proceeds to uniformity, these regions merge with nearby X-ray heated regions which leads to an increase in anisotropy. On the other hand $\beta^{ch}_{hole}$ shows a decrease in anisotropy initially, followed by an increase around $z \sim 18$.  The gradual transition from this initial increase to a decrease around $z \sim 15$ is due to a flip in the interpretation of holes as regions in low density voids where the effect of X-ray heating has not reached. These are not localized regions, unlike the cooler regions at earlier redshifts.

In Fig.~\ref{fig:Ts_z_morph1} we show the redshift evolution of $N_{hole,con}$, $r^{ch}_{con,hole}$ and $\beta^{ch}_{con,hole}$ for the model with increased X-ray heating efficiency ($\zeta_X = 1 \times 10^{57}$). We find that the overall shape of the plots is the same while there is a general shift towards higher redshifts. This is because an increased X-ray emmissivity leads to an early heating of the IGM. Note that the collapsed fraction is the same at a given redshift in both the cases, only the X-ray emmissivity is higher for a greater value of $\zeta_X$.

In Fig.~\ref{fig:Ts_z_morph2} we show the redshift evolution of $N_{hole,con}$, $r^{ch}_{con,hole}$ and $\beta^{ch}_{con,hole}$ for the models with different $T_{vir}$ values. The error bars denote the error in mean over 32 slices. Note that here the X-ray heating efficiency is the same for all the three cases (i.e. $\zeta_X=2 \times 10^{56}$). We observe that the shape of the plots is the same apart from a shift towards higher $z$ values for lower $T_{vir}$ (less massive) sources. Lower $T_{vir}$ leads to a higher collapse fraction at a given redshift relative to higher $T_{vir}$ values. Therefore there are more numerous sources which leads to this shift towards higher $z$. However the overall X-ray emissivity would be lower. Therefore the redshift evolution is more gradual for the lowest $T_{vir}$ values. This trend can be seen in both Fig.~\ref{fig:Ts_avg_sd} and Fig.~\ref{fig:Ts_z_morph2}.

	
%


%

\section{\bf{Morphology of the Brightness Temperature field: $\delta T_b$}}

\begin{figure}[h]
	\begin{center}
		
		\resizebox{1.8in}{2.4in}{{\includegraphics{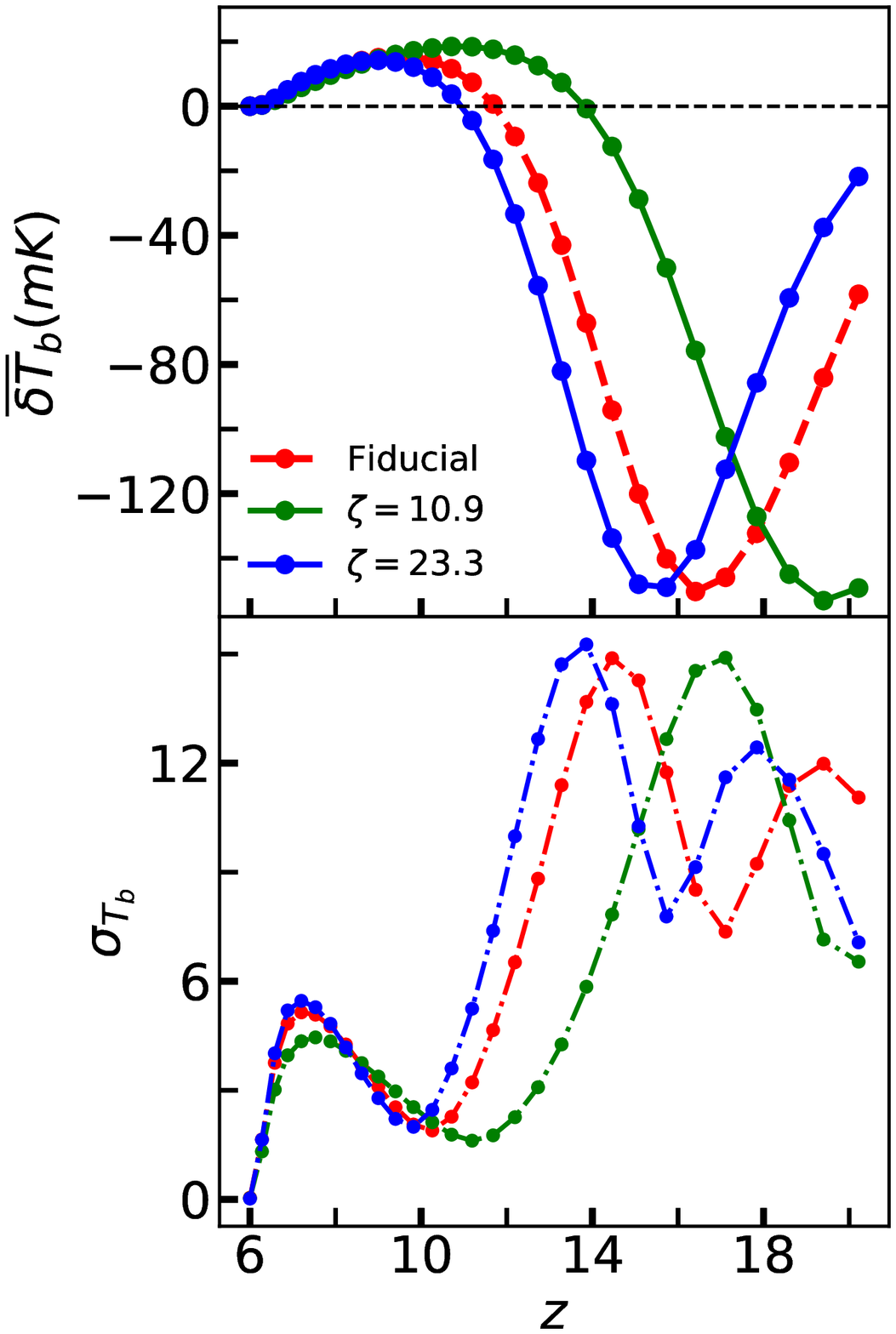}}}
		\resizebox{1.8in}{2.4in}{{\includegraphics{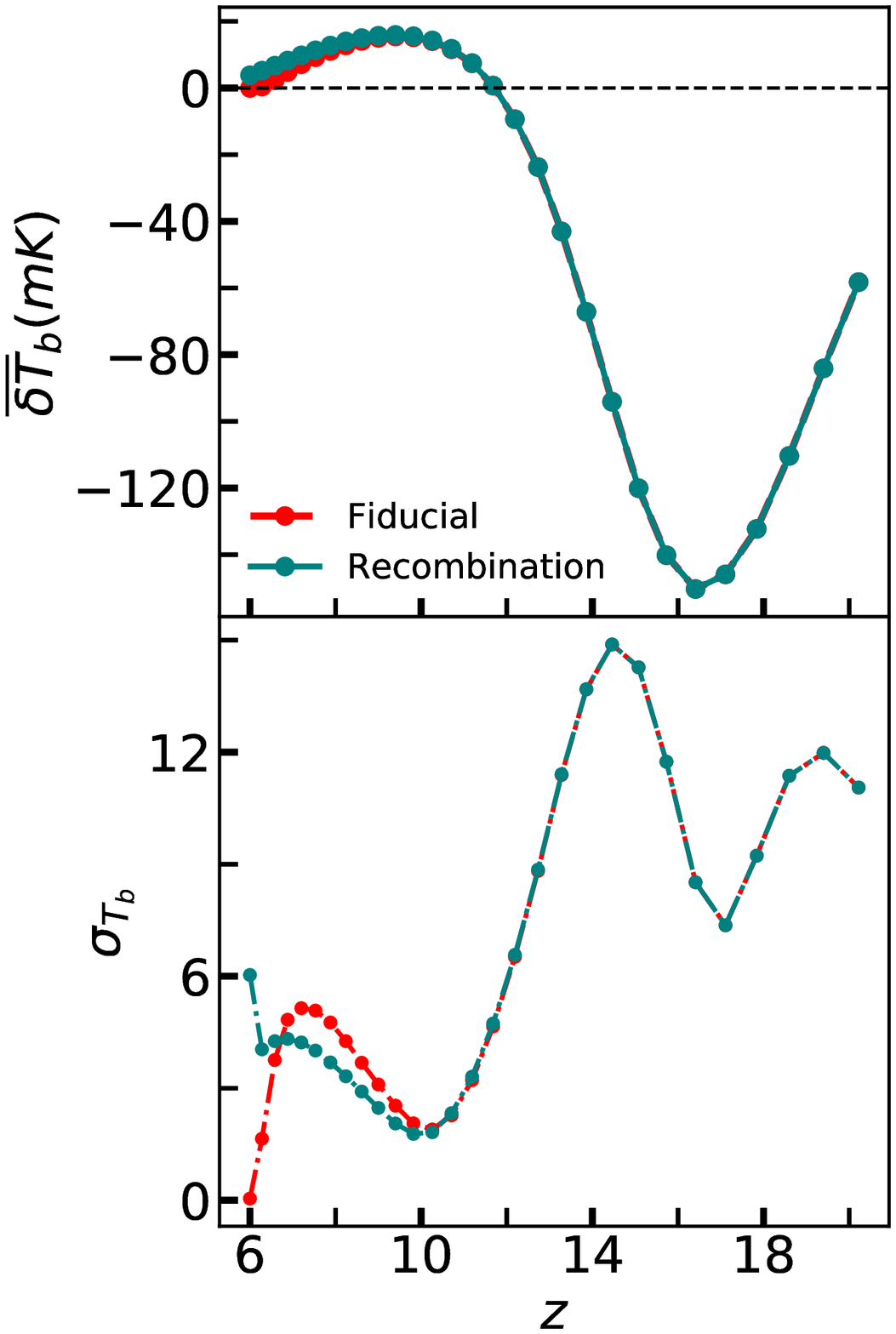}}}	\resizebox{1.8in}{2.4in}{{\includegraphics{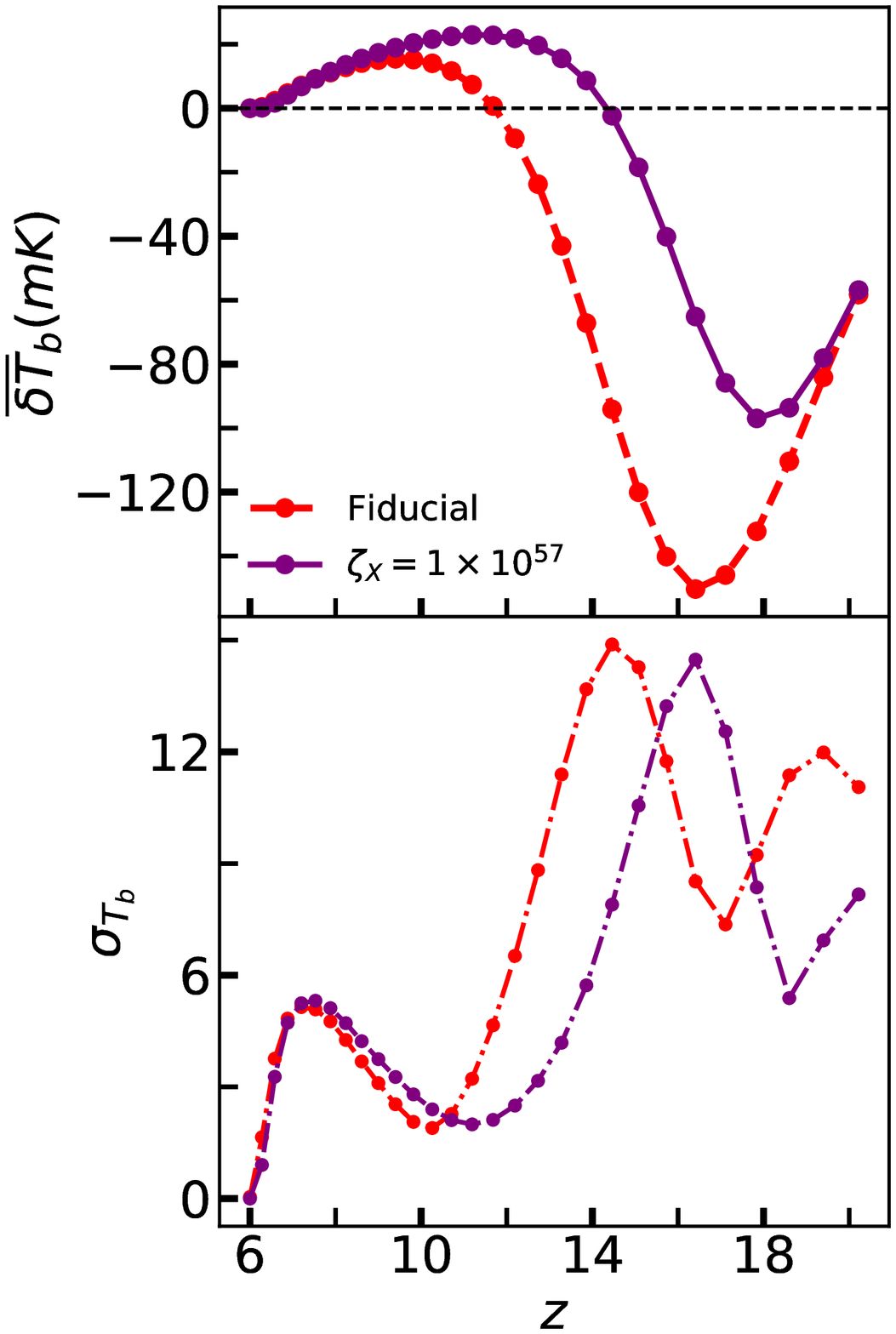}}}
		\caption{The evolution of the mean 21cm brightness temperature $\delta T_b$ for the different models, relative to the \textit{fiducial model} (in Red).}
		\label{fig:Tb_avg}
	\end{center}
\end{figure}

The evolution of the brightness temperature $\delta T_b$ is determined by the evolution of $x_{HI}$, $T_s$ and $\delta_{nl}$ fields. The fluctuation in $\delta T_b$ is sourced by those in $\delta_{nl}$ until the growing non linearities become important. However as the first objects form, which is a highly non linear process and reionization and X-ray heating progresses, the fluctuations are not directly sourced by the underlying density fluctuations but by the processes of heating and ionization. In Fig. \ref{fig:Tb_avg} we show the redshift evolution of the average brightness temperature $\delta \overline{T}_b$ and its standard deviation $\sigma_{T_b}$ for the various models under consideration. Transitions or turnovers are as expected for different models \cite{Furlanetto:2006tf}. The main transition points are the dip where X-ray heating dominates over Ly-$\alpha$ coupling, followed by the transition point where the fluctuations due to ionization dominates (i.e. where the plot crosses the horizontal dashed line to a positive $\delta T_b$ value). The evolution of $\sigma_{T_b}$ shows three peaks. The first peak at the highest $z$ values corresponds to the regime where the fluctuations in Ly-$\alpha$ coupling dominates and saturate. This is followed by the second peak which describes the regime where fluctuations due to X-ray heating take over and saturate. The third peak is due to the fluctuations in $x_{HI}$ field which dominate in this regime as reionization progresses.  

\begin{figure}
	\begin{center}
		
		\resizebox{3.in}{3.8in}{{\includegraphics{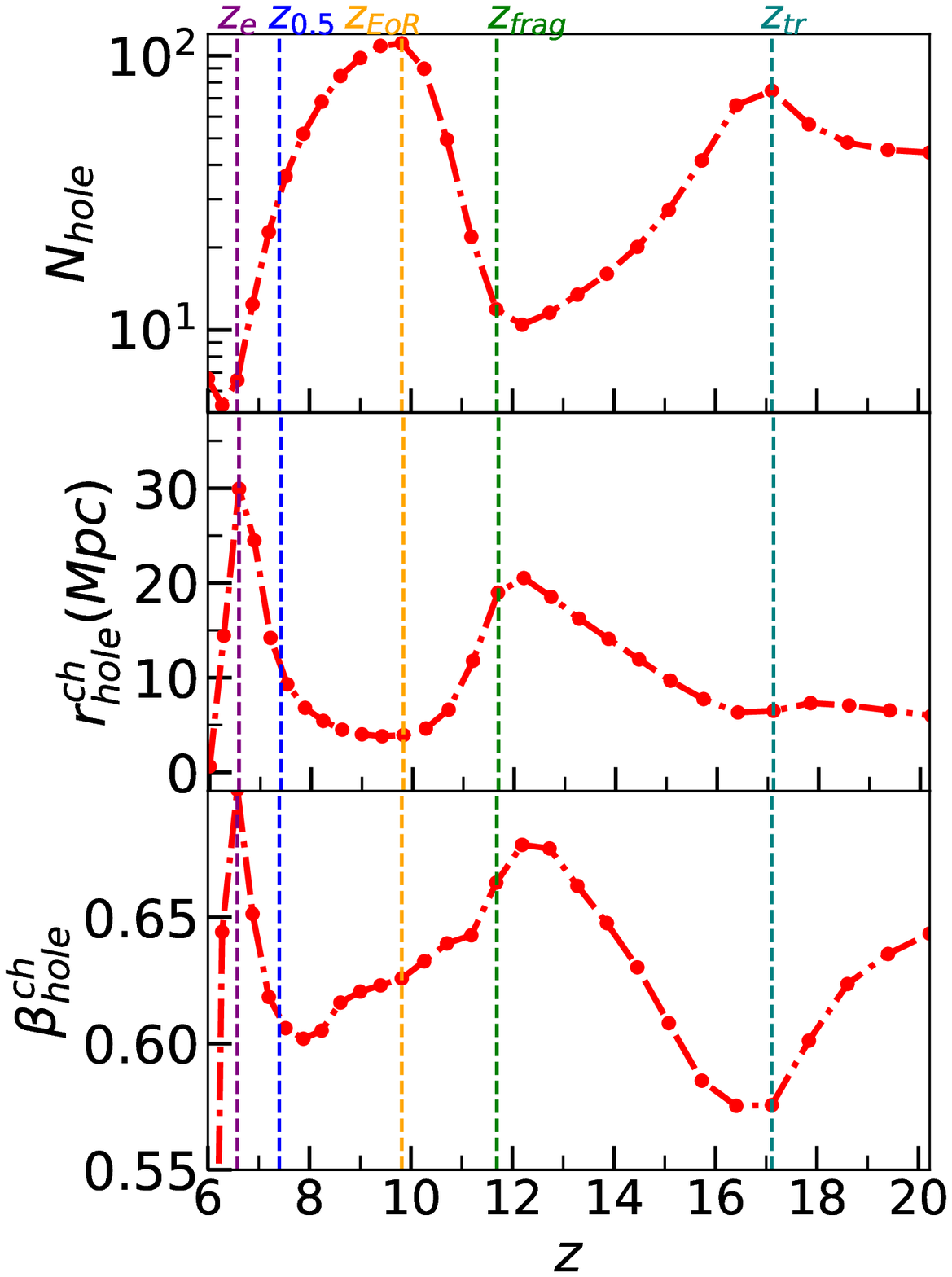}}}
		\resizebox{3.in}{3.8in}{{\includegraphics{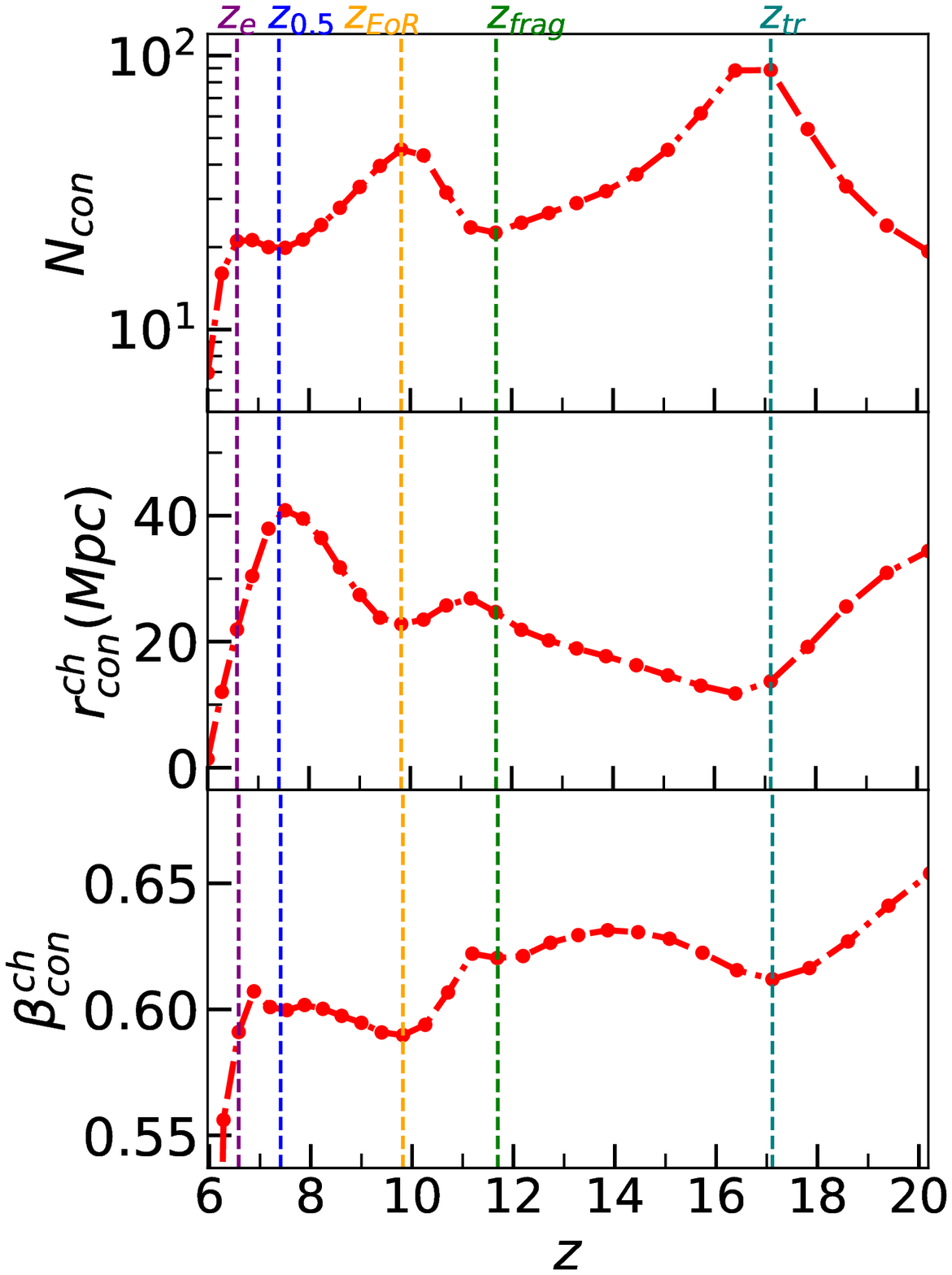}}}

	\end{center}
	\caption{The morphology of brightness temperature field $\delta {T_b}$, for \textit{fiducial model} with $\nu_{cut}=0$. The vertical lines (\textit{purple, blue and teal}) mark the transitions observed for the $x_{HI}$ field i.e. $z_{e}$, $z_{0.5}$ and $z_{frag}$. The redshift $z_{EoR}$ (\textit{orange}) marks the redshift below which the morphology of holes in the $\delta T_b$ field directly trace the morphology of holes in $x_{HI}$ field. The redshift $z_{tr}$ marks the epoch before which the $\delta T_b$ morphology is similar to $T_s$ morphology and is dominated by fluctuations in the Ly-$\alpha$ coupling.}
	\label{fig:Tb_z_morph}
\end{figure}

In Fig. {\ref{fig:Tb_z_morph}} we show the redshift evolution of the morphology of the brightness temperature field for the range of redshifts from $z=20.22$ to $z=6$ for our \textit{fiducial model}. We also mark the redshifts where the transition epochs were observed for the evolution of $x_{HI}$ morphology. We observe two more transition points and name them as $z_{EoR}$ and $z_{tr}$. The redshift $z_{EoR}$, is where the redshift evolution of the morphology of holes in the brightness temperature field is similar to those in the $x_{HI}$ field to $10 \%$ (elaborated further in the later part of the section). The redshift, $z_{tr}$ is where the morphology of the brightness temperature field transitions from being similar to the morphology of $T_s$ field to a regime where the morphology is an interplay between the morphology of $T_s$ and $x_{HI}$ fields.

In Fig.{\ref{fig:Tb_z_morph}} we focus on interpreting the evolution of $\delta T_b$ morphology for the \textit{fiducial model} to identify the transition redshifts mentioned above. Comparison of different models will be carried out later in the section. The fluctuations in  $\delta T_b$ arise from a product of fluctuations in $x_{HI}$, $(1+\delta_{nl})$, and $(1-T_{\gamma}/T_s)$. It is not straightforward to interpret the individual contributions. We can identify roughly three regimes from Fig.~\ref{fig:Tb_z_morph}. 
\begin{figure}[h]
	\begin{center}
		
		\resizebox{1.8in}{2.0in}{{\includegraphics{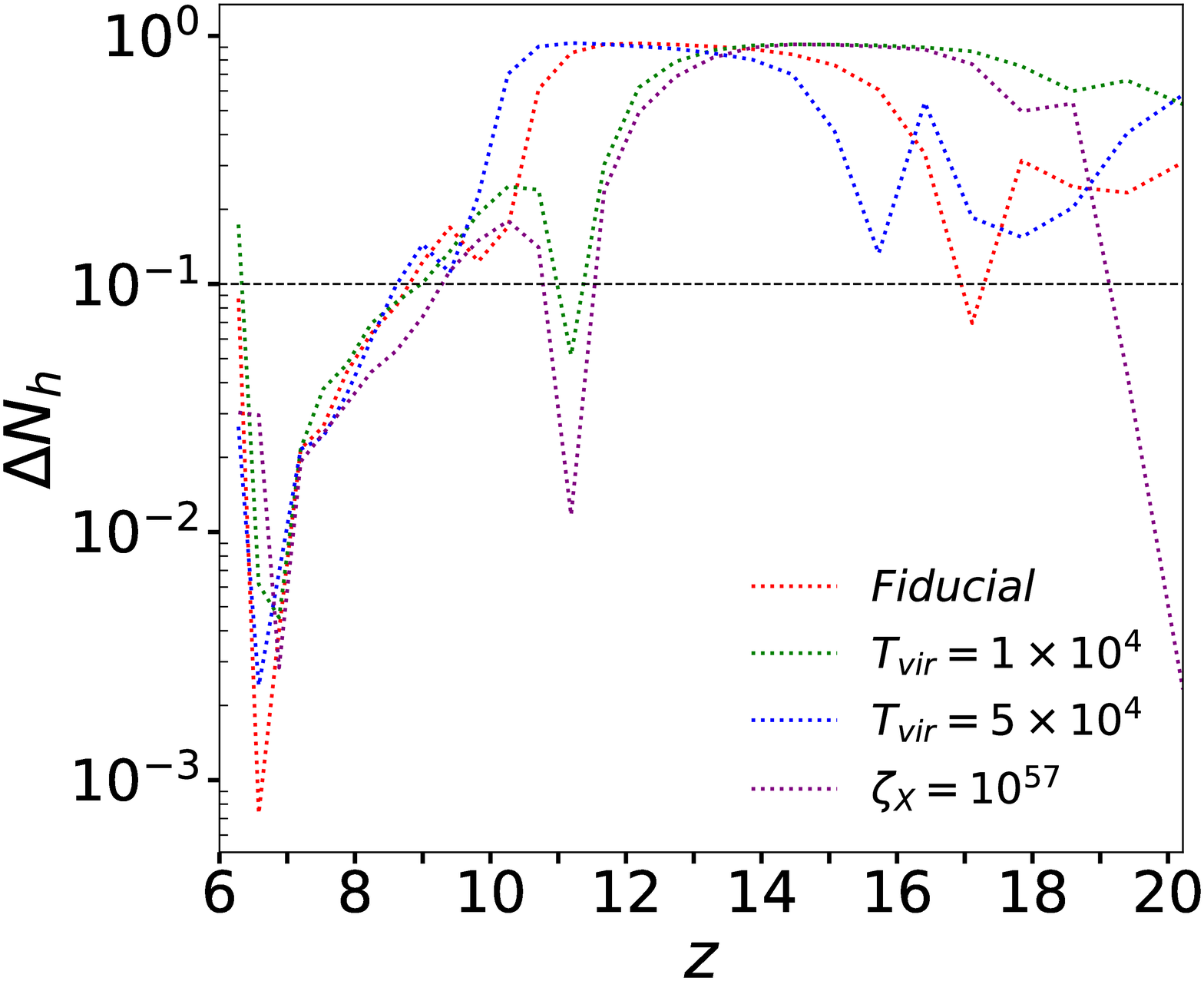}}}
		\resizebox{1.8in}{2.0in}{{\includegraphics{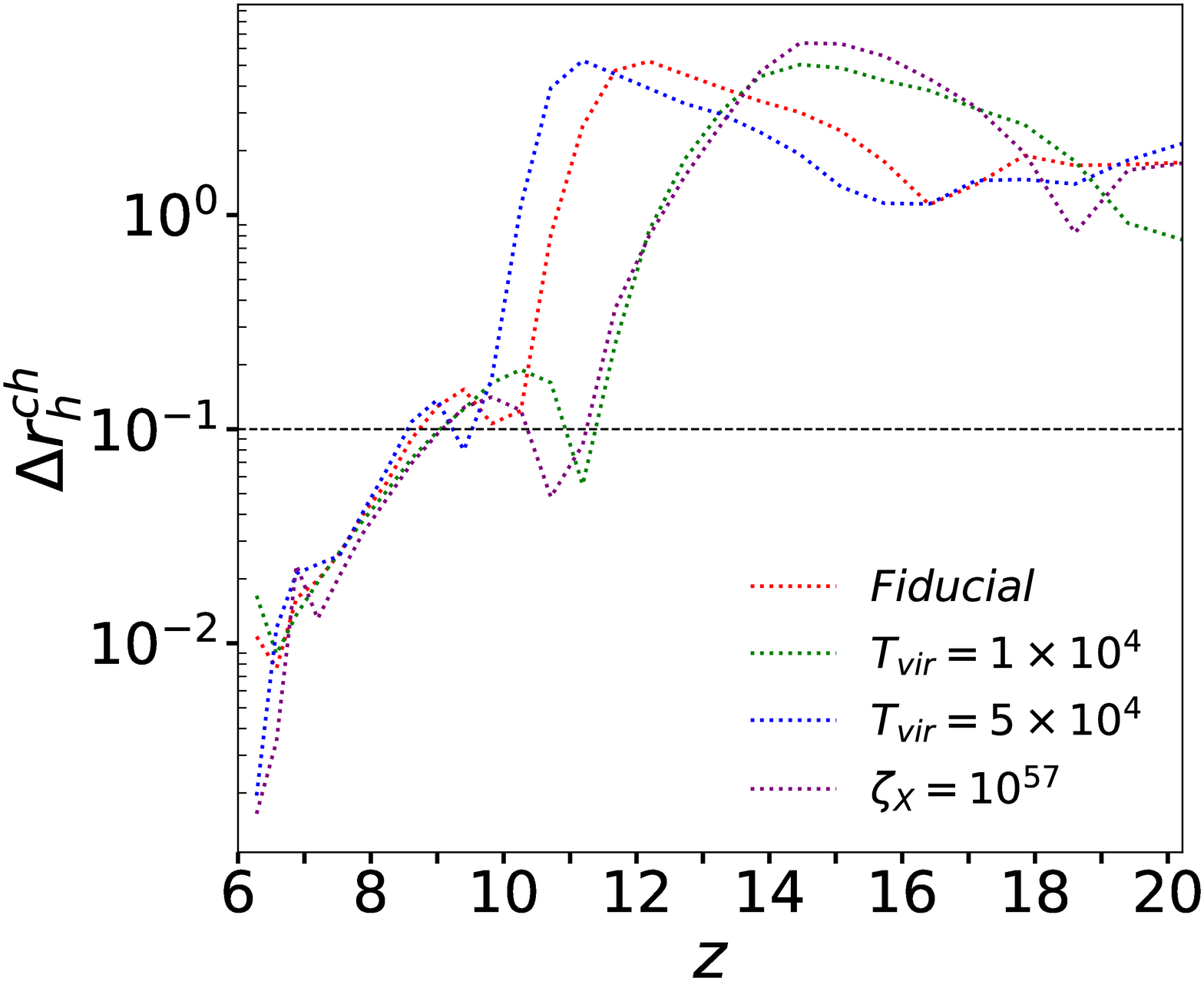}}}
		\resizebox{1.8in}{2.0in}{{\includegraphics{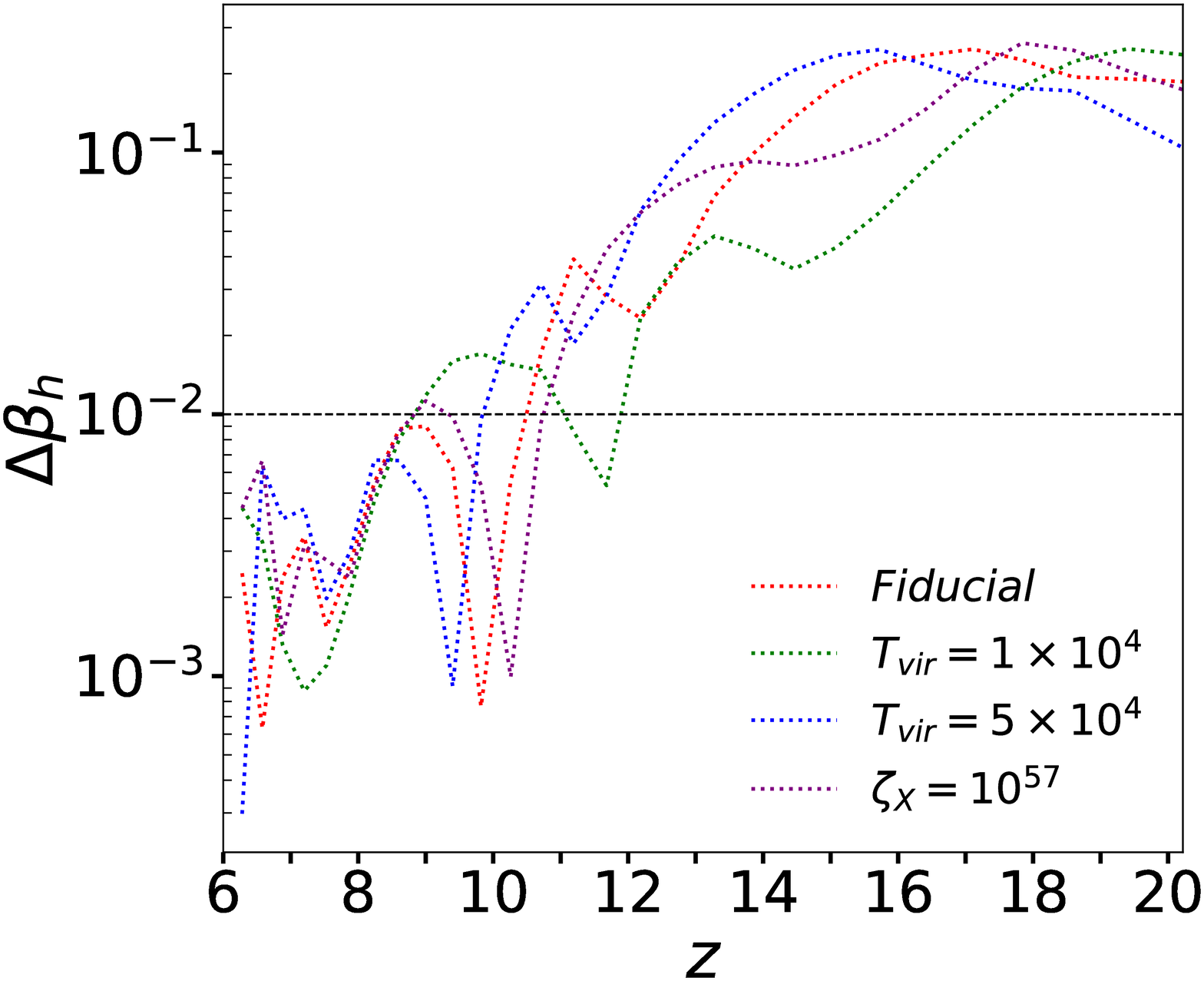}}}	
		\caption{The fractional difference for $N_{hole}$, $r^{ch}_h$ and $\beta^{ch}_{hole}$ between $\delta T_b$ and $x_{HI}$ relative to $x_{HI}$ for all the models under consideration. The horizontal line in the left and middle panel marks the point where the differences $\Delta N_{hole}$ and $\Delta r^{ch}_{hole}$ are $10\%$ respectively, while the horizontal line on the right panel marks the point where $\Delta \beta^{ch}_{hole}$ is $1\%$.}
		\label{fig:frac_diff}
	\end{center}
\end{figure}

\begin{itemize}
\item \textbf{Regime 1}: High redshift $z \gtrsim z_{tr}$ \\

As described in Sec.~\ref{sec:Ts} this is the regime where the fluctuations in $T_s$ is dominated by fluctuations in Ly-$\alpha$ coupling described by $x_{\alpha}$. The regions where coupling is more efficient correspond to valleys in $T_s$ and are high density peaks in the density field. However the density fluctuations are smaller scale fluctuations (see $r^{ch}_{con,hole}$ in Fig.~\ref{fig:delta_z}) compared to the fluctuations in $T_s$ (see $r^{ch}_{con,hole}$ in Fig.~\ref{fig:Ts_z_morph2}) and do not evolve much with redshift.  In this regime one can ignore $x_{HI}$ in the product $x_{HI}(1+\delta_{nl})(1-T_{\gamma}/T_s)$ because nearly all of the IGM is neutral and $\overline{x}_{HI} \sim 1$. Therefore the morphology of the fluctuations of  $\delta T_b$ is an interplay between the fluctuations of $(1+\delta_{nl})$ and  $(1-T_{\gamma}/T_s)$. Since fluctuations of $\delta_{nl}$ and hence that of $(1+\delta_{nl})$, do not show much variation with redshift, any evolution in the fluctuations of $(1-T_{\gamma}/T_s)$ will determine the evolution of $\delta T_b$ fluctuations, however the morphology will be affected by both. The difference in the scales of fluctuations and the fact that the $x_{\alpha}$ fluctuations are anti-correlated with those in the $\delta_{nl}$ field will reduce the overall value of $N_{con,hole}$ below that of $\delta_{nl}$ but more than that for $T_s$ (also notice that the fluctuations in $T_s$ in this regime have small variance as seen in Fig.~\ref{fig:Ts_avg_sd}). The numbers are however closer to the values for the $T_s$ field. The increase in both $N_{con}$ and $N_{hole}$ with redshift is due to the corresponding evolution in the values for $T_s$ field and has been described in Sec.~\ref{sec:Ts}. Therefore the morphology of $\delta T_b$ in this regime is dominated by that of $T_s$ field, more specifically by the Ly-$\alpha$ fluctuations. This is further corroborated by the plots for $r^{ch}_{con,hole}$ in Fig.~\ref{fig:Tb_z_morph}, where the values are similar to those for the $T_s$ field. The shape of $\beta_{con,hole}$, shows an initial decrease which is not straightforward to understand as both $\delta$ and $T_s$ field dominate in this regime. It is interesting to note that the dip in $\beta_{con,hole}$ corresponds to the peak in the $N_{con,hole}$ plots.

\item \textbf{Regime 2}: Intermediate redshift $z \lesssim z_{tr} $ and $z\gtrsim z_{EoR}$ \\

This is the regime where no single field is expected to dominate the morphology. This is a phase where the $\delta T_b$ morphology will transition from that which is determined completely by $T_s$ to the one which is determined completely by $x_{HI}$. Therefore, within this transition period one would expect that the morphology of $\delta T_b$ would go from a period where $T_s$ dominates more than $x_{HI}$ to a period where $x_{HI}$ dominates more than $T_s$.

Initially, for $z \gtrsim z_{frag}$ the morphology is dominated by $T_s$ but determined by a combination of fluctuations in X-ray heating and Ly-$\alpha$. In this regime Ly-$\alpha$ coupling is approaching saturation while several X-ray efficient sources start to appear. These correspond to highest peaks in the $T_s$ field and positively correlate with the density field. This erases the smaller scale fluctuations in the field caused by the density field because X-rays have high mean free path and X-ray heated peaks in $T_s$ are much higher than the very slowly evolving $\delta_{nl}$ peaks. Therefore we see a decrease in the number of structures $N_{con,hole}$ and a corresponding decrease in the size of holes and connected regions, $r^{ch}_{con,hole}$. Scattered ionized regions also start appearing at these $z$ values. These would correspond to holes in the $\delta T_b$ field. Therefore the morphology of holes in this regime is expected to be a combination of that of $T_s$ and $x_{HI}$ field. The number of holes, $N_{hole}$ for $x_{HI}$ is more than that for $T_s$. However the redshift evolution is closer to that for $T_s$ than that of $x_{HI}$ because the holes corresponding to the $x_{HI}$ field correspond to very small regions in the $T_b$ field and are fewer in number at these redshifts. Both $\beta_{con,hole}$ show an increase till $z_{frag}$. This increase is a trend observed in the $T_s$ field at these redshifts.  Since most of the region is a single connected neutral region, the morphology of connected regions for $\delta T_b$ is dominated by the $T_s$ field in such regions. Thus in this regime both $T_s$ and $x_{HI}$ affect the morphology but it is the evolution of $T_s$ morphology that is marginally dominant. 

At lower $z$ values, i.e. $z_{EoR} \lesssim z \lesssim z_{frag}$ the morphology is dominated by the morphology of the $x_{HI}$ and $1-T_{\gamma}/T_s$ field, but the evolution is dominated more by the morphology of $x_{HI}$ field. The increase in the number of small ionized regions and the fact that X-ray heating is saturated in most of the IGM leads to an increase in $N_{hole}$. Since the ultra violet radiation capable of ionizing neutral hydrogen has lower mean free path than X-rays, such numerous regions are smaller in size and would appear in hottest regions of the IGM. This is reflected in the decrease in the average size of holes, $r^{ch}_{hole}$. Post $z_{frag}$ there is an increase in the number of connected regions $N_{con}$ and a corresponding decrease in $r^{ch}_{con}$.  Post $z_{frag}$ the trend in the variation of $\beta_{con,hole}$ begins to transition to that towards $x_{HI}$ dominating over $T_s$ as described above. Note that this entire regime is a regime of transition.\\
	
	\item \textbf{Regime 3}: Low redshift $z < z_{EoR}$ \\
	
     This is the regime where the morphology of ionized regions is directly manifested in the morphology of the brightness temperature field $\delta T_b$. In Fig.~\ref{fig:frac_diff} we plot the fractional differences between the two fields as a function of redshift. The fractional difference $\Delta F=\frac{F_{\delta T_b} - F_{x_{HI}}}{F_{x_{HI}}}$, where F is the quantity of interest for holes, i.e. $N_{hole}$, $r^{ch}_{hole}$ and $\beta^{ch}_{hole}$. The reason why we compare only for holes is because once ionization begins, holes give a more physical picture as ionized regions in the morphology of brightness temperature field. Any fully ionized region would appear as a hole in the brightness temperature field excursion set. We define $z_{EoR}$ to be the lowest redshift where $\Delta r^{ch}_h$ is $10\%$. We observe that thereafter the difference decreases with decreasing $z$. For this choice of $z_{EoR}$, $\Delta \beta^{ch}_{hole}$ is always below $1 \%$. Therefore one can infer $z_{0.5}$ and $z_e$ to good accuracy from the $\delta T_b$ morphology. We also find that at $z=z_{0.5}$, $N_{con} \simeq N_{hole}$ for $\delta T_b$.
	  
\end{itemize}
 The evolution of brightness temperature morphology for the different models relative to the \textit{fiducial model} is shown in Fig.~{\ref{fig:TbErr}. The shape of the redshift evolution of the $\delta T_b$ morphology as encoded in $N_{con,hole}$, $r^{ch}_{con,hole}$ and $\beta^{ch}_{con,hole}$ is similar for all models under consideration except for the shifts in the various transitions described above.  
 The results are summarized in Table \ref{table:Tb_ev}. From the table we see that the shift in $z_{EoR}$ is a consequence of a general shift in the redshift at which EoR starts for these models and has been described in Sec.~\ref{Sec:xh}. The shifts in $z_{tr}$ can be traced to the differences in X-ray and Ly-$\alpha$ emmissivities for these models and has been described in detail in Sec.~\ref{sec:Ts}.  We observe that one model can be differentiated from another both from the morphology at a given redshift and from the shift in the transition  redshifts. The transition epoch $z=z_{EoR}$ can be obtained for different models. 
 \begin{table}[ht]
 	
 	\centering 
 	
 	\begin{tabular}{c c c c}
 		
 		\hline
 		\hline   
 		\\                     
 		Model & $z_{EoR}$& $\bar x_{HI}^{EoR}$& $z_{tr}$\\[0.6ex]
 		\hline
 		\\
 	 
     	Fiducial & $\sim 8.7$& $\sim$ 0.73 & $\sim 17.11$ \\
 		$T_{vir}=1\times 10^4 K$ &$\sim 9.1$& $\sim$ 0.71 &$\sim 19.4 $\\
 		$T_{vir}=5\times 10^4 K$ & $\sim 8.6$&$\sim$0.77& $\sim 15.7 $\\
 		$\zeta_X=1 \times 10^{57}$& $ \sim 9.12$&$\sim$0.77 &$\sim 18.6$\\
 	
 		\\
 		\hline
 		
 	\end{tabular}
 	\caption{The redshift $z_{EoR}$ below which the difference between $\delta T_b$ and $x_{HI}$ morphologies defined in terms of $\Delta r^{ch}_h$ $<10\%$ for different models. The last column shows the corresponding $\overline{x}_{HI}$ values at $z_{EoR}$.} 
 	\label{table:Tb_ev}
 	
 \end{table}
 \begin{figure}[h]
 	\begin{center}
 		\resizebox{2.8in}{2.8in}{{\includegraphics{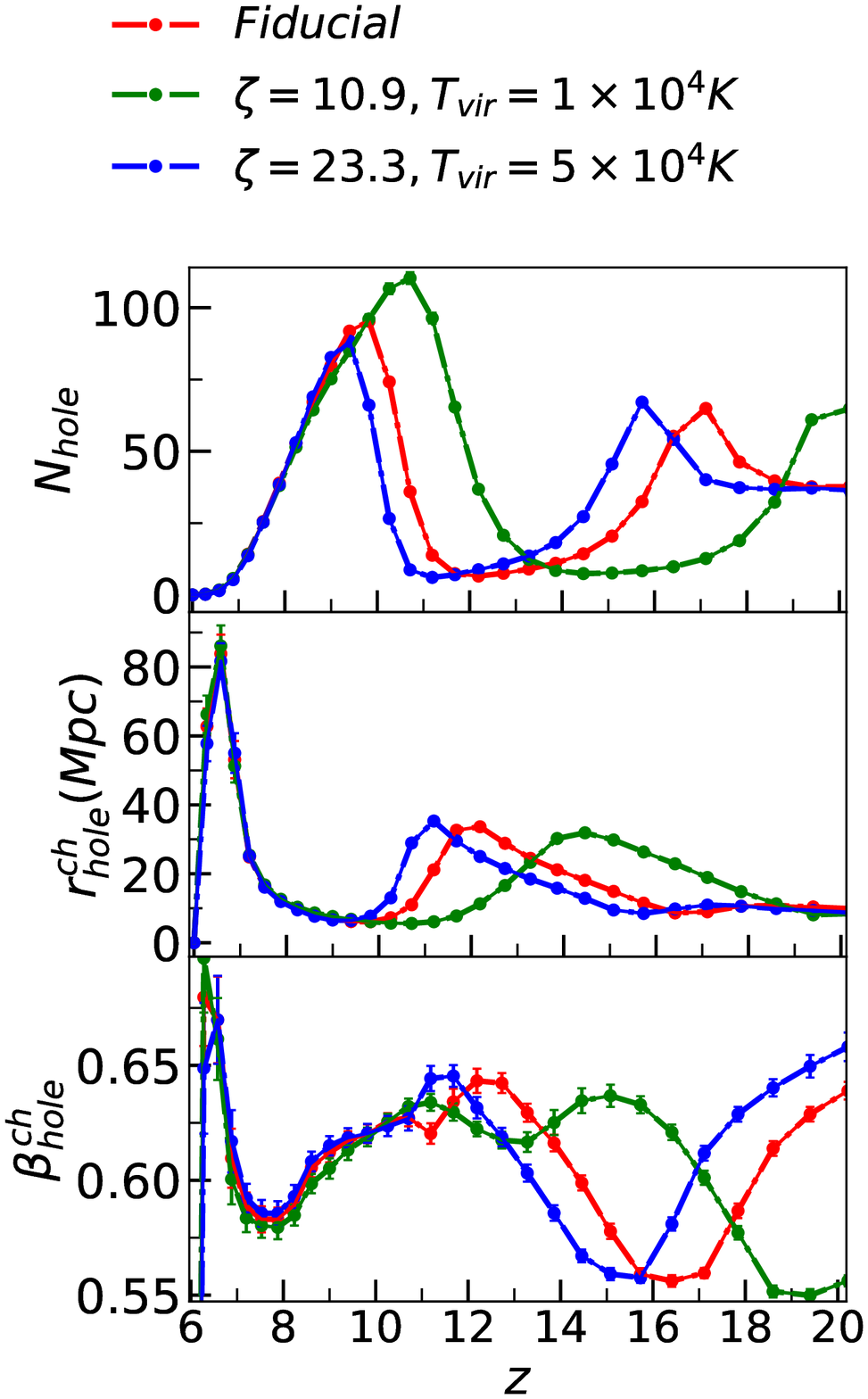}}}
 		\resizebox{2.8in}{2.8in}{{\includegraphics{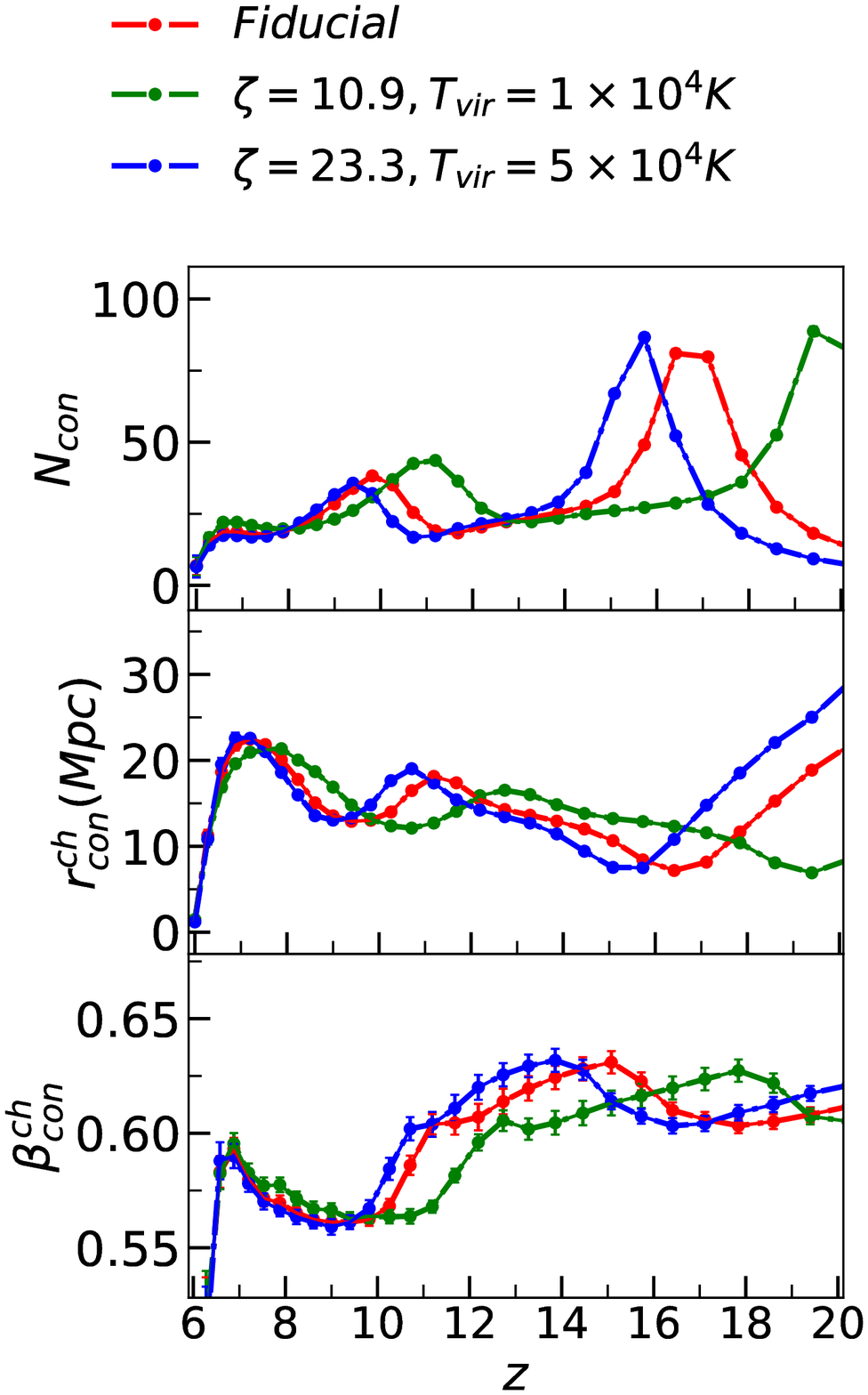}}}\\ 
 		\resizebox{2.8in}{2.8in}{{\includegraphics{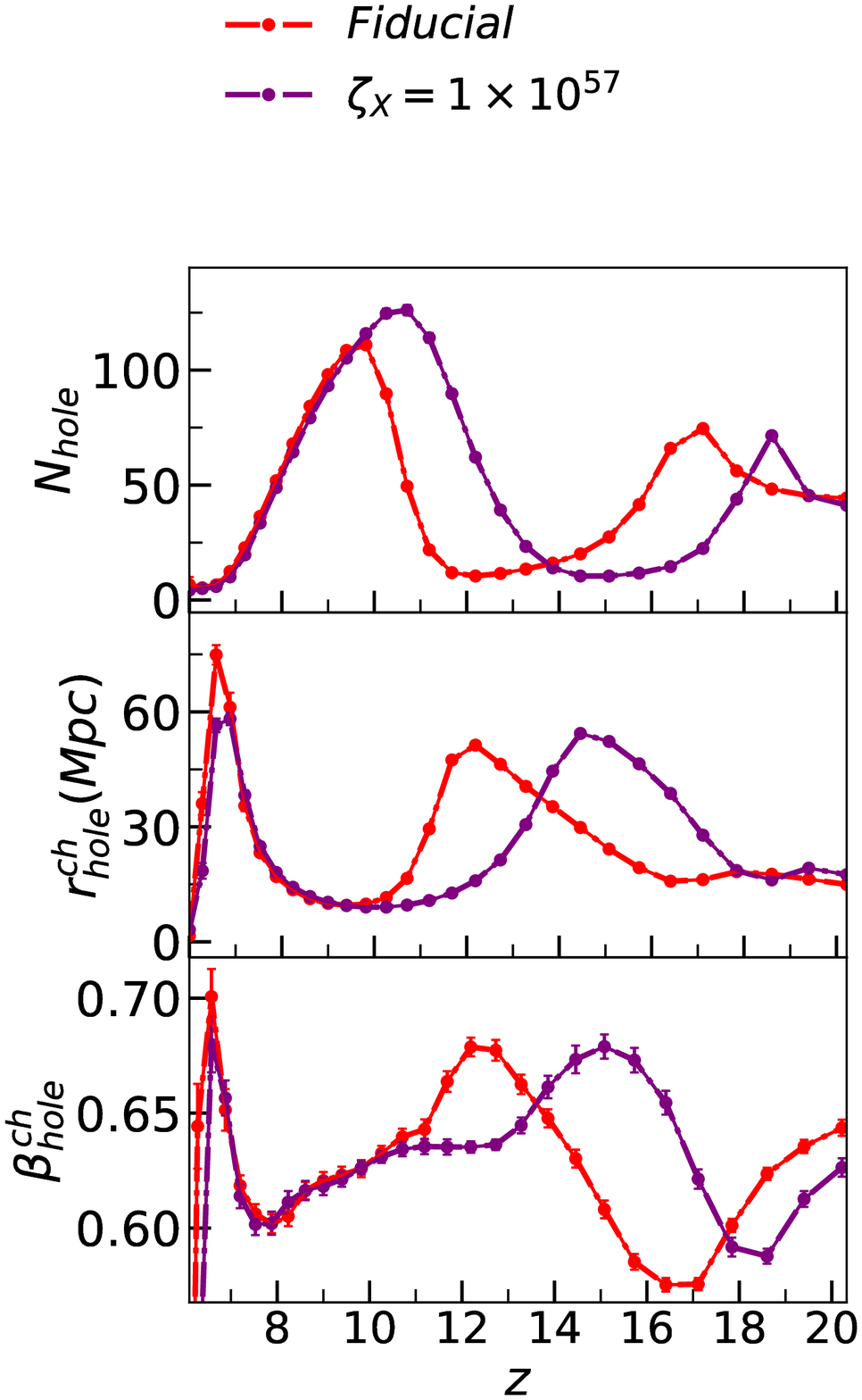}}}
 		\resizebox{2.8in}{2.8in}{{\includegraphics{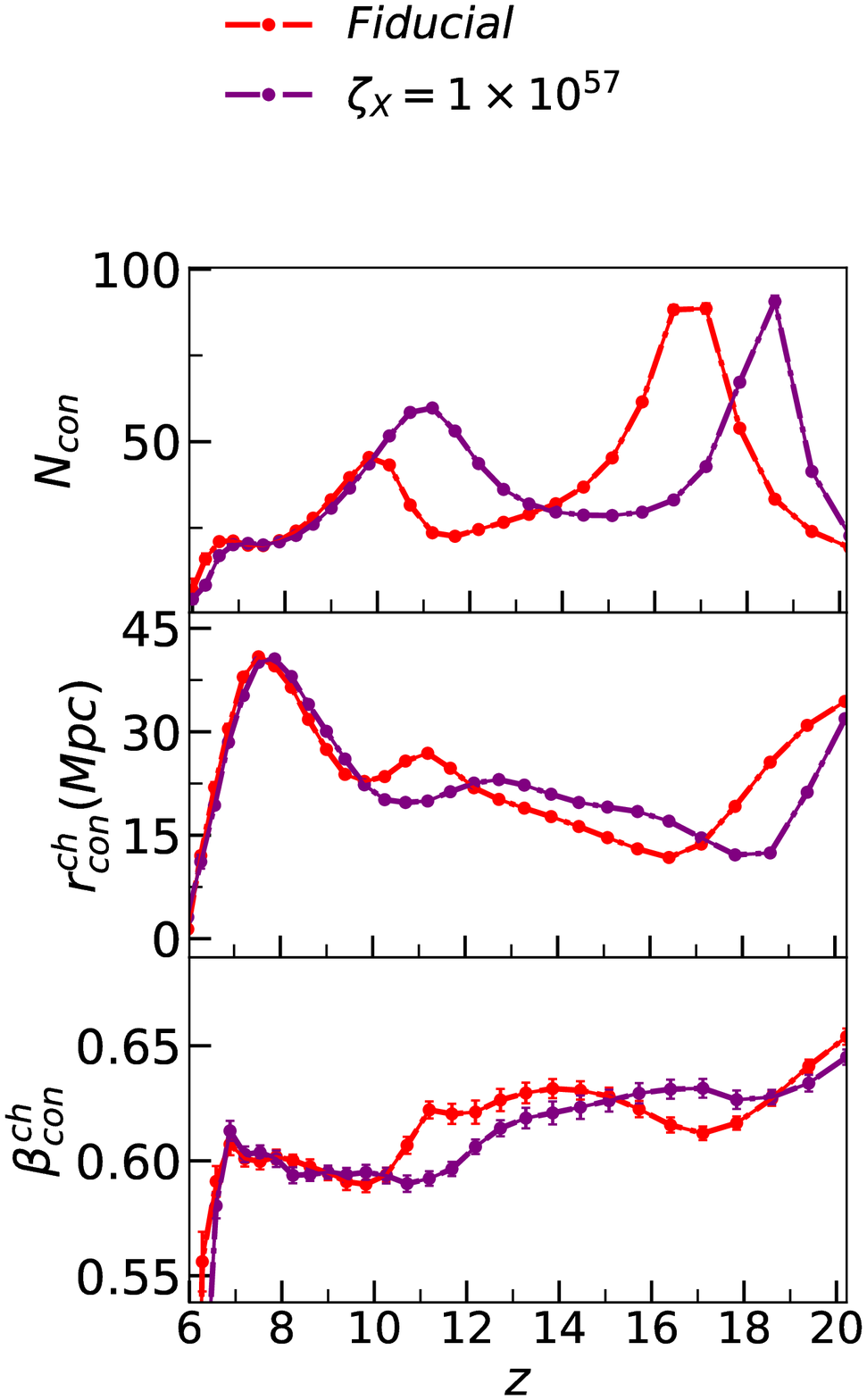}}}
 			
 		\caption{The morphology of $\delta T_b$ for models with different $T_{vir}$ values (\textit{left}) and model with an increased X-ray efficiency (\textit{right}), relative to the \textit{fiducial model}. The error bars are calculated over 32 slices from our 200 Mpc box. Each slice has a thickness of 6.25 Mpc.}
 		\label{fig:TbErr}
 	\end{center}
 \end{figure}
\section{Conclusion and Discussion}
In this paper we have extended our previous work and described how the Contour Minkowski Tensor can be used to differentiate various models of ionization and heating history of the IGM after the first collapsed structures form. We have studied the morphological properties of the individual fields and identified important transition redshifts in their respective evolution. Further, we have shown how these transitions are reflected in the evolution of the morphology of the brightness temperature field. 
We studied the evolution of the $\delta_{nl}$ field to study it's effect on the brighntess temperature field. We find that the morphology of the density field does not show any marked evolution with redshift at the high redshifts probed, where it's effect on the brightness temperature fluctuations is expected to be more.

We studied the evolution of morphology of the $x_{HI}$ field for different models of reionization. We find different signatures of the models on the $x_{HI}$ field.  We found that less efficient sources with lower values of $T_{vir}$ have higher $N_{hole}$ and smaller $r^{ch}_{hole}$ as compared to the sources with higher $T_{vir}$. We also observe a shift in the value of the redshift of transitions, redshift of fragmenation ($z_{frag}$),the redshift of equality of Betti numbers (which occurs at $z=z_{0.5}$, where $\bar x_{HI}=0.5$) and redshift of end of progress of ionization ($z_e$) to higher redshift values relative to the models with higher $T_{vir}$. We also studied the effect of inhomogenous recombination in comparison to our\textit{fiducial model}. We find that recombination delays the different transition redshifts and introduces more anisotropy in growth of ionized regions. We also observe that the size of ionized regions is smaller in the case of less efficient sources and it varies with $T_{vir}$ in a monotonic but non linear fashion.

The evolution of the morphology of the $T_s$ field does not show any marked transition but shows a shift in the evolution to higher $z$ values for sources with higher X-ray emissivity. 

We have shown that the evolution of the brightness temperature captures the various transitions for the ionization field and spin temperature evolution. We identified three regimes in the evolution of the morphology of $\delta T_b$. The first regime for $z>z_{tr}$ is where the morphology of $T_s$ field determines the evolution. This is the regime where the evolution in fluctuations in the Ly-$\alpha$ coupling in combination with the fluctuations in $\delta_{nl}$ dominate the evolution of the morphology of $T_s$. The second regime for $z_{EoR}<z<z_{tr}$ is where the morphology of $\delta T_b$ is an interplay between the $T_s$ and $x_{HI}$ morphology. We observe a transition around $z_{frag}$ in the redshift evolution of the morphology of $\delta T_b$. 
In the third regime at $z<z_{EoR}$, the morphology of $\delta T_b$ is similar to that of $x_{HI}$ in terms of the morphology of holes for the respective fields. The morphology of the brightness temperature captures most of the ionization history below $z_{EoR}$}. For our \textit{fiducial model} the average neutral hydrogen fraction at $z_{EoR}$ is $x_{HI} \sim 0.8$. Therefore $z_e$, $z_{0.5}$ and $z_{frag}$ are captured by $\delta T_b$ morphology. 

The calculations in this paper show how the contour minkowski tensor be used to differentiate models of EoR in an ideal scenario where there is no foreground or instrumental noise. Our results are very encouraging for application of our method to future data of the brightness temperature to constrain models of the EoR. The next obvious step is to carry out realistic analysis by including instrumental effects and foreground contamination in the simulations
and obtain constraints on model parameters using Bayesian analysis. We plan to pursue this as a follow up work. Further we plan to carry out our analysis using the more exact numerical simulations of EoR.
\section*{Acknowledgment}
The computation required for this work was carried out on the Nova
cluster at the Indian Institute of Astrophysics. 
We acknowledge use of the \texttt{21cmFAST} code~\cite{Mesinger:2010ne}. The work of P.C. is supported by the Science and Engineering Research Board of the Department of Science and Technology, India, under the \texttt{MATRICS} scheme, with reference no. \texttt{MTR/2018/000896}.




%

	
\setcounter{section}{1}

\end{document}